\documentclass[12pt,a4paper]{article}
\title{On the Mourre estimates\\
for three-body Floquet Hamiltonians}
\author{Tadayoshi ADACHI\\
{\footnotesize Division of Mathematical and Information Sciences,}\\
{\footnotesize Graduate School of
Human and Environmental Studies,
Kyoto University}\\
{\footnotesize Yoshida-Nihonmatsu-cho, Sakyo-ku, Kyoto-shi, Kyoto 606-8501, Japan}}
\date{ }
\usepackage{amssymb}
\usepackage{amsmath}
\usepackage{amsthm}
\usepackage{mathrsfs}
\usepackage{color}
\usepackage{times}
\theoremstyle{plain}
\newtheorem{thm}{Theorem}[section]
\newtheorem{prop}[thm]{Proposition}
\newtheorem{lem}[thm]{Lemma}
\newtheorem{cor}[thm]{Corollary}
\theoremstyle{definition}

\newtheorem{rem}{Remark}[section]
\numberwithin{equation}{section}
\DeclareMathOperator*{\slim}{s-lim}
\allowdisplaybreaks
\begin{document}
\maketitle

\begin{abstract} In this paper,
we consider the Floquet Hamiltonian $K$ associated with
a three-body Schr\"odinger operator
with time-periodic pair potentials $H(t)$.
By introducing a conjugate operator $A$ for
$K$ in the standard Mourre theory,
we prove the Mourre estimate for $K$.
As by-products of the Mourre estimate for $K$,
the minimal velocity estimates for the physical propagator $U(t,0)$
generated by $H(t)$ as well as the propagator $e^{-i\sigma K}$ generated by $K$
can be obtained.
\end{abstract}

\section{Introduction}

In this paper, we consider a three-body quantum system
with time-periodic pair interactions.
Since we would like to introduce some notation in many body scattering
theory, we denote the number of particles in the system by
$N$ for a while. Of course, we mainly consider the case where $N=3$.
The system under consideration
is governed by the following Schr\"odinger operator with
time-periodic potentials
\begin{equation}
\tilde{H}(t)=\sum_{j=1}^N\left(-\frac{1}{2m_j}\Delta_j\right)+V(t),\quad
V(t)=\sum_{1\le j<k\le N}V_{jk}(t,r_j-r_k)
\label{1.1}
\end{equation}
acting on $L^2(\boldsymbol{R}^{d\times N})$,
where $m_j$ and $r_j\in\boldsymbol{R}^d$ are
the mass and position vector of the $j$-th particle, respectively,
$\Delta_j=\sum_{l=1}^d\partial_{r_{j,l}}^2$
is the Laplacian with
respect to $r_j$, and $V_{jk}(t,r_j-r_k)$'s are pair potentials.
We suppose that $V_{jk}(t,y)$'s are real-valued functions
on $\boldsymbol{R}\times\boldsymbol{R}^d$
which are periodic in $t$ with a period $T>0$:
\begin{equation}
V_{jk}(t+T,y)=V_{jk}(t,y),\quad(t,y)\in\boldsymbol{R}\times
\boldsymbol{R}^d.
\label{1.2}
\end{equation}
We would like to watch the motion of the system in the
center-of-mass frame. To this end, we will introduce the
following configuration spaces:
We equip $\boldsymbol{R}^{d\times N}$ with the metric
$r\cdot\tilde{r}=\sum_{j=1}^Nm_j\langle r_j,\tilde{r}_j
\rangle$; $r=(r_1,\ldots,r_N),\,
\tilde{r}=(\tilde{r}_1,\ldots,\tilde{r}_N)
\in\boldsymbol{R}^{d\times N}$,
where $\langle\cdot,\cdot\rangle$ is the standard inner product
on $\boldsymbol{R}^d$.
We will denote this $\boldsymbol{R}^{d\times N}$ by $\bar{X}$.
We usually write $r\cdot r$
as $r^2$. We put $|r|=\sqrt{r^2}$.
$\tilde{H}(t)$ can be written as
$$\tilde{H}(t)=-\frac{1}{2}\Delta_{\bar{X}}+V(t)=\frac{1}{2}(p_{\bar{X}})^2+V(t)$$
acting on $L^2(\bar{X})$, where $\Delta_{\bar{X}}$
is the Laplace-Beltrami operator on $\bar{X}$ and $p_{\bar{X}}$ is
the velocity operator
$p_{\bar{X}}=(p_1/m_1,\ldots,p_N/m_N)=(-i\nabla_{r_1}/m_1,\ldots,
-i\nabla_{r_N}/m_N)$ on $\bar{X}$.
From now on, the kinetic energies will
be represented by the velocity operators, for simplicity.
We define two subspaces $X$ and $X_\mathrm{cm}$ of $\bar{X}$
as
$$X=\left\{r\in\bar{X}\,\middle|\,\sum_{j=1}^Nm_jr_j=0\right\},\quad
X_\mathrm{cm}=\left\{r\in\bar{X}\,\middle|\,r_1=\cdots=r_N\right\}.$$
Then $X$ and $X_\mathrm{cm}$ are perpendicular to each other,
and satisfy $\bar{X}=X\oplus X_\mathrm{cm}$.
$\pi:\bar{X}\to X$ and
$\pi_\mathrm{cm}:\bar{X}\to X_\mathrm{cm}$
denote the orthogonal projections
onto $X$ and $X_\mathrm{cm}$, respectively. We put
$x=\pi r$ and $x_\mathrm{cm}=\pi_\mathrm{cm}r$ for
$r\in\bar{X}$.
Now we introduce the time-dependent Hamiltonian
\begin{equation}
H(t)=\frac{1}{2}p^2+V(t)
\label{1.3}
\end{equation}
acting on $\mathscr{H}=L^2(X)$. Then $\tilde{H}(t)$ is represented
as
$$\tilde{H}(t)=H(t)\otimes\mathrm{Id}+\mathrm{Id}\otimes\left(
\frac{1}{2}(p_\mathrm{cm})^2\right)$$
on $L^2(\bar{X})=\mathscr{H}\otimes L^2(X_\mathrm{cm})$.
Here $p$ and $p_\mathrm{cm}$ are the velocity operators on $X$
and $X_\mathrm{cm}$, respectively.
We would like to study some scattering problems for this Hamiltonian $H(t)$
with $N=3$.

A non-empty subset of the set $\{1,\ldots,N\}$ is called
a cluster. Let $C_j$, $1\le j\le m$, be clusters.
If $\cup_{1\le j\le m}C_j=\{1,\ldots,N\}$ and
$C_j\cap C_k=\emptyset$ for $1\le j<k\le m$,
$a=\{C_1,\ldots,C_m\}$ is called a cluster decomposition.
$\#(a)$ denotes the number of clusters in $a$.
Let $\mathscr{A}$ be the set of all cluster decompositions. Suppose $a$,
$b\in\mathscr{A}$. If $b$ is obtained as a refinement of $a$,
that is, if each cluster in $b$ is a subset of a cluster in $a$,
we say $b\subset a$, and
its negation is denoted by $b\not\subset a$. Any $a$ is
regarded as a refinement of itself.
The one and $N$-cluster decompositions are denoted
by $a_\mathrm{max}$ and $a_\mathrm{min}$, respectively.
For the sake of brevity, we write
$$\mathscr{A}_0=\mathscr{A}\setminus\{a_\mathrm{max}\},\,\,\,
\mathscr{A}^0=\mathscr{A}\setminus\{a_\mathrm{min}\},\,\,\,
\mathscr{A}_0^0=\mathscr{A}\setminus\{a_\mathrm{max},a_\mathrm{min}\}=
\mathscr{A}_0\cap\mathscr{A}^0.$$
The pair $(j,k)$ is identified with the $(N-1)$-cluster
decomposition
$\{(j,k),(1),\ldots,$
$(\widehat{j}),\ldots,(\widehat{k}),\ldots,(N)\}$.
If $N=3$, then $\{(1,2),\,(1,3),\,(2,3)\}$ is the set of all two-cluster
decompositions, and is equal to $\mathscr{A}_0^0$.

Let $a\in\mathscr{A}$. We introduce two subspaces $X^a$ and $X_a$ of
$X$:
\begin{align*}
&X^a=\left\{r\in X\,\middle|\,\sum_{j\in C}m_jr_j=0\ \text{ for each cluster }
\ C\ \text{ in }\ a\right\},\\
&X_a=\left\{r\in X\,\middle|\,r_j=r_k\ \text{ for each pair }\ (j,k)\subset a\right\}.
\end{align*}
$\pi^a:X\to X^a$ and
$\pi_a:X\to X_a$ denote the orthogonal projections onto $X^a$ and $X_a$,
respectively. We put $x^a=\pi^a\,x$ and $x_a=\pi_a\,x$ for $x\in X$.
Since $X^{(j,k)}$ is identified with the
configuration space for the relative position of $j$-th and $k$-th
particles, one can put
$$V^{(j,k)}(t,x^{(j,k)})=V_{jk}(t,r_j-r_k).$$
As for $\{X^a\}_{a\in\mathscr{A}}$,
$X^a\subset X^b$ is equivalent to $a\subset b$.
On the other hand, as for $\{X_a\}_{a\in\mathscr{A}}$,
$X_a\subset X_b$ is equivalent to $a\supset b$.
For $a,\,b\in\mathscr{A}$,
$a\cup b$ stands for the smallest cluster decomposition $c\in\mathscr{A}$
with $a\subset c$ and $b\subset c$. Then we see that
$X^a+X^b=X^{a\cup b}$ and
$X_a\cap X_b=X_{a\cup b}$ hold.
We now define the cluster Hamiltonian
$$H_a(t)=\frac{1}{2}p^2+V^a(t),\quad
V^a(t)=\sum_{(j,k)\subset a}V^{(j,k)}(t,x^{(j,k)}),$$ 
which governs the motion of the system broken into non-interacting
clusters of particles.
Then $H_a(t)$ is represented as
$$H_a(t)=H^a(t)\otimes \mathrm{Id}+\mathrm{Id}\otimes\left(\frac{1}{2}(p_a)^2\right);\quad
H^a(t)=\frac{1}{2}(p^a)^2+V^a(t)$$
on $\mathscr{H}=\mathscr{H}^a\otimes \mathscr{H}_a=L^2(X^a)\otimes
L^2(X_a)$, where
$p^a$ and $p_a$ are the velocity operators on $X^a$
and $X_a$, respectively.
The intercluster potential $I_a(t)$ is given by
$$I_a(t,x)=V(t,x)-V^a(t,x)=\sum_{(j,k)\not\subset a}V^{(j,k)}(t,x^{(j,k)}).$$

Under some suitable conditions
on $V_{jk}(t)$, the existence and uniqueness of
the unitary propagator $U(t,s)$ generated by $H(t)$
can be guaranteed, even if $N\ge3$ (see e.g. Yajima~\cite{Ya2,Ya3}).
In the study of the asymptotic behavior of
$U(t,s)\phi$, $\phi\in\mathscr{H}$,
as $t\to\pm\infty$, we will frequently
utilize the so-called Floquet Hamiltonian
$K$ associated with $H(t)$ (see e.g. Howland~\cite{Ho1,Ho2},
Yajima~\cite{Ya1}): Let
$\boldsymbol{T}=\boldsymbol{R}/(T\boldsymbol{Z})$ be the torus.
Set $\mathscr{K}=L^2(\boldsymbol{T};\mathscr{H})\cong
L^2(\boldsymbol{T})\otimes \mathscr{H}$,
and introduce a strongly continuous one-parameter unitary group
$\{\hat{U}(\sigma)\}_{{\sigma\in\boldsymbol{R}}}$
on $\mathscr{K}$ given by
\begin{equation}
(\hat{U}(\sigma)\Phi)(t)=U(t,t-\sigma)\Phi(t-\sigma)
\label{1.4}
\end{equation}
for $\Phi\in\mathscr{K}$. By virtue of Stone's theorem,
$\hat{U}(\sigma)$ is written as
\begin{equation}
\hat{U}(\sigma)=e^{-i\sigma K}
\label{1.5}
\end{equation}
with a unique self-adjoint operator $K$ on $\mathscr{K}$.
$K$ is called the Floquet Hamiltonian associated with $H(t)$,
and is equal to the natural self-adjoint realization of
$-i\partial_t+H(t)$.
Here we denote by $D_t$ the operator $-i\partial_t$
with domain $AC(\boldsymbol{T})$, which is the
space of absolutely continuous functions on $\boldsymbol{T}$
with their derivatives being square integrable (following
the notation in Reed-Simon~\cite{RS}).
As is well-known, $D_t$ is self-adjoint on $L^2(\boldsymbol{T})$,
and its spectrum $\sigma(D_t)$ is equal to $\omega\boldsymbol{Z}$
with $\omega=2\pi/T$. For $a\in\mathscr{A}$, we also
introduce the cluster Floquet Hamiltonian $K_a$ associated with
the cluster Hamiltonian $H_a(t)$ by
$K_a=D_t+H_a(t)$
on $\mathscr{K}$. In particular, we have
$K_{a_\mathrm{max}}=K$. We also write $K_{a_\mathrm{min}}$ as $K_0=D_t+H_0$,
which is the free Floquet Hamiltonian associated with
the free Hamiltonian $H_0=p^2/2$ acting on $\mathscr{H}$,
for the sake of simplicity. Moreover, we introduce
the subsystem Floquet Hamiltonian $K^a$ associated with
the subsystem Hamiltonian $H^a(t)$ by
$K^a=D_t+H^a(t)$
on $\mathscr{K}^a=L^2(\boldsymbol{T};\mathscr{H}^a)\cong
L^2(\boldsymbol{T})\otimes\mathscr{H}^a$.
In particular, we see that
$K^{a_\mathrm{max}}=K$, $K^{a_\mathrm{min}}=D_t$, and
for $a\in\mathscr{A}$, $K_a$ is represented as
$$K_a=K^a\otimes\mathrm{Id}+\mathrm{Id}\otimes
\left(\frac{1}{2}(p_a)^2\right)$$
on $\mathscr{K}=\mathscr{K}^a\otimes\mathscr{H}_a=
L^2(\boldsymbol{T};\mathscr{H}^a)\otimes\mathscr{H}_a$.

If $H(t)$ is strictly time-dependent,
the lack of energy conservation becomes a barrier in the study of
the asymptotic behavior of $U(t,s)\phi$.
Howland~\cite{Ho1} proposed the stationary
scattering theory for time-dependent Hamiltonians,
by introducing a new Hamiltonian $-i\partial_t+H(t)$
acting on $L^2(\boldsymbol{R};\mathscr{H})$. 
As mentioned above, its
formulation was the quantum analogue to the procedure
in the classical mechanics in order to \textit{recover} the
conservation of energy. Yajima~\cite{Ya1} applied this Howland method
to the two-body quantum system with a time-periodic
short-range potential, and studied the problem of the asymptotic
completeness of the wave operators
$$W^\pm(s)=\slim_{t\to\pm\infty}U(t,s)^*e^{-i(t-s)H_0}$$
(see also Howland~\cite{Ho2}). In fact, he proved the asymptotic
completeness of 
$$\mathscr{W}^\pm=\slim_{\sigma\to\pm\infty}e^{i\sigma K}e^{-i\sigma K_0}$$
firstly, and deduced that of $W^\pm(s)$ from this result.
Such a method is called the Howland-Yajima method.
This method, together with the Faddeev method, was applied to
the three-body case under some \textit{very short-range} conditions
by Korotyaev~\cite{Ko} and Nakamura~\cite{N} later.
However, even in the case where $N=3$, the problem of
the asymptotic completeness under a general short-range
condition on pair potentials $V^{(j,k)}(t,x^{(j,k)})$
has not been solved yet, unlike in the case of
time-independent many body Schr\"odinger operators
(see e.g. Sigal-Soffer~\cite{SS1}, Graf~\cite{Gr}, Yafaev~\cite{Yaf},
Derezi\'nski~\cite{D}, Derezi\'nski-G\'erard~\cite{DG}, and so on),
as far as we know. Thus it is worth
continuing the study of the spectral and scattering theory for
$N$-body Floquet Hamiltonians, even in the case where $N=3$.

Here we recall known results in the case where $N=2$.
Yokoyama~\cite{Yo} introduced the self-adjoint operator
\begin{equation}
\bar{\bar{A}}_{0,1}=\frac{1}{2}\{x\cdot p\langle p\rangle^{-2}+\langle p\rangle^{-2}p\cdot x\};\quad \langle p\rangle^{-2}=(1+p^2)^{-1},
\label{1.06}
\end{equation}
on $\mathscr{K}$ as a conjugate operator for $K$.
For the sake of brevity, for an operator $T$, we will use the notation
$\mathrm{Re}\,T$,
which is defined by
$\mathrm{Re}\,T=(T+T^*)/2$.
Then $\bar{\bar{A}}_{0,1}$ can be written as
$\mathrm{Re}\,(x\cdot p\langle p\rangle^{-2})$.
Roughly speaking, $\bar{\bar{A}}_{0,1}$ is defined
by multiplying the generator of dilations
\begin{equation}
\hat{A}_0=\frac{1}{2}(x\cdot p+p\cdot x)=\mathrm{Re}\,(x\cdot p)
\label{1.07}
\end{equation}
and the resolvent $\langle p\rangle^{-2}$ of $p^2$.
He established the following Mourre estimate
under some suitable conditions
on $V$, by using the commutation relation
$i[K_0,\bar{\bar{A}}_{0,1}]=p^2\langle p\rangle^{-2}=p^2(1+p^2)^{-1}$
by simple calculation.
Put
$$d_0(\lambda)=\mathrm{dist}(\lambda,\omega\boldsymbol{Z}),\quad
d_1(\lambda)=\mathrm{dist}(\lambda,\omega\boldsymbol{Z}\cap(-\infty,\lambda])$$
for $\lambda\in\boldsymbol{R}$.
Here we note that $\omega\boldsymbol{Z}=\sigma(D_t)$ is equal to
$\varTheta=\sigma(K^{a_\mathrm{min}})$, which is the threshold set for $K$ with $N=2$.
Suppose $\lambda_0\in\boldsymbol{R}\setminus\omega\boldsymbol{Z}$,
that is, $\lambda_0$ is a non-threshold energy, and $0<\delta<\mathrm{dist}(\lambda_0,\omega\boldsymbol{Z})=d_0(\lambda_0)$.
Then, for any $f_{\lambda_0,\delta}
\in C_0^\infty(\boldsymbol{R};\boldsymbol{R})$
supported in $[\lambda_0-\delta,\lambda_0+\delta]$, the Mourre estimate
\begin{equation}
f_{\lambda_0,\delta}(K)i[K,\bar{\bar{A}}_{0,1}]f_{\lambda_0,\delta}(K)
\ge\frac{2(d_1(\lambda_0)-\delta)}{1+2(d_1(\lambda_0)-\delta)}f_{\lambda_0,\delta}(K)^2+C_1
\label{1.08}
\end{equation}
holds with some compact operator $C_1$ on $\mathscr{K}$.
This estimate \eqref{1.08} is slightly better than
the one obtained in \cite{Yo}
$$f_{\lambda_0,\delta}(K)i[K,\bar{\bar{A}}_{0,1}]f_{\lambda_0,\delta}(K)
\ge\frac{2(d_0(\lambda_0)-\delta)}{1+2(d_0(\lambda_0)-\delta)}f_{\lambda_0,\delta}(K)^2+C_1'
$$
with some compact operator $C_1'$ on $\mathscr{K}$,
since $d_0(\lambda_0)\le d_1(\lambda_0)$.
Here we note that the positive constant of the Mourre estimate \eqref{1.08}
depends on $\lambda_0$ strictly but the conjugate
operator $\bar{\bar{A}}_{0,1}$ is independent of $\lambda_0$.
However, its extension to the case where $N\ge3$ has not
been obtained yet, as far as we know (see also M{\o}ller-Skibsted~\cite{MS}).
Recently, Adachi-Kiyose~\cite{AK} proposed
an alternative conjugate operator for $K$ with $N=2$
at a non-threshold energy $\lambda_0$:
Let $\lambda_0\in\boldsymbol{R}\setminus\omega\boldsymbol{Z}$.
Then there exists a unique $n_{\lambda_0}\in\boldsymbol{Z}$
such that $\lambda_0\in I_{n_{\lambda_0}}$.
Take $\delta$ as $0<\delta<\mathrm{dist}(\lambda_0,\omega\boldsymbol{Z})=d_0(\lambda_0)$.
Since $\lambda_0-\delta\in I_{n_{\lambda_0}}$,
it is obvious that $\lambda_0-\delta
\in\boldsymbol{R}\setminus\omega\boldsymbol{Z}=\rho(D_t)$. Then
we introduce the self-adjoint operator
\begin{equation}
A_{\lambda_0,\delta}=R_{-D_t,\lambda_0-\delta}\otimes \hat{A}_0;\quad
R_{-D_t,\lambda_0-\delta}=(\lambda_0-\delta-D_t)^{-1}
\label{1.09}
\end{equation}
on $\mathscr{K}\cong L^2(\boldsymbol{T})\otimes \mathscr{H}$,
by multiplying $\hat{A}_0$ and the resolvent $R_{-D_t,\lambda_0-\delta}$ of $D_t$. Here we note that $R_{-D_t,\lambda_0-\delta}$
is bounded and self-adjoint, and $A_{\lambda_0,\delta}$
satisfies the commutation relation
$$i[K_0,A_{\lambda_0,\delta}]=p^2R_{-D_t,\lambda_0-\delta}=
2(K_0-D_t)(\lambda_0-\delta-D_t)^{-1}.$$
Then the Mourre estimate
\begin{equation}
f_{\lambda_0,\delta}(K)i[K,A_{\lambda_0,\delta}]f_{\lambda_0,\delta}(K)
\ge2f_{\lambda_0,\delta}(K)^2+C_{f_{\lambda_0,\delta}}
\label{1.010}
\end{equation}
holds with some compact operator $C_{f_{\lambda_0,\delta}}$ on $\mathscr{K}$.
Here we note that the positive constant of the Mourre estimate \eqref{1.010}
is independent of $\lambda_0$ but the conjugate
operator $A_{\lambda_0,\delta}$ depends on $\lambda_0$ strictly.
Its extension to the case where $N\ge3$ has not
been obtained generally yet, except in the case where all the pair potentials
are independent of $t$.

The aim of this paper is that we will introduce
a conjugate operator for $K$ with $N=3$.
As is pointed out by M{\o}ller-Skibsted~\cite{MS}, it is important in obtaining the
Mourre estimates for time-independent many body Schr\"odinger operators
that the generator of dilations
$\hat{A}_0$ in \eqref{1.07} can be decomposed into the sum
$$(\hat{A}_0)^a\otimes\mathrm{Id}+\mathrm{Id}\otimes(\hat{A}_0)_a$$
acting on $\mathscr{H}\cong\mathscr{H}^a\otimes\mathscr{H}_a$, for $a\in\mathscr{A}$, where
\begin{equation}
\begin{split}
&(\hat{A}_0)^a=\frac{1}{2}(x^a\cdot p^a+p^a\cdot x^a)=\mathrm{Re}\,(x^a\cdot p^a),\\
&(\hat{A}_0)_a=\frac{1}{2}(x_a\cdot p_a+p_a\cdot x_a)=\mathrm{Re}\,(x_a\cdot p_a).
\end{split}\label{1.011}
\end{equation}
Unfortunately, the conjugate operator
$\bar{\bar{A}}_{0,1}$ in \eqref{1.06} does not have such a property.
This seems one of the reasons why its extension to the
case where $N\ge3$ has not been given yet.
On the other hand, the conjugate operator 
$A_{\lambda_0,\delta}$ in
\eqref{1.09} can be decomposed into the sum
$$R_{-D_t,\lambda_0-\delta}\otimes\{(\hat{A}_0)^a\otimes\mathrm{Id}+\mathrm{Id}\otimes(\hat{A}_0)_a\}$$
acting on $\mathscr{K}\cong L^2(\boldsymbol{T})\otimes\mathscr{H}^a\otimes\mathscr{H}_a$, for $a\in\mathscr{A}$. If $N=3$ and $a\in\mathscr{A}$ is a pair,
that is, $a\in\mathscr{A}_0^0=\mathscr{A}\setminus\{a_\mathrm{max},a_\mathrm{min}\}$,
then one can regard the operator
$R_{-D_t,\lambda_0-\delta}\otimes(\hat{A}_0)^a$ as a
conjugate operator for $K^a=D_t+H^a(t)$
acting on
$\mathscr{K}^a=L^2(\boldsymbol{T};\mathscr{H}^a)\cong L^2(\boldsymbol{T})\otimes\mathscr{H}^a$,
by virtue of a result of Adachi-Kiyose~\cite{AK}.
However, we cannot interpret the operator $R_{-D_t,\lambda_0-\delta}
\otimes(\hat{A}_0)_a$ as a conjugate operator for the
intercluster Hamiltonian $(p_a)^2/2$ acting on $\mathscr{H}_a$,
unfortunately. The operator $R_{-D_t,\lambda_0-\delta}\otimes(\hat{A}_0)_a$ can be a conjugate operator
for the Floquet Hamiltonian $D_t+(p_a)^2/2$ acting on $L^2(\boldsymbol{T};\mathscr{H}_a)\cong
L^2(\boldsymbol{T})\otimes\mathscr{H}_a$ associated with $(p_a)^2/2$. But,
since $K_a=K^a\otimes\mathrm{Id}+\mathrm{Id}\otimes
(p_a)^2/2$, what we need here is
not a conjugate operator for $D_t+(p_a)^2/2$
but that just for $(p_a)^2/2$.
We think that this is
one of the reasons why any extension of $A_{\lambda_0,\delta}$ to the
case where $N\ge3$ has not been given yet.

In order to overcome the difficulty mentioned above, roughly speaking,
we will regard the operator
$$\bar{A}_{a,1}=\mathrm{Re}\,(x_a\cdot p_a\langle p_a/\sqrt{2}\rangle^{-2});\quad \langle p_a/\sqrt{2}\rangle^{-2}=(1+(p_a)^2/2)^{-1}$$
acting on $\mathscr{H}_a$ as a conjugate operator for $(p_a)^2/2$,
and the sum 
$$R_{-D_t,\lambda_0-\delta}\otimes(\hat{A}_0)^a\otimes\mathrm{Id}+
\mathrm{Id}\otimes\mathrm{Id}\otimes\bar{A}_{a,1}$$
as a conjugate operator $A_a$ for $K_a=D_t+H_a(t)$ acting on $\mathscr{K}\cong L^2(\boldsymbol{T})\otimes\mathscr{H}^a\otimes\mathscr{H}_a$.
After introducing $A_a$'s, we will glue these together by using a
partition of unity of $X$.
This is our strategy of introducing a conjugate operator $A$ for $K$
with $N=3$.
Obtaining a conjugate operator for $K$ with $N=3$ is the
first step for the definition of conjugate operators for $K$ with $N\ge4$.
In our strategy, for example, to define a conjugate operator for
$K$ with $N=4$, we need conjugate operators for not only two-body subsystem
Floquet Hamiltonians but also three-body subsystem Floquet Hamiltonians.
The latter has not been obtained until now.

Now we will give the precise definition of $A$. Without loss of generality,
we may assume that an energy $\lambda_0$ belongs to
the interval $[0,\omega)$, because the spectrum $\sigma(K)$
of $K$ is $\omega$-periodic (see Proposition \ref{prop3.4}), as is well-known.
As for a conjugate operator for the free Floquet Hamiltonian
$K_0=D_t+p^2/2$, we introduce
\begin{equation}
\tilde{A}_0=\langle D_t\rangle^{-1/2}\hat{A}_0\langle D_t\rangle^{-1/2}=\mathrm{Re}\,(x\cdot p\langle D_t\rangle^{-1})
\label{1.012}
\end{equation}
acting on $\mathscr{K}$, where $\langle D_t\rangle=(1+D_t^2)^{1/2}$.
In this work, we have found that
one can utilize the positive weight
$\langle D_t\rangle^{-1}$ in place of
the signed weight $R_{-D_t,\lambda_0-\delta}$
in the definition of a conjugate operator $\tilde{A}_0$ for $K_0$.
We think that $\langle D_t\rangle^{-1}$
is more suitable for $\tilde{A}_0$ than $R_{-D_t,\lambda_0-\delta}$.
For instance, $i[K_0,\tilde{A}_0]=\langle D_t\rangle^{-1}
p^2$
$=\langle D_t\rangle^{-1/2}p^2\langle D_t\rangle^{-1/2}$
holds (see Lemma \ref{lem4.1} in \S4), which makes
its non-negativity available. This is a big advantage of
$\langle D_t\rangle^{-1}$.
For $a\in\mathscr{A}_0^0$, we also introduce
\begin{equation}
\begin{split}
(\tilde{A}_0)^a={}&\langle D_t\rangle^{-1/2}(\hat{A}_0)^a\langle D_t\rangle^{-1/2}=\mathrm{Re}\,(x^a\cdot p^a\langle D_t\rangle^{-1}),\\
(\tilde{A}_0)_a={}&\langle D_t\rangle^{-1/2}(\hat{A}_0)_a\langle D_t\rangle^{-1/2}=\mathrm{Re}\,(x_a\cdot p_a\langle D_t\rangle^{-1}),\\
\bar{A}_a={}&\mathrm{Re}\,(x_a\cdot p_a\langle p_a/\sqrt{2}\rangle^{-2});\quad
\langle p_a/\sqrt{2}\rangle^{-2}=(1+(p_a)^2/2)^{-1}.
\end{split}\label{1.013}
\end{equation}
As will be seen in \S4, $(\tilde{A}_0)^a$ can be regarded
as a conjugate operator for the two-body subsystem Floquet
Hamiltonian $K^a$ on $\mathscr{K}^a$.
$\tilde{A}_0$, $(\tilde{A}_0)^a$'s and
$\bar{A}_a$'s are all self-adjoint.
For the sake of glueing these together, we have
mainly three choices of partitions of unity of $X$ until now;
the reducing partition of unity (see e.g.
Cycon-Froese-Kirsch-Simon~\cite{CFKS}), which
was used in Froese-Herbst~\cite{FH} for
the sake of showing the Mourre estimate
for time-independent $N$-body Schr\"odinger operators,
the Graf partition of unity (see e.g. Graf~\cite{Gr},
Derezi\'nski~\cite{D}, Derezi\'nski-G\'erard~\cite{DG} and G\'erard-{\L}aba~\cite{GL}), and
the Yafaev partition of unity (see e.g. Yafaev~\cite{Yaf},
Hunziker-Sigal~\cite{HuS} and G\'erard-{\L}aba~\cite{GL}).
In this paper, we will
construct the Yafaev-type partition of unity
$\{\hat{J}_a\}_{a\in\mathscr{A}}=\{\tilde{J}_a(L\langle D_t\rangle;x)\}_{a\in\mathscr{A}}$
dependent on $\langle D_t\rangle$, by utilizing the construction of
the Yafaev partition
of unity. We do not mention the way of construction here because of its complexity.
In \S2, we will give its definition
and state its useful properties.
By using this $\{\hat{J}_a\}_{a\in\mathscr{A}}$,
we will introduce
\begin{equation}
\begin{split}
A={}&\sum_{a\in\mathscr{A}\setminus\mathscr{A}_0^0}
\hat{J}_a\tilde{A}_0\hat{J}_a+\sum_{a\in\mathscr{A}_0^0}\hat{J}_a\{(\tilde{A}_0)^a+\bar{A}_a\}\hat{J}_a\\
={}&\tilde{A}_0
+\sum_{a\in\mathscr{A}_0^0}
\hat{J}_a\{-(\tilde{A}_0)_a+\bar{A}_a\}\hat{J}_a.
\end{split}\label{1.014}
\end{equation}
We note $\langle D_t\rangle^{-1/4}x\langle D_t\rangle^{-1/4}=\langle D_t\rangle^{-1/2}\otimes x$.
As will be seen in \S2, 
the second representation of $A$ can be derived
directly from the first one, by
$\sum_{a\in\mathscr{A}}\hat{J}_a^2=1$.
In our dealing with the errors like
$i[K_0,\hat{J}_a]\{-(\tilde{A}_0)_a
+\bar{A}_a\}\hat{J}_a$,
which come from
the \textit{overlap width} of glueing 
$\tilde{A}_0$, $(\tilde{A}_0)^a$'s and
$\bar{A}_a$'s by some partition of unity of $X$,
the Yafaev-type partition of unity is much handier
than the others. In fact, we see that
$i[K_0,\hat{J}_a]\{-(\tilde{A}_0)_a+\bar{A}_a\}\hat{J}_a\langle K_0\rangle^{-1}$
is bounded, and that
$\langle K_0\rangle^{-1}i[K_0,\hat{J}_a]\{-(\tilde{A}_0)_a+\bar{A}_a\}\hat{J}_a\langle K_0\rangle^{-1}=O(L^{-1/2})+C_a$ with some compact $C_a$,
where  $O(L^{-1/2})$ is not compact but very small for sufficiently large $L$. Moreover,
this $\{\hat{J}_a\}_{a\in\mathscr{A}}$ has a great effect on
dealing with $i[V,A]$.
We will impose the following {\it well-regulated}
condition $(V_\mathrm{WR})$ on $V$:
\medskip

\noindent
$(V_\mathrm{WR})$ Suppose $\rho_0>0$, $\rho_1>1$, $\rho_2\ge2$, $\rho_3\ge2$
and $\rho_4\ge2$. $V_{jk}(t,y)$, $(j,k)\in\mathscr{A}$,
belongs to $C^4(\boldsymbol{R}\times
\boldsymbol{R}^d;\boldsymbol{R})$,
is $T$-periodic in $t$, 
and satisfies the decaying conditions
\begin{equation}
\sup_{t\in\boldsymbol{R}}|(\partial_t^m\partial_y^\alpha
V_{jk})(t,y)|\le C_{m,\alpha}\langle y\rangle^{-\rho_m-|\alpha|},\quad
0\le m+|\alpha|\le4.
\label{1.015}
\end{equation}

\noindent
In fact, in order to make $\{\hat{J}_a\}_{a\in\mathscr{A}}$
work well, we need the
following conditions on $V_{jk}$'s based on Fourier analysis
(as for the detail, see \S2):
It is very important that the functions which will be taken up below
possess the regularity of $C^2(\boldsymbol{T})$.
For $m,\,\ell\in\boldsymbol{N}\cup\{0\}$ such that $0\le m\le2$
and $0\le\ell\le2-m$,
we introduce
$\rho_{m,\ell}'=\min\{\rho_k+\ell\,|\,m\le k\le4-\ell\}$ and
$\rho_{m,\ell}''=\min\{\rho_{m,\ell}',4\}$.
As mentioned above, we write $V_{jk}(t,r_j-r_k)$ as $V^{(j,k)}(t,x^{(j,k)})$.
One can regard $V^{(j,k)}$ as an
$O(\langle x^{(j,k)}\rangle^{-\rho_{0,0}''})$-valued $C^4(\boldsymbol{T})$ function.
Then one can show that for $a\in\mathscr{A}_0$ such that $(j,k)\not\subset a$, $V^{(j,k)}\hat{J}_a$
is of order $O(\langle x\rangle^{-\rho_{0,0}''})$ (see Lemma \ref{lem2.006}), which implies that  $V^{(j,k)}\hat{J}_a$
is $K_0$-compact, by $\rho_{0,0}''>0$. One can also regard $\nabla^{(j,k)}V^{(j,k)}$ as an
$O(\langle x^{(j,k)}\rangle^{-\rho_{0,1}''})$-valued $C^3(\boldsymbol{T})$ function
and $\partial_tV^{(j,k)}$ as an
$O(\langle x^{(j,k)}\rangle^{-\rho_{1,0}''})$-valued $C^3(\boldsymbol{T})$ function,
with $\rho_{0,1}''>1$ and $\rho_{1,0}''>1$.
By the latter, one can also show that $i[V^{(j,k)},\hat{J}_a]$
is of order $O(\langle x\rangle^{-\rho_{1,0}''})$  even if $(j,k)=a$ (see Lemma \ref{lem2.005}),
which implies that  $\langle x\rangle i[V^{(j,k)},\hat{J}_a]$ is $K_0$-compact,
by $\rho_{1,0}''>1$.
It follows from these that
$i[V,A]\langle K_0\rangle^{-1}$ is bounded, and
\eqref{2.0019} stated in \S2
$$\langle K_0\rangle^{-1}i[V,A]\langle K_0\rangle^{-1}=\sum_{a\in\mathscr{A}_0^0}
\langle K_0\rangle^{-1}\hat{J}_ai[V^a,(\tilde{A}_0)^a]\hat{J}_a\langle K_0\rangle^{-1}+C_0$$
holds with some compact operator $C_0$ on $\mathscr{K}$.
This is one of the keys in the proof of the Mourre estimate for $K$.
One can also regard $(\nabla^{(j,k)})^2V^{(j,k)}$ as an
$O(\langle x^{(j,k)}\rangle^{-\rho_{0,2}''})$-valued $C^2(\boldsymbol{T})$ function,
$\partial_t\nabla^{(j,k)}V^{(j,k)}$ as an
$O(\langle x^{(j,k)}\rangle^{-\rho_{1,1}''})$-valued $C^2(\boldsymbol{T})$ function,
and $\partial_t^2V^{(j,k)}$ as an
$O(\langle x^{(j,k)}\rangle^{-\rho_{2,0}''})$-valued $C^2(\boldsymbol{T})$ function,
with $\rho_{0,2}''>2$, $\rho_{1,1}''>2$ and $\rho_{2,0}''\ge2$.
Then one can see that $\langle K_0\rangle^{-1}i[i[V,A],A]\langle K_0\rangle^{-1}$
is bounded (see Lemma \ref{lem2.007}). These properties will be shown in \S2,
by using the results stated in \S3.
\medskip

Now we state the main results of this paper.

\begin{thm}\label{thm1.1}
Suppose $N=3$. Assume $V$ satisfies $(V_\mathrm{WR})$. 
Put
$$\varTheta=\bigcup_{a\in\mathscr{A}_0}\sigma_{\mathrm{pp}}(K^a),\quad
\widehat{\varTheta}=\bigcup_{a\in\mathscr{A}}\sigma_{\mathrm{pp}}(K^a)=
\varTheta\cup\sigma_\mathrm{pp}(K).$$
Let $\lambda_0\in[0,\omega)\setminus\varTheta$ and $\epsilon>0$.
Then there exists a Yafaev-type partition of unity
$\{\hat{J}_a\}_{a\in\mathscr{A}}$ in the definition \eqref{1.014} of
$A$, the following hold:

\noindent
$(1)$ Put
$$d_0(\lambda)=\mathrm{dist}(\lambda,\varTheta),\quad
d_1(\lambda)=\mathrm{dist}(\lambda,\varTheta\cap(-\infty,\lambda])$$
for $\lambda\in\boldsymbol{R}$. Note $d_0(\lambda)\le d_1(\lambda)$.
Then there exists a small $\delta_{\epsilon,0}>0$
such that $\delta_{\epsilon,0}\le d_0(\lambda_0)/2$ and
the following holds: Let $0<\delta\le\delta_{\epsilon,0}$.
Then, for any
$f_{\lambda_0,\delta}\in C_0^\infty(\boldsymbol{R};\boldsymbol{R})$
supported in $[\lambda_0-\delta,\lambda_0+\delta]$,
\begin{equation}
f_{\lambda_0,\delta}(K)i[K,A]f_{\lambda_0,\delta}(K)
\ge\frac{2(d_1(\lambda_0)-\delta)-\epsilon}{1+\omega}f_{\lambda_0,\delta}(K)^2+C_0
\label{1.16}
\end{equation}
holds with some compact operator
$C_0$ on $\mathscr{K}$. 

Hence, for any $\hat{\delta}$ such that
$0<\hat{\delta}<\delta$,
$\sigma_\mathrm{pp}(K)\cap(\lambda_0-\hat{\delta},\lambda_0+\hat{\delta})$
is finite, and the eigenvalues of $K$
in $(\lambda_0-\hat{\delta},\lambda_0+\hat{\delta})$
are of finite multiplicity.

\noindent
$(2)$ Suppose $\lambda_0\in[0,\omega)\setminus\widehat{\varTheta}$.
Take $\epsilon$ as
$0<\epsilon<d_1(\lambda_0)-d_0(\lambda_0)/2$.
Then there exists a small $\delta_{\epsilon,1}>0$ such that
$\delta_{\epsilon,1}\le\delta_{\epsilon,0}\le d_0(\lambda_0)/2$, and
for any 
$f_{\lambda_0,\delta}\in C_0^\infty(\boldsymbol{R};\boldsymbol{R})$
supported in $[\lambda_0-\delta,\lambda_0+\delta]$ with $0<\delta\le
\delta_{\epsilon,1}$,
\begin{equation}
f_{\lambda_0,\delta}(K)i[K,A]f_{\lambda_0,\delta}(K)
\ge\frac{2(d_1(\lambda_0)-\delta)-2\epsilon}{1+\omega}f_{\lambda_0,\delta}(K)^2
\label{1.17}
\end{equation}
holds. Suppose $s>1/2$ and $0<\hat{\delta}<\delta_{\epsilon,1}$. Then
\begin{equation}
\sup_{\substack{\mathrm{Re}\,z\in[\lambda_0-\hat{\delta},\lambda_0+\hat{\delta}]\\
\mathrm{Im}\,z\not=0}}\|
\langle A\rangle^{-s}(K-z)^{-1}\langle A\rangle^{-s}
\|_{\mathscr{B}(\mathscr{K})}<\infty
\label{1.18}
\end{equation}
holds. Moreover, $\langle A\rangle^{-s}(K-z)^{-1}\langle A\rangle^{-s}$
is a $\mathscr{B}(\mathscr{K})$-valued $\theta(s)$-H\"older continuous
function on $z\in S_{\lambda_0,\hat{\delta},\pm}$ with some $0<\theta(s)<1$, where
$S_{\lambda_0,\hat{\delta},\pm}=\bigl\{\zeta\in\boldsymbol{C}\bigm|
\mathrm{Re}\,\zeta\in[\lambda_0-\hat{\delta},\lambda_0+\hat{\delta}],\ 
0<\pm\mathrm{Im}\,\zeta\le1\bigr\}$.
And, there exist the norm limits
$$\langle A\rangle^{-s}(K-(\lambda\pm i0))^{-1}\langle A\rangle^{-s}=
\lim_{\varepsilon\to+0}\langle A\rangle^{-s}(K-(\lambda\pm i\varepsilon))^{-1}\langle A\rangle^{-s}$$
in $\mathscr{B}(\mathscr{K})$ for any
$\lambda\in[\lambda_0-\hat{\delta},\lambda_0+\hat{\delta}]$.
$\langle A\rangle^{-s}(K-(\lambda\pm i0))^{-1}\langle A\rangle^{-s}$
are also $\theta(s)$-H\"older continuous in $\lambda$.
\end{thm}

If $\lambda_0\in[0,\omega)\cap\varTheta$, then
a weaker estimate like
$$
f_{\lambda_0,\delta}(K)i[K,A]f_{\lambda_0,\delta}(K)\\
\ge\frac{-\epsilon}{1+\omega}f_{\lambda_0,\delta}(K)^2+C_0
$$
holds with some compact operator
$C_0$ on $\mathscr{K}$. When we will
introduce a conjugate operator for the Floquet Hamiltonian $K$ with $N=4$,
not only \eqref{1.16} but also
this estimate will be needed.

\begin{cor}\label{cor1.2}
Assume $V$ satisfies $(V_\mathrm{WR})$. Then the following hold:

\noindent
$(1)$ The eigenvalues of $K$ in $\boldsymbol{R}\setminus\varTheta$
can accumulate only at $\varTheta$.
Moreover, $\widehat{\varTheta}$ is a countable
closed set.

\noindent
$(2)$ Let $I$ be a compact interval in
$\boldsymbol{R}\setminus\widehat{\varTheta}$.
Suppose $1/2<s\le1$. Then
\begin{equation}
\sup_{\substack{\mathrm{Re}\,z\in I\\
\mathrm{Im}\,z\not=0}}\|
\langle x\rangle^{-s}(K-z)^{-1}\langle x\rangle^{-s}
\|_{\mathscr{B}(\mathscr{K})}<\infty
\label{1.19}
\end{equation}
holds. Moreover, $\langle x\rangle^{-s}(K-z)^{-1}\langle x\rangle^{-s}$
is a $\mathscr{B}(\mathscr{K})$-valued $\theta(s)$-H\"older continuous
function on $z\in S_{I,\pm}$, where
$S_{I,\pm}=\bigl\{\zeta\in\boldsymbol{C}\bigm|
\mathrm{Re}\,\zeta\in I,\ 
0<\pm\mathrm{Im}\,\zeta\le1\bigr\}$.
And, there exist the norm limits
$$\langle x\rangle^{-s}(K-(\lambda\pm i0))^{-1}\langle x\rangle^{-s}=
\lim_{\varepsilon\to+0}\langle x\rangle^{-s}(K-(\lambda\pm i\varepsilon))^{-1}
\langle x\rangle^{-s}$$
in $\mathscr{B}(\mathscr{K})$ for $\lambda\in I$.
$\langle x\rangle^{-s}(K-(\lambda\pm i0))^{-1}
\langle x\rangle^{-s}$ are also $\theta(s)$-H\"older continuous in $\lambda$.
\end{cor}

In order to obtain Corollary \ref{cor1.2}, we use
the argument of Perry-Sigal-Simon
\cite{PSS},
and the boundedness of 
$A(K-\lambda_0-i)^{-1}\langle x\rangle^{-1}$,
which can be given by that
$\langle D_t\rangle^{-1}(K-\lambda_0-i)^{-1}\langle p\rangle^2$ is bounded
(see Proposition \ref{prop3.5} in \S3).
By virtue of this, one can show that
$A(K-\lambda_0-i)^{-1}\langle p\rangle\langle x\rangle^{-1}$
and $A(K-\lambda_0-i)^{-1}\langle D_t\rangle^{1/2}
\langle x\rangle^{-1}$
are also bounded. Then the limiting absorption
principle like 
$$\sup_{\substack{\mathrm{Re}\,z\in I\\
\mathrm{Im}\,z\not=0}}\|
\langle x\rangle^{-s}\mathcal{D}^{s'}(K-z)^{-1}\mathcal{D}^{s'}
\langle x\rangle^{-s}
\|_{\mathscr{B}(\mathscr{K})}<\infty
$$
with $s>1/2$ and some $s'>0$ may be expected,
where $\mathcal{D}=\langle p\rangle+\langle D_t\rangle^{1/2}$ is equivalent to
$\mathscr{D}^{1/2}=(\langle p\rangle^4+\langle D_t\rangle^2)^{1/4}$
as weights,
which was introduced in Kuwabara-Yajima~\cite{KuY}
for the sake of obtaining a refined limiting absorption principle for
$K$ with $N=2$ in the Besov space setting. In fact,
the result of \cite{KuY} yields the estimate
$$\sup_{\substack{\mathrm{Re}\,z\in I\\
\mathrm{Im}\,z\not=0}}\|
\langle x\rangle^{-s}\mathscr{D}^{1/2}(K-z)^{-1}\langle x\rangle^{-s}
\|_{\mathscr{B}(\mathscr{K})}<\infty
$$
with $s>1/2$. By virtue of complex interpolation, one can also obtain
a refined limiting absorption principle
$$\sup_{\substack{\mathrm{Re}\,z\in I\\
\mathrm{Im}\,z\not=0}}\|
\langle x\rangle^{-s}\mathscr{D}^{1/4}(K-z)^{-1}\mathscr{D}^{1/4}\langle x\rangle^{-s}
\|_{\mathscr{B}(\mathscr{K})}<\infty$$
with $s>1/2$.
But this has not been given by our analysis yet.
It is caused by the unboundedness of
$(K-\lambda_0-i)^{-1}\langle p\rangle\langle x\rangle^{-1}$
and $(K-\lambda_0-i)^{-1}\langle D_t\rangle^{1/2}\langle x\rangle^{-1}$.
As for general $N$-body Floquet Hamiltonians,
a refined limiting absorption principle for $K$
$$\sup_{\substack{\mathrm{Re}\,z\in I\\
\mathrm{Im}\,z\not=0}}\|
\langle x\rangle^{-s}\langle p\rangle^r(K-z)^{-1}\langle p\rangle^r
\langle x\rangle^{-s}\|_{\mathscr{B}(\mathscr{K})}<\infty$$
with $0\le r<1/2<s\le1$ was obtained by M\o ller-Skibsted~\cite{MS}.
They used an extended Mourre theory due to
Skibsted~\cite{Sk}, and took a conjugate operator for $K$ in
the extended Mourre theory as $\hat{A}_0$.
However, we would like to stick to find
an option of a conjugate operator for $K$ not in an extended
but in the standard Mourre theory.

As is well-known, the limiting absorption principle
\eqref{1.19} yields the local $K$-smoothness of
$\langle x\rangle^{-s}$ with $s>1/2$
\begin{equation}
\int_{-\infty}^\infty\|\langle x\rangle^{-s}e^{-i\sigma K}f_{\lambda_0,\delta}(K)\varPhi\|_{\mathscr{K}}^2\,d\sigma\le C\|\varPhi\|_{\mathscr{K}}^2
\label{1.20}
\end{equation}
for $\lambda_0\in[0,\omega)\setminus\widehat{\varTheta}$.
\eqref{1.20} was already obtained by \cite{MS},
even if $N>3$. However, \eqref{1.20} is not enough for the
proof of the asymptotic completeness in the case where $N\ge3$,
unlike in the case where $N=2$.
For instance, when $K$ is a time-independent many body Schr\"odinger operator
$p^2/2+V$ with short-range pair interactions,
instead of the propagation estimate \eqref{1.20}
with $\mathscr{K}=L^2(X)$, the so-called minimal velocity estimate
$$\int_1^\infty\left\|F\left(\frac{|x|}{\sigma}\le \tilde{c}_{\lambda_0}\right)e^{-i\sigma K}f_{\lambda_0,\delta}(K)\varPhi\right\|_{\mathscr{K}}^2\,\frac{d\sigma}{\sigma}\le C\|\varPhi\|_{\mathscr{K}}^2$$
with $\tilde{c}_{\lambda_0}>0$ and $\mathscr{K}=L^2(X)$ 
has been used as a key propagation estimate
in the proof of the asymptotic completeness
(see e.g. Graf~\cite{Gr}). Here we used the following
convention for smooth cut-off functions
$F$ with $0\le F\le1$: For sufficiently small $\delta>0$, we define
\begin{align*}
&F(s\le d)=1\;\;\;\text{for}\;\;s\le d-\delta,\quad=0\;\;\;\text{for}\;\;
s\ge d,\\
&F(s\ge d)=1\;\;\;\text{for}\;\;s\ge d+\delta,\quad=0\;\;\;\text{for}\;\;
s\le d,
\end{align*}
and $F(d_1\le s\le d_2)=F(s\ge d_1)\,F(s\le d_2)$.
The above minimal velocity estimate
can be obtained by the Mourre estimate for $K=p^2/2+V$.
At the present stage, we have gotten the Mourre estimate
\eqref{1.17} for the Floquet Hamiltonian $K=D_t+p^2/2+V(t)$.
Hence, it can be expected that the Mourre estimate \eqref{1.17} will
yield the minimal velocity estimate for $e^{-i\sigma K}$.
In fact, there is an abstract theory for
getting the minimal velocity estimate from the Mourre estimate,
which was initiated by Sigal-Soffer (see e.g. \cite{SS1,SS2}).
For the detail, see e.g.
Sect. 4.4 ``Minimal velocity estimates'' of G\'erard-{\L}aba~\cite{GL}.
By virtue of the abstract theory,
one can obtain the following minimal velocity estimate
for $e^{-i\sigma K}$:

\begin{thm}\label{thm1.3}
Suppose $N=3$. Assume $V$ satisfies $(V_\mathrm{WR})$. Let $\lambda_0\in[0,\omega)\setminus\widehat{\varTheta}$.
Put
\begin{equation}
\begin{split}
&B=(1+\tilde{B}_0+\tilde{B}_1)^{1/2};\quad
\tilde{B}_0=\langle D_t\rangle^{-1/2}Q_0
\langle D_t\rangle^{-1/2},\\
&\tilde{B}_1=\sum_{a\in\mathscr{A}_0^0}\tilde{B}_{1,a},\quad
\tilde{B}_{1,a}=\langle p_a/\sqrt{2}\rangle^{-1}Q_{1,a}\langle p_a/\sqrt{2}\rangle^{-1},
\end{split}\label{1.21}
\end{equation}
where
\begin{align*}
&Q_0=\sum_{a\in\mathscr{A}\setminus\mathscr{A}_0^0}
x^2\hat{J}_a^2
+\sum_{a\in\mathscr{A}_0^0}
(x^a)^2\hat{J}_a^2=x^2-\sum_{a\in\mathscr{A}_0^0}
(x_a)^2\hat{J}_a^2,\\
&Q_{1,a}=(x_a)^2\hat{J}_a^2,
\end{align*}
with $a\in\mathscr{A}_0^0$.
Then there exists an $\varepsilon_0(\lambda_0)=\varepsilon_0(\lambda_0;\epsilon,\delta)>0$,
which is determined by 
the positive constant $\{2(d_1(\lambda_0)-\delta)-2\epsilon\}/(1+\omega)$
in the Mourre estimate
\eqref{1.17}
such that
\begin{align}\
&\int_1^\infty\left\|F\left(\frac{B}{\sigma}\le\varepsilon_0(\lambda_0)\right)e^{-i\sigma K}f_{\lambda_0,\delta}(K)\varPhi\right\|_{\mathscr{K}}^2\,\frac{d\sigma}{\sigma}\le C\|\varPhi\|_{\mathscr{K}}^2,
\label{1.22}\\
&\slim_{\sigma\to\infty}
F\left(\frac{B}{\sigma}\le\varepsilon_0(\lambda_0)\right)e^{-i\sigma K}f_{\lambda_0,\delta}(K)=0\label{1.23}
\end{align}
hold. In particular, these yield
\begin{align}
&\int_1^\infty\left\|F\left(\frac{\langle x\rangle}{\sigma}\le\varepsilon_0(\lambda_0)\right)e^{-i\sigma K}f_{\lambda_0,\delta_{\epsilon,1}}(K)\varPhi\right\|_{\mathscr{K}}^2\,\frac{d\sigma}{\sigma}\le C\|\varPhi\|_{\mathscr{K}}^2,
\label{1.24}\\
&\slim_{\sigma\to\infty}F\left(\frac{\langle x\rangle}{\sigma}\le\varepsilon_0(\lambda_0)\right)e^{-i\sigma K}f_{\lambda_0,\delta_{\epsilon,1}}(K)=0.
\label{1.25}
\end{align}
\end{thm}

Here we note that
$$B^2=1+\tilde{B}_0+\tilde{B}_1\le1+Q_0+\sum_{a\in\mathscr{A}_0^0}Q_{1,a}=1+x^2=\langle x\rangle^2$$
holds,
in deriving \eqref{1.24} from \eqref{1.22}.

By using the arguments of Yajima-Kitada~\cite{YK}
and M{\o}ller-Skibsted~\cite{MS}, one
can translate \eqref{1.24} into the
minimal velocity estimate for the physical propagator
$U(t,s)$ generated by $H(t)$:

\begin{cor}\label{cor1.4}
Suppose that the hypotheses of Theorem \ref{thm1.3} are satisfied.
Let $g_{\lambda_0,\delta}$ be
the function on $\bigl\{z\in\boldsymbol{C}\bigm||z|=1\bigr\}$
such that $g_{\lambda_0,\delta}(e^{-i2\pi\lambda/\omega})=
f_{\lambda_0,\delta}(\lambda)$ for $\lambda\in[0,\omega)$.
Then
\begin{align}
&\int_1^\infty\left\|F\left(\frac{\langle x\rangle}{t}\le\varepsilon_0(\lambda_0)\right)U(t,0)g_{\lambda_0,\delta}(U(T,0))\phi\right\|_{\mathscr{H}}^2\,\frac{dt}{t}\le C\|\phi\|_{\mathscr{H}}^2,
\label{1.26}\\
&\slim_{t\to\infty}F\left(\frac{\langle x\rangle}{t}\le\varepsilon_0(\lambda_0)\right)U(t,0)g_{\lambda_0,\delta}(U(T,0))=0
\label{1.27}
\end{align}
hold, where $U(T,0)$ is the Floquet operator
associated with $H(t)$.
\end{cor}

One can utilize the above results in the study of quantum
systems of $N$ particles under the so-called AC Stark
effect with $N=3$: 
We suppose that $\mathscr{E}(t)\in C^2(\boldsymbol{R};\boldsymbol{R}^d)$,
$\mathscr{E}(t)$ has a period $T>0$, that is,
$\mathscr{E}(t+T)=\mathscr{E}(t)$ for any $t\in\boldsymbol{R}$,
and
\begin{equation}
\mathscr{E}_\mathrm{m}=\frac{1}{T}\int_0^T\mathscr{E}(s)\,ds=0,
\end{equation}
which is the condition for the AC Stark effect.
Let $m_j>0$, $q_j\in\boldsymbol{R}$ and $r_j \in \boldsymbol{R}^d$,
$1\le j\le N$, denote the mass, charge and
position vector of the $j$-th particle, respectively.
We suppose that the particles under consideration
interact with one another through the time-independent pair
potentials $\bar{V}_{jk}(r_j-r_k)$, $1\le j<k\le N$.
The system under consideration is governed by the total Hamiltonian
in the laboratory frame
\begin{equation}
\begin{split}
&\bar{H}_\mathrm{LF}(t)=\sum_{j=1}^N\left(-\frac{1}{2m_j}\Delta_j-q_j\langle \mathscr{E}(t),
r_j\rangle\right)+\bar{V}(r);\\
&\bar{V}(r)=\sum_{1\le j<k\le N}\bar{V}_{jk}(r_j-r_k).
\end{split}
\end{equation}
In the same way as
in M{\o}ller~\cite{Mo} and Adachi~\cite{A1,A2} (as for the detail, see \S7),
we have only to consider the Hamiltonian in the moving frame
\begin{equation}
\hat{H}_\mathrm{MF}(t)=\frac{1}{2}p^2+\bar{V}(x+c(t))
\end{equation}
acting on $\mathscr{H}=L^2(X)$, where
$c(t)$ is defined as follows:
\begin{align*}
&\mathscr{E}_\mathrm{os}(t)=
\mathscr{E}(t)-\mathscr{E}_\mathrm{m},\quad
\bar{\mathscr{E}}(t)=\int_0^t\mathscr{E}_\mathrm{os}(s)\,ds,\quad
\bar{\mathscr{E}}_\mathrm{m}=\frac{1}{T}\int_0^T\bar{\mathscr{E}}(s)\,ds,\\
&\bar{\mathscr{E}}_\mathrm{os}(t)=\bar{\mathscr{E}}(t)-\bar{\mathscr{E}}_\mathrm{m},\quad
\bar{\bar{\mathscr{E}}}(t)=\int_0^t\bar{\mathscr{E}}_\mathrm{os}(s)\,ds,\\
&c(t)=\pi((q_1/m_1)\bar{\bar{\mathscr{E}}}(t),\ldots,(q_N/m_N)\bar{\bar{\mathscr{E}}}(t)).
\end{align*}
Here we note that $c(t)$ is of $C^4(\boldsymbol{T})$, because
$\mathscr{E}(t)$ is of $C^2(\boldsymbol{T})$.

Now we impose the following condition $(V_\mathrm{ST})_{\bar{\rho}}$ on $\bar{V}_{jk}$'s with $\bar{\rho}>0$:
\medskip

\noindent
$(V_\mathrm{ST})_{\bar{\rho}}$ $\bar{V}_{jk}(y)$, $(j,k)\in\mathscr{A}$,
belongs to $C^4(\boldsymbol{R}^d;\boldsymbol{R})$,
is independent of $t$, 
and satisfies the decaying conditions
\begin{equation}
|(\partial_y^\alpha
\bar{V}_{jk})(y)|\le C_{\alpha}\langle y\rangle^{-\bar{\rho}-|\alpha|},\quad
0\le|\alpha|\le4.
\end{equation}

As is shown in \S7, if $\bar{\rho}\ge1$, then one can regard 
$V_{jk}(t,y)=
\bar{V}_{jk}(y+\tilde{e}_{jk}\bar{\bar{\mathscr{E}}}(t))$
with $\tilde{e}_{jk}=q_j/m_j-q_k/m_k$
as a time-periodic potential satisfying
the condition $(V_\mathrm{WR})$.
Therefore, in the case where $N=3$, the following theorem
is a direct consequence of Theorem \ref{thm1.1},
Corollary \ref{cor1.2}, Theorem \ref{thm1.3},
and Corollary \ref{cor1.4}.

\begin{thm}\label{thm1.5}
Suppose $\mathscr{E}_\mathrm{m}=0$ and $N=3$.
Introduce $K=D_t+\hat{H}_\mathrm{MF}(t)$
acting on $\mathscr{K}$.
Assume $\bar{V}_{jk}$'s satisfy $(V_\mathrm{ST})_{\bar{\rho}}$
with some $\bar{\rho}\ge1$. 
Put
$$\varTheta=\bigcup_{a\in\mathscr{A}_0}\sigma_{\mathrm{pp}}(K^a),\quad
\widehat{\varTheta}=\bigcup_{a\in\mathscr{A}}\sigma_{\mathrm{pp}}(K^a)=
\varTheta\cup\sigma_\mathrm{pp}(K).$$
Then, the statements of Theorem \ref{thm1.1},
Corollary \ref{cor1.2}, Theorem \ref{thm1.3},
and Corollary \ref{cor1.4} hold.
\end{thm}

As for the asymptotic completeness for
$\hat{H}_\mathrm{MF}(t)$ with $N=2$,
Yajima~\cite{Ya1} proved it in the short-range case
via the Howland-Yajima method,
and Kitada-Yajima~\cite{KY} proved it in the long-range case
via the Enss method.
On the other hand, for
$\hat{H}_\mathrm{MF}(t)$ with $N=3$,
Korotyaev~\cite{Ko} and Nakamura~\cite{N} gave some
partial results on it in the very short-range case
via the Howland-Yajima and the Faddeev methods.
The study of the problem of the asymptotic completeness
for $\hat{H}_\mathrm{MF}(t)$ with $N=3$
should be done by using some useful propagation estimates like
\eqref{1.22} and \eqref{1.24} in future research.
\bigskip

The plan of this paper is as follows: 
In \S2, we will give the definition of the Yafaev-type
partition of unity and its properties
which are useful for our analysis.
In \S3, we collect frequently used propositions
which are useful for our analysis.
In \S4, we will revisit
the case where $N=2$. The construction of $A$ in \eqref{1.014}
is based on the arguments and results in \S4.
In \S5, we will give
the proof of Theorem \ref{thm1.1}, in particular, \eqref{1.16}.
In \S6, we will obtain the minimal velocity estimates for $e^{-i\sigma K}$
and $U(t,s)$. In \S7, we will deal with the AC Stark effect case.
\medskip

\leftline{\textbf{Acknowledgement}}
The author
is partially supported by the Grant-in-Aid for Scientific Research
(C) \#25K07031 from JSPS. The author is grateful to the referee
for many valuable comments and suggestions.

\section{Yafaev-type partition of unity}

In this section, we give the definition
of the Yafaev-type partition of unity $\{\hat{J}_a\}_{a\in\mathscr{A}}$
of $X$ dependent on $\langle D_t\rangle$, and its properties
which are
powerful in the proof of the Mourre estimate for $K$.

First of all, for references,
we will give an outline of the original construction due to Yafaev~\cite{Yaf} by
following the arguments in \cite{Yaf}, Hunziker-Sigal~\cite{HuS}
and G\'erard-{\L}aba~\cite{GL}
(also refer to Graf~\cite{Gr}, Derezi\'nski~\cite{D},
Derezi\'nski-G\'erard~\cite{DG} and G\'erard-{\L}aba~\cite{GL}
on the construction
of the Graf partition of unity, for comparison):
Let $\sigma=\{\sigma_a\}_{a\in\mathscr{A}}$ be a sequence of 
numbers greater than $1$ indexed with the elements of $\mathscr{A}$
such that $1<\sigma_a<\sigma_b$ holds if $a\subsetneq b$
with $a,\,b\in\mathscr{A}_0=\mathscr{A}\setminus\{a_\mathrm{max}\}$.
The argument below is based on the fact that
there exists an $M>0$ such that
$$M|x^{a\cup b}|\le|x^a|+|x^b|$$
holds for any $a,\,b\in\mathscr{A}$,
because $X^{a\cup b}=X^a+X^b$.
Now we suppose that for sufficiently small $\epsilon>0$,
$\sigma_a=1+O(\epsilon^{d_a})$ holds for $a\in\mathscr{A}_0$,
 where $d_a=\dim X_a$.
For each $x\in X\setminus\{0\}$, we consider the family
$\{\sigma_c|x_c|\}_{c\in\mathscr{A}_0}$, and watch
its maximum $\max_{c\in\mathscr{A}_0}\{\sigma_c|x_c|\}$.
Since $x_{a_\mathrm{max}}=0$, when watching
$\max_{c\in\mathscr{A}_0}\{\sigma_c|x_c|\}$,
$\sigma_{a_\mathrm{max}}$ is insignificant.
This $\sigma_{a_\mathrm{max}}$ will be used
when guaranteeing the smoothness of the
functions defined later on a neighborhood of the
origin $0$ of $X$. We put
\begin{equation}
U_a^\sigma=\bigl\{x\in X\bigm|\sigma_a|x_a|>
\sigma_{a_\mathrm{min}}|x_{a_\mathrm{min}}|
=\sigma_{a_\mathrm{min}}|x|\bigr\}\quad
(a\in\mathscr{A}).
\label{2.1}
\end{equation}
We see that 
$U_{a_\mathrm{max}}^\sigma=\emptyset$
since $x_{a_\mathrm{max}}=0$,
$U_{a_\mathrm{min}}^\sigma=\emptyset$
trivially, and that
for $a\in\mathscr{A}_0^0=\mathscr{A}\setminus\{a_\mathrm{max},a_\mathrm{min}\}$,
$$U_a^\sigma=\bigl\{x\in X\bigm||x^a|<(1-\sigma_{a_\mathrm{min}}^2/\sigma_a^2)^{1/2}|x|\bigr\}$$
is a conical
neighborhood of $X_a\setminus\{0\}$.
Since $(1-\sigma_{a_\mathrm{min}}^2/\sigma_a^2)^{1/2}=O(\epsilon^{d_a/2})$,
$U_a^\sigma$ is considerably sharp.
Let $a\in\mathscr{A}_0^0$ be
given, $x\in U_a^\sigma$,
and $b\in\mathscr{A}$ be such that $b\not\supset a$,
that is, $X_b\not\subset X_a$.
By virtue of Lemma 3.1 of \cite{Yaf},
we see that there exist sequences $\sigma=\{\sigma_a\}_{a\in\mathscr{A}}$'s
such that
$$\sigma_b|x_b|<\max_{\substack{f\in\mathscr{A}_0\\ f\supset a}}\{\sigma_f|x_f|\}$$
holds, because $a\cup b\supsetneq a$ by assumption,
and $d_a-d_{a\cup b}\ge d=\dim\boldsymbol{R}^d$.
Such sequences $\sigma$'s
are called admissible.
The above inequality implies that for $x\in U_a^\sigma$,
$$\max_{c\in\mathscr{A}_0}\{\sigma_c|x_c|\}=
\max_{\substack{f\in\mathscr{A}_0\\ f\supset a}}\{\sigma_f|x_f|\}$$
holds. Taking account of these, we put
\begin{equation}
C_a^\sigma=U_a^\sigma\setminus\left(
\bigcup_{f\supsetneq a}U_f^\sigma\right)
\quad
(a\in\mathscr{A}_0^0),\quad
C_{a_\mathrm{min}}^\sigma=(X\setminus\{0\})\setminus\left(
\bigcup_{a\in\mathscr{A}_0^0}C_a^\sigma\right).
\label{2.2}
\end{equation}
This definition yields that $\{C_a^\sigma\}_{a\in\mathscr{A}_0}$ is a disjoint covering of $X\setminus\{0\}$, and
that for $x\in C_a^\sigma$
with $a\in\mathscr{A}_0$,
$$\max_{c\in\mathscr{A}_0}\{\sigma_c|x_c|\}=
\sigma_a|x_a|$$
holds. In particular,
for $a\in\mathscr{A}_0^0$,
if the pair $(j,k)$ satisfies $(j,k)\subsetneq a$, then
$C_a^\sigma\cap C_{(j,k)}^\sigma=\emptyset$.
By following the argument of \cite{HuS}, we introduce
the intercluster distance
\begin{equation}
|x|_a=\min_{\substack{b\in\mathscr{A}_0\\ b\not\subset a}}
|x^b|\label{2.3}
\end{equation}
for $a\in\mathscr{A}_0$. By virtue of Lemma 3.1 of \cite{HuS},
it is known that if $a\in\mathscr{A}_0$, then
\begin{equation}
|x|_a\ge\left(\min_{\substack{b\in\mathscr{A}_0\\ b\not\subset a}}
(1-\sigma_{a_\mathrm{min}}^2/\sigma_b^2)^{1/2}\right)|x|\label{2.4}
\end{equation}
holds for $x\in C_a^\sigma$.
Now we will take a family of sets $\{C_a^\sigma\}_{a\in\mathscr{A}_0}$ with an admissible sequence $\sigma=\{\sigma_a\}_{a\in\mathscr{A}}$, and put
\begin{equation}
\widetilde{C}_a^\sigma=\bigl\{x\in C_a^\sigma\bigm|\sigma_a|x_a|>\sigma_{a_\mathrm{max}}\bigr\}\quad (a\in\mathscr{A}_0),\quad
\widetilde{C}_{a_\mathrm{max}}^\sigma=X\setminus
\bigcup_{a\in\mathscr{A}_0}\widetilde{C}_a^\sigma.
\label{2.5}
\end{equation}
This definition yields that $\{\widetilde{C}_a^\sigma\}_{a\in\mathscr{A}}$
is a disjoint covering of $X$, and that
for $x\in \widetilde{C}_{a_\mathrm{max}}^\sigma$,
$$\max\left\{\sigma_{a_\mathrm{max}},\max_{c\in\mathscr{A}_0}\{\sigma_c|x_c|\}\right\}=
\sigma_{a_\mathrm{max}}$$
holds, and that for $x\in \widetilde{C}_a^\sigma$
with $a\in\mathscr{A}_0$,
$$\max\left\{\sigma_{a_\mathrm{max}},\max_{c\in\mathscr{A}_0}\{\sigma_c|x_c|\}\right\}=
\sigma_a|x_a|$$
holds.
Then we will introduce
\begin{equation}
J_a^\sigma(x)=\chi_{\widetilde{C}_a^\sigma}(x)\quad(a\in\mathscr{A}),
\label{2.6}
\end{equation}
where $\chi_{\widetilde{C}_a^\sigma}$ is the characteristic function
of $\widetilde{C}_a^\sigma$. It is known that
$$\sum_{a\in\mathscr{A}}J_a^\sigma(x)\equiv1$$
holds.
By the arguments of \cite{Yaf} and \cite{HuS},
those are known that for sufficiently small $\epsilon>0$,
both $\sigma^-=\{\sigma_a^-\}_{a\in\mathscr{A}}$ with $\sigma_a^-=1+2\epsilon^{d_a}$
and $\sigma^+=\{\sigma_a^+\}_{a\in\mathscr{A}}$  with
$\sigma_a^+=1+3\epsilon^{d_a}$ are admissible, and that
if $\sigma=\{\sigma_a\}_{a\in\mathscr{A}}$ satisfies
$\sigma_a^-\le \sigma_a\le \sigma_a^+$ for any $a\in\mathscr{A}$,
then $\sigma$ is also admissible. As for $a=a_\mathrm{max}$,
we put $\sigma_{a_\mathrm{max}}^-=1+2=3$
and $\sigma_{a_\mathrm{max}}^+=1+3=4$, of course, because $d_{a_\mathrm{max}}=0$.
For $b\in\mathscr{A}$,
we introduce a smoothing function $s_b$, that is,
a function $s_b\in C_0^\infty(
\boldsymbol{R};\boldsymbol{R})$ such that
$$\mathrm{supp}\,s_b\subset[\sigma_b^-,\sigma_b^+],\quad
s_b\ge 0,\quad \int_{\boldsymbol{R}}s_b(\sigma_b)\,d\sigma_b=1.$$
For $b\in\mathscr{A}_0$, we take a function $t_b\in C_0^\infty(\boldsymbol{R};\boldsymbol{R})$,
which is independent of $\epsilon$, such that
$$\mathrm{supp}\,t_b\subset[2,3],\quad
t_b\ge 0,\quad \int_{\boldsymbol{R}}t_b(\tau_b)\,d\tau_b=1,$$
and put
$$s_b(\sigma_b)=\epsilon^{-d_b}t_b(\epsilon^{-d_b}(\sigma_b-1)).$$
This is an example of the smoothing function $s_b$.
Without loss of generality, one can assume that $t_b(\tau_b)=O(|\tau_b-2|^{N-1})$ holds in a small
neighborhood of $\tau_b=2$, and 
$t_b(\tau_b)=O(|\tau_b-3|^{N-1})$ holds in a small
neighborhood of $\tau_b=3$.
Putting $\tilde{s}_0(\sigma_0)=\prod_{b\in\mathscr{A}_0}s_b(\sigma_b)$,
it satisfies
$$\mathrm{supp}\,\tilde{s}_0\subset\mathbf{S}_0=\prod_{b\in\mathscr{A}_0}[\sigma_b^-,\sigma_b^+],\quad
\tilde{s}_0\ge 0,\quad \int_{\mathbf{S}_0}\tilde{s}_0(\sigma_0)\,d\sigma_0=1,$$
where $d\sigma_0=\otimes_{b\in\mathscr{A}_0}d\sigma_b$.
Now we define
\begin{equation}
J_a(x)=\int_{\sigma_{a_\mathrm{max}}^-}^{\sigma_{a_\mathrm{max}}^+}
\int_{\mathbf{S}_0}\tilde{s}_0(\sigma_0)s_{a_\mathrm{max}}(\sigma_{a_\mathrm{max}})J_a^\sigma(x)\,d\sigma_0d\sigma_{a_\mathrm{max}}\quad(a\in\mathscr{A}).
\label{2.7}
\end{equation}
$\{J_a\}_{a\in\mathscr{A}}$ is called a Yafaev partition of unity.
Then we have the following:

\begin{prop}\label{prop2.1}
The family of functions $\{J_a\}_{a\in\mathscr{A}}$ on $X$
satisfies the following: Each $J_a(x)$ is a smooth function
on $X$. If $a\in\mathscr{A}_0$,
then $J_a(x)$ is homogeneous of degree $0$ outside some compact
neighborhood of the origin $0$ of $X$; while, 
$J_{a_\mathrm{max}}(x)$ is supported in some compact
neighborhood of the origin $0$ of $X$.
Moreover,
\begin{equation}
\sum_{a\in\mathscr{A}}J_a(x)\equiv1
\label{2.8}
\end{equation}
holds.
\end{prop}

\begin{proof}
The proof was already given in \cite{Yaf} and \cite{HuS}.
As for the detail, see these papers.
Here we will give an outline of the proof of the smoothness
of $J_a(x)$ with $a\in\mathscr{A}_0^0=\mathscr{A}
\setminus\{a_\mathrm{max},a_\mathrm{min}\}$ only,
since we need to know the properties
of $(\nabla J_a)(x)$: For the sake of simplicity,
we consider the case where $x\in X$ satisfies
$\sigma_a^-|x_a|>\sigma_{a_\mathrm{max}}^+$ only.
If $x\in X$ satisfies
$\sigma_a^-|x_a|>\sigma_{a_\mathrm{min}}^+|x_{a_\mathrm{min}}|
=\sigma_{a_\mathrm{min}}^+|x|$, then $x\in U_a^\sigma$
for any $\sigma\in\mathbf{S}$, because $\sigma_a\ge\sigma_a^-$
and $\sigma_{a_\mathrm{min}}^+\ge\sigma_{a_\mathrm{min}}$.
Then $J_a(y)=1$ holds in a neighborhood of $x$. 
If $x\in X$ satisfies
$\sigma_a^+|x_a|<\sigma_{a_\mathrm{min}}^-|x|$,
then $x\not\in U_a^\sigma$
for any $\sigma\in\mathbf{S}$, because $\sigma_a\le\sigma_a^+$
and $\sigma_{a_\mathrm{min}}^-\le\sigma_{a_\mathrm{min}}$.
Then $J_a(y)=0$ holds in a neighborhood of $x$.
Thus we have only to watch the case where
$x\in X$ satisfies
$\sigma_a^-|x_a|\le\sigma_{a_\mathrm{min}}^+|x|$
and $\sigma_a^+|x_a|\ge\sigma_{a_\mathrm{min}}^-|x|$.
Consider the rectangle $[\sigma_a^-,\sigma_a^+]\times
[\sigma_{a_\mathrm{min}}^-,\sigma_{a_\mathrm{min}}^+]$
in the $\sigma_a\sigma_{a_\mathrm{min}}$-plane.
Since $d_{a_\mathrm{min}}=(\#(a_\mathrm{min})-1)d=(N-1)d$ and
$d_a=(\#(a)-1)d$ with $2\le\#(a)\le N-1$,
$\epsilon^{d_{a_\mathrm{min}}}<\epsilon^{d_a}$ holds
for sufficiently small $\epsilon>0$,
without loss of generality, we may consider
the following three cases: (a) $x$ satisfies 
$\sigma_a^-|x_a|<\sigma_{a_\mathrm{min}}^-|x|$
and $\sigma_{a_\mathrm{min}}^-|x|\le\sigma_a^+|x_a|\le\sigma_{a_\mathrm{min}}^+|x|$;
(b) $x$ satisfies 
$\sigma_a^-|x_a|<\sigma_{a_\mathrm{min}}^-|x|$
and $\sigma_a^+|x_a|>\sigma_{a_\mathrm{min}}^+|x|$;
(c) $x$ satisfies 
$\sigma_{a_\mathrm{min}}^-|x|\le\sigma_a^-|x_a|\le\sigma_{a_\mathrm{min}}^+|x|$
and $\sigma_a^+|x_a|>\sigma_{a_\mathrm{min}}^+|x|$.
In the case (a), we have
\begin{align*}
J_a(x)={}&\int_{\sigma_a^-}^{\sigma_{a_\mathrm{min}}^-|x|/|x_a|}
\int_{\sigma_{a_\mathrm{min}}^-}^{\sigma_{a_\mathrm{min}}^+}
s_a(\sigma_a)s_{a_\mathrm{min}}(\sigma_{a_\mathrm{min}})\,d\sigma_{a_\mathrm{min}}d\sigma_a\\
&+\int_{\sigma_{a_\mathrm{min}}^-|x|/|x_a|}^{\sigma_a^+}
\int_{\sigma_a|x_a|/|x|}^{\sigma_{a_\mathrm{min}}^+}
s_a(\sigma_a)s_{a_\mathrm{min}}(\sigma_{a_\mathrm{min}})
\,d\sigma_{a_\mathrm{min}}d\sigma_a\\
={}&\int_{\sigma_a^-}^{\sigma_{a_\mathrm{min}}^-|x|/|x_a|}
s_a(\sigma_a)\,d\sigma_a\\
&+\int_{\sigma_{a_\mathrm{min}}^-|x|/|x_a|}^{\sigma_a^+}
\int_{\sigma_a|x_a|/|x|}^{\sigma_{a_\mathrm{min}}^+}
s_a(\sigma_a)s_{a_\mathrm{min}}(\sigma_{a_\mathrm{min}})
\,d\sigma_{a_\mathrm{min}}d\sigma_a.
\end{align*}
In particular, we obtain
\begin{equation}
\begin{split}
(\nabla J_a)(x)={}&\nabla\left(\frac{\sigma_{a_\mathrm{min}}^-|x|}{|x_a|}\right)s_a\left(\frac{\sigma_{a_\mathrm{min}}^-|x|}{|x_a|}\right)\\
&-\nabla\left(\frac{\sigma_{a_\mathrm{min}}^-|x|}{|x_a|}\right)
s_a\left(\frac{\sigma_{a_\mathrm{min}}^-|x|}{|x_a|}\right)
\int_{\sigma_{a_\mathrm{min}}^-}^{\sigma_{a_\mathrm{min}}^+}
s_{a_\mathrm{min}}(\sigma_{a_\mathrm{min}})\,d\sigma_{a_\mathrm{min}}\\
&\quad-\int_{\sigma_{a_\mathrm{min}}^-|x|/|x_a|}^{\sigma_a^+}
\nabla\left(\frac{\sigma_a|x_a|}{|x|}\right)
s_a(\sigma_a)s_{a_\mathrm{min}}\left(\frac{\sigma_a|x_a|}{|x|}\right)
\,d\sigma_a\\
={}&-\nabla\left(\frac{|x_a|}{|x|}\right)\int_{\sigma_{a_\mathrm{min}}^-|x|/|x_a|}^{\sigma_a^+}
\sigma_as_a(\sigma_a)s_{a_\mathrm{min}}\left(\frac{\sigma_a|x_a|}{|x|}\right)
\,d\sigma_a.
\end{split}\label{2.09}
\end{equation}
For $\sigma_a\in[\sigma_{a_\mathrm{min}}^-|x|/|x_a|,\sigma_a^+]
\subset[\sigma_{a_\mathrm{min}}^-|x|/|x_a|,\sigma_{a_\mathrm{min}}^+|x|/|x_a|]$,
$|\sigma_a-\sigma_a^+|$ is of order $O(\epsilon^{d_{a_\mathrm{min}}})=O(\epsilon^{(N-1)d})$,
and 
$$s_a(\sigma_a)=s_a(\sigma_a)-s_a(\sigma_a^+)=e^{-d_a}O((e^{-d_a+d_{a_\mathrm{min}}})^{N-1})\le O(\epsilon^d)$$
holds, because $-d_a+(-d_a+d_{a_\mathrm{min}})(N-1)=-N(\#(a)-1)d+(N-1)^2d=\{N(N-2-(\#(a)-1))+1\}d\ge d$ for
$2\le\#(a)\le N-1$. This implies
\begin{equation}\int_{\sigma_{a_\mathrm{min}}^-|x|/|x_a|}^{\sigma_a^+}
	\sigma_as_a(\sigma_a)s_{a_\mathrm{min}}\left(\frac{\sigma_a|x_a|}{|x|}\right)
	\,d\sigma_a=O(\epsilon^d).\label{2.010}
\end{equation}
The cases (b) and (c) can be also treated similarly.
These yield the smoothness of $J_a(x)$.
\end{proof}

We put
\begin{equation}
\tilde{J}_a(x)=J_a(x)\left\{\sum_{b\in\mathscr{A}}J_b(x)^2\right\}^{-1/2}\quad
(a\in\mathscr{A}).
\end{equation}
We also call $\{\tilde{J}_a\}_{a\in\mathscr{A}}$ a
Yafaev partition of unity:

\begin{cor}\label{cor2.2}
The family of functions $\{\tilde{J}_a\}_{a\in\mathscr{A}}$ on $X$
satisfies the following: Each $\tilde{J}_a(x)$ is a smooth function
on $X$. If $a\in\mathscr{A}_0$,
then $\tilde{J}_a(x)$ is homogeneous of degree $0$ outside some compact
neighborhood of the origin $0$ of $X$; while, 
$\tilde{J}_{a_\mathrm{max}}(x)$ is supported in some compact
neighborhood of the origin $0$ of $X$.
Moreover,
\begin{equation}
\sum_{a\in\mathscr{A}}\tilde{J}_a(x)^2\equiv1
\label{2.012}
\end{equation}
holds.
\end{cor}

By virtue of \eqref{2.4}, the following lemma can be obtained immediately:

\begin{lem}\label{lem2.3}
Let $a\in\mathscr{A}_0$. Then
\begin{equation}
|x|_a\ge c_a|x|;
\quad c_a=\left(\min_{\substack{b\in\mathscr{A}_0\\ b\not\subset a}}
(1-(\sigma_{a_\mathrm{min}}^+)^2/(\sigma_b^-)^2)^{1/2}\right)>0,\label{2.013}
\end{equation}
holds for $x\in\mathrm{supp}\,\tilde{J}_a$.
In particular, if $(j,k)\in\mathscr{A}$ satisfies
$(j,k)\not\subset a$, then
$$|x^{(j,k)}|\ge c_a|x|$$
holds for $x\in\mathrm{supp}\,\tilde{J}_a$.
\end{lem}

Now suppose $N=3$. For the sake of constructing the Yafaev-type partition of unity
in our analysis, we will take the above small parameter $\epsilon^d$ in
the estimates like \eqref{2.010} as $(L\langle D_t\rangle)^{-1/2}$ with
sufficiently large $L\ge1$: For the sake of displaying the dependence
on $\epsilon$ of $\tilde{J}_a$
in a simple way, we denote $\tilde{J}_a(x)$ by $\tilde{J}_a(\epsilon^{-2d};x)$.
For $n\in\boldsymbol{Z}$,  put $\varphi_n(t)=T^{-1/2}e^{in\omega t}$,
which is called a Fourier mode. It is well-known that
$\{\varphi_n\}_{n\in\boldsymbol{Z}}$ forms a complete orthonormal system of $L^2(\boldsymbol{T})$.
Then  the Fourier transform
$\widehat{\mathscr{F}}:L^2(\boldsymbol{T})\to\ell^2(\boldsymbol{Z})$ is defined by
$$\widehat{\mathscr{F}}[g](n)=(g,\varphi_n)_{L^2(\boldsymbol{T})}=T^{-1/2}\int_0^Tg(t)e^{-in\omega t}\,dt,\quad
n\in\boldsymbol{Z};\quad g\in L^2(\boldsymbol{T}).$$
$\widehat{\mathscr{F}}[g](n)$ is called a Fourier coefficient of $g$.
Via $\widehat{\mathscr{F}}$, $D_t$ becomes the multiplication by $n\omega$ on $\ell^2(\boldsymbol{Z})$
for $n\in\boldsymbol{Z}$, that is, $D_t$ can be decomposed into
the direct sum
$\bigoplus_{n\in\boldsymbol{Z}}n\omega$.
By using $\widehat{\mathscr{F}}$, we define $\hat{J}_a=\tilde{J}_a(L\langle D_t\rangle;x)$ as
$\bigoplus_{n\in\boldsymbol{Z}}\tilde{J}_a(L\langle n\omega\rangle;x)$.
For any $n\in\boldsymbol{Z}$, we have
$$\sum_{a\in\mathscr{A}}\tilde{J}_a(L\langle n\omega\rangle;x)^2\equiv1$$
by \eqref{2.012}, which yields
\begin{equation}
\sum_{a\in\mathscr{A}}\hat{J}_a^2=\sum_{a\in\mathscr{A}}\tilde{J}_a(L\langle D_t\rangle;x)^2=1
\label{2.014}
\end{equation}

The following lemma is a motivation of adopting the definition of $\{\hat{J}_a\}_{a\in\mathscr{A}}$.
\begin{lem}\label{lem2.04}
Let $a\in\mathscr{A}_0^0$. Then $F(|x|\ge\sigma_{a_\mathrm{max}}^+)\langle x\rangle(\nabla\hat{J}_a)\cdot p\langle K_0\rangle^{-1/2}$
is bounded, and is of order $O(L^{-1/2})$.
\end{lem}

\begin{proof}
The estimates like \eqref{2.09} and \eqref{2.010} in the proof of Proposition \ref{prop2.1} imply that $(L\langle D_t\rangle)^{1/2}F(|x|\ge\sigma_{a_\mathrm{max}}^+)\langle x\rangle(\nabla \hat{J}_a)$ is
bounded uniformly in $L$.  This yields that $F(|x|\ge\sigma_{a_\mathrm{max}}^+)\langle x\rangle(\nabla\hat{J}_a)\cdot p\langle K_0\rangle^{-1/2}=O(L^{-1/2})$,
because $\langle D_t\rangle^{-1/2} p\langle K_0\rangle^{-1/2}$ is bounded,
by Proposition \ref{prop3.5} $(1)$ stated in \S3.
\end{proof}

By using the Yafaev-type partition of unity $\{\hat{J}_a\}_{a\in\mathscr{A}}$,
we will introduce a conjugate operator
$A$ for $K$ by the first representation of \eqref{1.014}
$$A=\sum_{a\in\mathscr{A}\setminus\mathscr{A}_0^0}
\hat{J}_a\tilde{A}_0\hat{J}_a+\sum_{a\in\mathscr{A}_0^0}
\hat{J}_a\{(\tilde{A}_0)^a+\bar{A}_a\}\hat{J}_a.$$
By using \eqref{2.014}, we have
$$\sum_{a\in\mathscr{A}}
\hat{J}_a\tilde{A}_0\hat{J}_a=\tilde{A}_0,$$
because $\hat{J}_a\tilde{A}_0\hat{J}_a=\hat{J}_a[\tilde{A}_0,\hat{J}_a]+
\hat{J}_a^2\tilde{A}_0$, $[\tilde{A}_0,\hat{J}_a]=-i\langle D_t\rangle^{-1}
\{x\cdot\nabla \hat{J}_a\}$ and
$$\sum_{a\in\mathscr{A}}
\hat{J}_a\{x\cdot\nabla \hat{J}_a\}
=\left\{x\cdot\nabla\left(\frac{1}{2}
\sum_{a\in\mathscr{A}}
\hat{J}_a^2\right)\right\}
=0.$$
Since
$(\tilde{A}_0)^a-\tilde{A}_0
=-(\tilde{A}_0)_a$,
we obtain the second representation of \eqref{1.014}
$$A=\tilde{A}_0
+\sum_{a\in\mathscr{A}_0^0}
\hat{J}_a\{-(\tilde{A}_0)_a+\bar{A}_a\}\hat{J}_a.$$
By this representation of $A$, we have
\begin{equation}
i[K_0,A]=
i[K_0,\tilde{A}_0]+\sum_{a\in\mathscr{A}_0^0}S_{1,a}
\label{2.015}
\end{equation}
with $S_{1,a}=i[K_0,\hat{J}_a\{-(\tilde{A}_0)_a+\bar{A}_a\}\hat{J}_a]$.
In our analysis, we will regard
\begin{equation}
i[K_0,A]^\mathrm{m}=i[K_0,\tilde{A}_0]
+\sum_{a\in\mathscr{A}_0^0}S_{1,a}^\mathrm{m}
\label{2.016}
\end{equation}
as the main part of
$i[K_0,A]$, where $S_{1,a}^\mathrm{m}=\hat{J}_ai[K_0,-(\tilde{A}_0)_a+\bar{A}_a]\hat{J}_a$.
Then we see that $\bar{S}_{1,a}=S_{1,a}-S_{1,a}^\mathrm{m}=2\mathrm{Re}\,(i[K_0,\hat{J}_a]\{-(\tilde{A}_0)_a+\bar{A}_a\}\hat{J}_a)$
is $K_0$-bounded such that $\langle K_0\rangle^{-1}\bar{S}_{1,a}\langle K_0\rangle^{-1}=O(L^{-1/2})+C_{1,a}$
with some compact operator $C_{1,a}$ on $\mathscr{K}$.
In fact, as for $\Delta \hat{J}_a$ of $i[K_0,\hat{J}_a]=(\nabla \hat{J}_a)\cdot p-i\Delta \hat{J}_a/2$,
one can see easily that $(\Delta \hat{J}_a)\{-(\tilde{A}_0)_a+\bar{A}_a\}\hat{J}_a\langle K_0\rangle^{-1}$
is compact, by $\Delta \hat{J}_a=O(\langle x\rangle^{-2})$.
We used that $\langle D_t\rangle^{-1/2}p_a\langle K_0\rangle^{-1/2}$
is bounded by Proposition \ref{prop3.5} $(1)$, and that
$p_a\langle p_a/\sqrt{2}\rangle^{-2}$ is also bounded.
We also used that $\langle x\rangle^{-1}\langle K_0\rangle^{-1/2}$ is compact,
which can be proved by Corollary \ref{cor3.7} $(2)$ and the almost analytic
extension method (see e.g. the proof of Proposition \ref{prop3.8}).
As for $(\nabla \hat{J}_a)\cdot p$,
one can see similarly that
$(\nabla \hat{J}_a)\cdot p\{-(\tilde{A}_0)_a+\bar{A}_a\}\hat{J}_a\langle K_0\rangle^{-1}$
is bounded, and
$\langle K_0\rangle^{-1}(\nabla \hat{J}_a)\cdot p\{-(\tilde{A}_0)_a+\bar{A}_a\}\hat{J}_a\langle K_0\rangle^{-1}
=O(L^{-1/2})+C_{1,a}'$ with some compact operator $C_{1,a}'$ on $\mathscr{K}$, by Lemma \ref{lem2.04} and Proposition \ref{prop3.6}. Here we note that where $\nabla \hat{J}_a=\nabla_a \hat{J}_a\not=0$ holds, $|x|$ satisfies $|x|\le\sigma_{a_\mathrm{max}}^+$,
by virtue of the construction of the Yafaev-type partition of unity.
Then, as for the factor such that $(\nabla \hat{J}_a)\cdot p=(\nabla_a \hat{J}_a)\cdot p_a$,
we used that $\langle D_t\rangle^{-1}(p_a)^2\langle K_0\rangle^{-1}$
is bounded by Proposition \ref{prop3.5} $(1)$, and that
$(p_a)^2\langle p_a/\sqrt{2}\rangle^{-2}$ is also bounded.

Actually, $\{\hat{J}_a\}_{a\in\mathscr{A}}$ has a great effect on
dealing with $i[V,A]$. Hence, we will watch $i[V,A]$:  We introduce the Fourier coefficients of
$V=V(t,x)$
$$\widehat{V}_n(x)=(V(\cdot,x),\varphi_n)_{L^2(\boldsymbol{T})}=\widehat{\mathscr{F}}[V(\cdot,x)](n),\quad
n\in\boldsymbol{Z}.$$
Then we obtain the Fourier series expansion of $V$
\begin{equation}
V(t,x)=\sum_{n\in\boldsymbol{Z}}\varphi_n(t)\widehat{V}_n(x).
\end{equation}
Let $\varPhi=\varPhi(t,x)\in\mathscr{K}$. By using the Fourier coefficients of $\varPhi$
$$\widehat{\varPhi}_n(x)=(\varPhi(\cdot,x),\varphi_n)_{L^2(\boldsymbol{T})}=\widehat{\mathscr{F}}[\varPhi(\cdot,x)](n),\quad
n\in\boldsymbol{Z},$$
we also have the Fourier series expansion of $\varPhi$
$$\varPhi(t,x)=\sum_{n\in\boldsymbol{Z}}\varphi_n(t)\widehat{\varPhi}_n(x).$$
Then we have
\begin{equation}
\begin{split}
V(t,x)\varPhi(t,x)={}&T^{-1/2}\sum_{n\in\boldsymbol{Z}}\varphi_n(t)\left(\sum_{k\in\boldsymbol{Z}}\widehat{V}_{n-k}(x)\widehat{\varPhi}_k(x)\right)\\
={}&T^{-1/2}\sum_{k\in\boldsymbol{Z}}\varphi_k(t)\widehat{\varPhi}_k(x)\left(\sum_{n\in\boldsymbol{Z}}\varphi_{n-k}(t)\widehat{V}_{n-k}(x)\right).
\end{split}\label{2.0018}
\end{equation}
We first suppose $a\in\mathscr{A}_0^0$. Because $N=3$, $a$ is a pair. In
$i[V,\hat{J}_a\{(\tilde{A}_0)^a+\bar{A}_a\}\hat{J}_a]=2\mathrm{Re}\,(i[V,\hat{J}_a]\{(\tilde{A}_0)^a+\bar{A}_a\}\hat{J}_a])+\hat{J}_ai[V,(\tilde{A}_0)^a+\bar{A}_a]\hat{J}_a$,
we have to deal with the factor $[V,\hat{J}_a]$, because in general,
$V$ does not commute with $\hat{J}_a$, unlike in the case where
$V$ is independent of $t$. Hence, we will consider $[V,\hat{J}_a]$.
By \eqref{2.0018}, we have
\begin{align*}
&-T^{1/2}[V,\hat{J}_a]\varPhi=T^{1/2}[\hat{J}_a,V]\varPhi\\
={}&
\sum_{n\in\boldsymbol{Z}}\tilde{J}_a(L\langle n\omega\rangle;x)\varphi_n(t)\left(\sum_{k\in\boldsymbol{Z}}\widehat{V}_{n-k}(x)\widehat{\varPhi}_k(x)\right)\\
&-\sum_{n\in\boldsymbol{Z}}\varphi_n(t)\left(\sum_{k\in\boldsymbol{Z}}\widehat{V}_{n-k}(x)\tilde{J}_a(L\langle k\omega\rangle;x)\widehat{\varPhi}_k(x)\right)\\
={}&\sum_{k\in\boldsymbol{Z}}\varphi_k(t)\widehat{\varPhi}_k(x)\left(\sum_{n\in\boldsymbol{Z}}\{\tilde{J}_a(L\langle n\omega\rangle;x)-\tilde{J}_a(L\langle k\omega\rangle;x)\}\varphi_{n-k}(t)\widehat{V}_{n-k}(x)\right).
\end{align*}
We rewrite $\tilde{J}_a(L\langle n\omega\rangle;x)-\tilde{J}_a(L\langle k\omega\rangle;x)$
as $\tilde{J}_a'(n,k;x)(n-k)\omega$ with
$$\tilde{J}_a'(n,k;x)
=\int_0^1\frac{L(\partial_\kappa\tilde{J}_a)(L\langle k\omega+\mu(n-k)\omega\rangle;x)}{\langle k\omega+\mu(n-k)\omega\rangle}
\,d\mu.
$$
Since 
$$(\partial_tV)(t,x)=\sum_{n\in\boldsymbol{Z}}in\omega\varphi_n(t)\widehat{V}_n(x)=\sum_{n\in\boldsymbol{Z}}\varphi_n(t)\widehat{(\partial_tV)}_n(x),$$
we have
$$iT^{1/2}[V,\hat{J}_a]\varPhi=\sum_{k\in\boldsymbol{Z}}\varphi_k(t)\widehat{\varPhi}_k(x)\left(\sum_{n\in\boldsymbol{Z}}\tilde{J}_a'(n,k;x)\varphi_{n-k}(t)\widehat{(\partial_tV)}_{n-k}(x)\right).$$
Here we note that $\tilde{J}_a(\epsilon^{-2d};x)$
is supported in
$\bigl\{x\in X\bigm||x|\ge\sigma_{a_\mathrm{max}}^-=3,\ |x^a|\le(1-(\sigma_{a_\mathrm{min}}^-)^2/(\sigma_a^+)^2)^{1/2}|x|\bigr\}$,
and that
$\tilde{J}_a(\epsilon^{-2d};x)=1$ holds for
$x\in X$ such that $|x|\ge\sigma_{a_\mathrm{max}}^+=4$ and
$|x^a|\le(1-(\sigma_{a_\mathrm{min}}^+)^2/(\sigma_a^-)^2)^{1/2}|x|$.
For sufficiently small $\epsilon>0$, one can obtain
$(1-(\sigma_{a_\mathrm{min}}^+)^2/(\sigma_a^-)^2)^{1/2}
\ge\sqrt{2}\epsilon^{d/2}$ and $(1-(\sigma_{a_\mathrm{min}}^-)^2/(\sigma_a^+)^2)^{1/2}
\le\sqrt{7}\epsilon^{d/2}$ easily. By virtue of this,
$\tilde{J}_a(L\langle n\omega\rangle;x)-\tilde{J}_a(L\langle k\omega\rangle;x)$
is supported in
$\bigl\{x\in X\bigm||x|\ge3,\ |x^a|\le\sqrt{7}L^{-1/4}|x|\bigr\}$,
and $\tilde{J}_a(L\langle n\omega\rangle;x)-\tilde{J}_a(L\langle k\omega\rangle;x)=0$
holds for $x\in X$ such that $|x|\ge4$ and
$|x^a|\le\sqrt{2}(L\max\{\langle n\omega\rangle,\langle k\omega\rangle\})^{-1/4}|x|$.
This implies that $\langle x^b\rangle^{-\nu}$ with $b\in\mathscr{A}_0^0$
and $\nu>0$ can be estimated on the support of
$\tilde{J}_a(L\langle n\omega\rangle;x)-\tilde{J}_a(L\langle k\omega\rangle;x)$
as $O((L\max\{\langle n\omega\rangle,\langle k\omega\rangle\})^{\nu/4}\langle x\rangle^{-\nu})$,
\textit{even if} $b=a$.
Take $j_a(n,k;x)\in C^\infty(X;\boldsymbol{R})$ such that
$0\le j_a(n,k;x)\le1$,  $j_a(n,k;x)$ is supported in
$\bigl\{x\in X\bigm||x|\ge2,\ (L\max\{\langle n\omega\rangle,\langle k\omega\rangle\})^{-1/4}|x|\le|x^a|\le3L^{-1/4}|x|\bigr\}$,
and $j_a(n,k;x)=1$ on the support of 
$\tilde{J}_a(L\langle n\omega\rangle;x)-\tilde{J}_a(L\langle k\omega\rangle;x)$.
Here, for $m,\,\ell\in\boldsymbol{N}\cup\{0\}$ such that $0\le m\le2$
and $0\le\ell\le2-m$,
we introduce
$\rho_{m,\ell}'=\min\{\rho_k+\ell\,|\,m\le k\le4-\ell\}$ and
$\rho_{m,\ell}''=\min\{\rho_{m,\ell}',4\}$.
By $(V_\mathrm{WR})$, there exists a positive sequence
$\{a_{n-k}\}_{n\in\boldsymbol{Z}}$ such that
\begin{align*}
&|j_a(n,k;x)\widehat{(\partial_tV)}_{n-k}(x)|\le a_{n-k}
(L\max\{\langle n\omega\rangle,\langle k\omega\rangle\})^{\rho_{1,0}''/4}\langle x\rangle^{-\rho_{1,0}''},\\
&\sum_{n\in\boldsymbol{Z}}|\langle(n-k)\omega\rangle^3a_{n-k}|^2<\infty,
\end{align*}
with $1<\rho_{1,0}''\le4$. As mentioned in \S1, in our analysis,
it is very important that the functions which we have to watch
possess the regularity of $C^2(\boldsymbol{T})$.
In fact, $\partial_t V$ possesses the regularity of $C^3(\boldsymbol{T})$,
which yields the estimate $\sum_{n\in\boldsymbol{Z}}|\langle(n-k)\omega\rangle^3a_{n-k}|^2<\infty$.
When
$\max\{\langle n\omega\rangle,\langle k\omega\rangle\}=\langle n\omega\rangle$,
by Peetre's inequality, we have
$$\left(\frac{\langle n\omega\rangle}{\langle k\omega+\mu(n-k)\omega\rangle}\right)^{\rho_{1,0}''/4}\le
2^{\rho_{1,0}''/8}\langle (1-\mu)(n-k)\omega\rangle^{\rho_{1,0}''/4};$$
while, when $\max\{\langle n\omega\rangle,\langle k\omega\rangle\}=\langle k\omega\rangle$,
we have
$$\left(\frac{\langle k\omega\rangle}{\langle k\omega+\mu(n-k)\omega\rangle}\right)^{\rho_{1,0}''/4}\le
2^{\rho_{1,0}''/8}\langle \mu(n-k)\omega\rangle^{\rho_{1,0}''/4}.$$
Since 
\begin{align*}
&\sum_{n\in\boldsymbol{Z}}|\langle (n-k)\omega\rangle^{\rho_{1,0}''/4}a_{n-k}|\\
\le{}&\left(\sum_{n\in\boldsymbol{Z}}\langle (n-k)\omega\rangle^{-2}\right)^{1/2}\left(\sum_{n\in\boldsymbol{Z}}|\langle (n-k)\omega\rangle^{\rho_{1,0}''/4+1}a_{n-k}|^2\right)^{1/2}<\infty
\end{align*}
because of $\rho_{1,0}''/4+1\le2<3$, there exists a constant $C>0$ independent of
$k\in\boldsymbol{Z}$ and $t\in\boldsymbol{T}$ such that
$$\left|\sum_{n\in\boldsymbol{Z}}\{\tilde{J}_a(L\langle n\omega\rangle;x)-\tilde{J}_a(L\langle k\omega\rangle;x)\}\varphi_{n-k}(t)\widehat{V}_{n-k}(x)\right|\le C\langle x\rangle^{-\rho_{1,0}''},$$
which implies $[V,\hat{J}_a]=O(\langle x\rangle^{-\rho_{1,0}''})$. 
Similarly,
one can show $[[V,\hat{J}_a],\hat{J}_a]=O(\langle x\rangle^{-\rho_{2,0}''})$
with $2\le\rho_{2,0}''\le4$, by watching $\partial_t^2V$.
One can see easily that these hold also in the case where
$a=a_\mathrm{min}$. Thus we have obtained the following lemma,
which is also one of key properties of $\{\hat{J}_a\}_{a\in\mathscr{A}}$.

\begin{lem}\label{lem2.005}
Let $a\in\mathscr{A}_0$. Then
$$[V,\hat{J}_a]=O(\langle x\rangle^{-\rho_{1,0}''}),\quad
[[V,\hat{J}_a],\hat{J}_a]=O(\langle x\rangle^{-\rho_{2,0}''})$$
hold.
\end{lem}

\begin{rem}
If one may show $[V,\hat{J}_a]=O(\langle x\rangle^{-\rho_{1,0}''})$ only,
one has only to assume the following condition on $V_{jk}$:
$V_{jk}(t,y)$ belongs to $C^3(\boldsymbol{R}\times
\boldsymbol{R}^d;\boldsymbol{R})$,
is $T$-periodic in $t$, 
and satisfies the decaying conditions
$$\sup_{t\in\boldsymbol{R}}|(\partial_t^m\partial_y^\alpha
V_{jk})(t,y)|\le C_{m,\alpha}\langle y\rangle^{-\rho_m-|\alpha|},\quad
0\le m+|\alpha|\le3.$$
However, since we also need the boundedness of
$\langle K_0\rangle^{-1}i[i[V,A],A]\langle K_0\rangle^{-1}$
based on the standard Mourre theory, we impose
the condition $(V_\mathrm{WR})$.
\end{rem}

In the same way, one can also obtain the following easily:

\begin{lem}\label{lem2.006}
Let $a\in\mathscr{A}_0$, and $m,\,\ell\in\boldsymbol{N}\cup\{0\}$
such that $0\le m\le2$ and $0\le\ell\le2-m$.
Then, for any $(j,k)\in\mathscr{A}$ such that $(j,k)\not\subset a$,
$$(\partial_t^m(\nabla^{(j,k)})^\ell V^{(j,k)})\hat{J}_a=O(\langle x\rangle^{-\rho_{m,\ell}''})$$
holds.
\end{lem}

By using Propositions \ref{prop3.5}, \ref{prop3.6} and Corollary \ref{cor3.7} which will be stated in \S3,
we watch
$$i[V,A]=\sum_{a\in\mathscr{A}\setminus\mathscr{A}_0^0}
i[V,\hat{J}_a\tilde{A}_0\hat{J}_a]+\sum_{a\in\mathscr{A}_0^0}
i[V,\hat{J}_a\{(\tilde{A}_0)^a+\bar{A}_a\}\hat{J}_a].$$
For $a\in\mathscr{A}\setminus\mathscr{A}_0^0$, 
$i[V,\hat{J}_a\tilde{A}_0\hat{J}_a]\langle K_0\rangle^{-1}$
is bounded, and $\langle K_0\rangle^{-1}i[V,\hat{J}_a\tilde{A}_0\hat{J}_a]\langle K_0\rangle^{-1}$
is compact. In fact, for $a=a_\mathrm{max}$,
this can be shown by the compactness of the support of
$\hat{J}_{a_\mathrm{max}}$.
For $a=a_\mathrm{min}$,  since $i[V,\hat{J}_{a_\mathrm{min}}]=O(\langle x\rangle^{-\rho_{1,0}''})$ by
Lemma \ref{lem2.005},
$i[V,\hat{J}_{a_\mathrm{min}}]\tilde{A}_0\hat{J}_{a_\mathrm{min}}
\langle K_0\rangle^{-1}$
is bounded, because $\langle x\rangle^{-1}\tilde{A}_0\hat{J}_{a_\mathrm{min}}
\langle K_0\rangle^{-1}$ is bounded by Proposition \ref{prop3.5} $(1)$,
and $i[V,\hat{J}_{a_\mathrm{min}}]
\langle x\rangle=O(\langle x\rangle^{-\rho_{1,0}''+1})$ with $\rho_{1,0}''>1$.
Moreover, $\langle K_0\rangle^{-1}i[V,\hat{J}_{a_\mathrm{min}}]\tilde{A}_0\hat{J}_{a_\mathrm{min}}\langle K_0\rangle^{-1}$
is compact because $\langle K_0\rangle^{-1}i[V,\hat{J}_{a_\mathrm{min}}]\langle x\rangle$
is compact by Corollary \ref{cor3.7} $(2)$. By Lemma \ref{lem2.006}, we have
$\hat{J}_{a_\mathrm{min}}i[V,\langle D_t\rangle^{-1}]\langle D_t\rangle=O(\langle x\rangle^{-\rho_{1,0}''})$
and $\hat{J}_{a_\mathrm{min}}i[V,\hat{A}_0]=O(\langle x\rangle^{-\rho_{0,1}''+1})$.
Hence,
$\hat{J}_{a_\mathrm{min}}i[V,\tilde{A}_0]\hat{J}_{a_\mathrm{min}}
\langle K_0\rangle^{-1}$
is bounded, and $\langle K_0\rangle^{-1}\hat{J}_{a_\mathrm{min}}i[V,\tilde{A}_0]\hat{J}_{a_\mathrm{min}}\langle K_0\rangle^{-1}$
is compact, by $\rho_{0,1}''>1$.
For $a\in\mathscr{A}_0^0$, 
$i[V,\hat{J}_a]\{(\tilde{A}_0)^a+\bar{A}_a\}\hat{J}_a\langle K_0\rangle^{-1}$
is bounded, and $\langle K_0\rangle^{-1}i[V,\hat{J}_a]\{(\tilde{A}_0)^a+\bar{A}_a\}\hat{J}_a\langle K_0\rangle^{-1}$
is compact, since $i[V,\hat{J}_a]=O(\langle x\rangle^{-\rho_{1,0}''})$ by
Lemma \ref{lem2.005}. In the same way as above,
$\hat{J}_ai[I_a,(\tilde{A}_0)^a+\bar{A}_a]\hat{J}_a
\langle K_0\rangle^{-1}$
is bounded, and $\langle K_0\rangle^{-1}\hat{J}_ai[I_a,(\tilde{A}_0)^a+\bar{A}_a]\hat{J}_a
\langle K_0\rangle^{-1}$
is compact, by
Lemma \ref{lem2.006}. Since $i[V^a,(\tilde{A}_0)^a+\bar{A}_a]=i[V^a,(\tilde{A}_0)^a]$,
which is one of the keys of the definition of $A$, we finally obtain
\begin{equation}
\langle K_0\rangle^{-1}i[V,A]\langle K_0\rangle^{-1}=\sum_{a\in\mathscr{A}_0^0}
\langle K_0\rangle^{-1}\hat{J}_ai[V^a,(\tilde{A}_0)^a]\hat{J}_a\langle K_0\rangle^{-1}+C_0
\label{2.0019}
\end{equation}
with some compact operator $C_0$ on $\mathscr{K}$.
This is the key estimate in the proof of the Mourre estimate for $K$.
Similarly, the boundedness of $\langle K_0\rangle^{-1}i[i[V,A],A]\langle K_0\rangle^{-1}$
can be shown.

\begin{lem}\label{lem2.007}
As for $i[V,A]$, \eqref{2.0019} holds with some compact operator $C_0$ on $\mathscr{K}$.
And, as for $i[i[V,A],A]$,
$\langle K_0\rangle^{-1}i[i[V,A],A]\langle K_0\rangle^{-1}$ is bounded.
\end{lem}

\section{Collection of frequently used propositions}

In this section, we collect frequently used propositions
which are useful for the proof of the Mourre estimate for $K$.

First of all, we state the Nelson's commutator theorem,
which guarantees the self-adjointness of $A$
in \eqref{1.014} (as for the proof,
see e.g. Reed-Simon~\cite{RS} and G{\'e}rard-{\L}aba~\cite{GL}). 

\begin{thm}\label{thm3.1} Let $\mathcal{K}$ be a Hilbert space.
Suppose that $N_0\ge c>0$ is a self-adjoint operator
on $\mathcal{K}$ and $A$ is a symmetric operator on $\mathcal{K}$
such that $D(N_0)\subset D(A)$ and
there exists a constant $C>0$ such that
\begin{align*}
&\|Au\|\le C\|N_0u\|\quad\textrm{for}\,\,u\in D(N_0),\\
&|(Au,N_0u)-(N_0u,Au)|\le C\|{N_0}^{1/2}u\|^2\quad\textrm{for}\,\,
u\in D(N_0)
\end{align*}
hold. Then $A$ is essentially self-adjoint on $D(N_0)$.
Denoting by $\bar{A}$ the unique self-adjoint extension
of $A$, if $u\in D(\bar{A})$, then $(1+i\epsilon N_0)^{-1}u$
converges to $u$ in the graph topology of $D(\bar{A})$
as $\epsilon\to0$.
\end{thm}

Applying Theorem \ref{thm3.1} with $\mathcal{K}=\mathscr{K}$ and
$N_0=\langle D_t\rangle+p^2/2+x^2/2$,
we see that $A$ in \eqref{1.014} has its unique
self-adjoint extension, which is also denoted by $A$.
\bigskip

In the usual proof of the Mourre estimate for $K$, one of the points
to be checked is that the condition
\begin{equation}
\sup_{|\kappa|\le1}\|Ke^{i\kappa A}(K+i)^{-1}\|_{\mathscr{B}(\mathscr{K})}<\infty
\label{3.1}
\end{equation}
is satisfied by a conjugate operator $A$ (see e.g. Mourre~\cite{M}).
However, it seems not easy to verify directly that $A$ satisfies \eqref{3.1}.
In order to overcome this difficulty,
we need the following proposition
(see e.g. Lemma 3.2.2 and Proposition 3.2.3 of \cite{GL};
see also Amrein-Boutet de Monvel-Georgescu~\cite{ABG}):

\begin{prop}\label{prop3.2} Let $\mathcal{K}$ be a Hilbert space.
Suppose that $K$, $K_0$ and $N_0$ are self-adjoint operators
on $\mathcal{K}$ such that $N_0\ge c>0$, $D(K)=D(K_0)$ as Banach spaces,
and for $z\in\boldsymbol{C}\setminus\sigma(K)$,
$(K-z)^{-1}$ preserves $D(N_0)$.
Let $A$ be a symmetric operator on $\mathcal{K}$. Suppose
that $K_0$ and $A$ satisfy
$D(N_0)\subset D(K_0)$, $D(N_0)\subset D(A)$,
\begin{align*}
&\|K_0u\|\le C\|N_0u\|\quad\textrm{for}\,\,u\in D(N_0),\\
&|(K_0u,N_0u)-(N_0u,K_0u)|\le C\|{N_0}^{1/2}u\|^2\quad\textrm{for}\,\,
u\in D(N_0),\\
&\|Au\|\le C\|N_0u\|\quad\textrm{for}\,\,u\in D(N_0),\\
&|(Au,N_0u)-(N_0u,Au)|\le C\|{N_0}^{1/2}u\|^2\quad\textrm{for}\,\,
u\in D(N_0).
\end{align*}
Denote the unique self-adjoint extension of $A$ also by $A$.
Assume moreover that
$$|(Au,Ku)-(Ku,Au)|\le C(\|Ku\|^2+\|u\|^2)\quad\textrm{for}\,\,u\in D(N_0)
$$
holds. Then the following hold:

\noindent
$(1)$ $D(N_0)$ is dense in $D(K)\cap D(A)$
with the norm $\|Ku\|+\|Au\|+\|u\|$.

\noindent
$(2)$ The commutator $i[K,A]$, defined as a quadratic form
on $D(K)\cap D(A)$, is the unique extension of the quadratic form
$i[K,A]$ on $D(N_0)$.

\noindent
$(3)$ $K\in C^1(A)$, that is, for some $z\in\boldsymbol{C}
\setminus\sigma(K)$, the map
$$\boldsymbol{R}\ni \kappa\mapsto e^{i\kappa A}
(K-z)^{-1}e^{-i\kappa A}\in \mathscr{B}(\mathcal{K})$$
is $C^1$ in the strong topology of $\mathscr{B}(\mathcal{K})$,
which is the algebra of bounded linear operators in $\mathcal{K}$.

\noindent
$(4)$ $D(K)\cap D(A)$ is a core for $K$, and the quadratic
form $i[K,A]$ on $D(K)\cap D(A)$ extends uniquely to a
bounded operator from $D(K)$ to its dual space $D(K)^*$,
which is denoted also by $i[K,A]$.

\noindent
$(5)$ The virial relation holds:\ \ For any $\lambda\in\boldsymbol{R}$,
$$E_K(\{\lambda\})i[K,A]E_K(\{\lambda\})=0$$
holds. Here $E_K(S)$ stands for the
spectral projection for $K$ onto $S\subset\boldsymbol{R}$.

\noindent
$(6)$ For $z\in\boldsymbol{C}\setminus\sigma(K)$,
$i[(K-z)^{-1},A]=-(K-z)^{-1}i[K,A](K-z)^{-1}$ holds.

\noindent
$(7)$ For $z\in\boldsymbol{C}\setminus\sigma(K)$,
$(K-z)^{-1}$ preserves $D(A)$.
\end{prop}

As for the characterization of the operators in $C^1(A)$,
we refer to the following proposition
(see e.g. Proposition 3.2.1 of \cite{GL};
see also \cite{ABG}):

\begin{prop}\label{prop3.3} Let $\mathcal{K}$ be a Hilbert space.
Suppose that $K$ and $A$ are self-adjoint operators
on $\mathcal{K}$. Then the following are equivalent:

\noindent
$(1)$ $K\in C^1(A)$.

\noindent
$(2)$ For some $z\in\boldsymbol{C}\setminus\sigma(K)$,
$$|(Au,(K-\bar{z})^{-1}u)-((K-z)^{-1}u,Au)|\le C\|u\|^2\quad\textrm{for}\,\,u\in D(A).$$

\noindent
$(3)$ For any $z\in\boldsymbol{C}\setminus\sigma(K)$,
$$|(Au,(K-\bar{z})^{-1}u)-((K-z)^{-1}u,Au)|\le C\|u\|^2\quad\textrm{for}\,\,u\in D(A).$$

\noindent
$(4)$ The following two conditions hold:

$(i)$ There exists some $C>0$ such that
$$|(Au,Ku)-(Ku,Au)|\le C\|(K+i)u\|^2\quad\textrm{for}\,\,u\in D(K)\cap D(A);$$

$(ii)$ There exists some $z\in\boldsymbol{C}\setminus\sigma(K)$
such that
$$\bigl\{u\in D(A)\bigm|(K-z)^{-1}u\in D(A),\ (K-\bar{z})^{-1}u\in D(A)\bigr\}$$

\quad\,\, is a core for $A$.
\end{prop}

\bigskip

From now on, we will give several propositions about the
Floquet Hamiltonians under consideration,
which are frequently used in this paper:

\begin{prop}\label{prop3.4}
Assume $V$ satisfies $(V_\mathrm{WR})$. Then the spectrum $\sigma(K)$ of the Floquet Hamiltonian
$K$ is $\omega$-periodic, that is,
$\sigma(K)=\sigma(K)+\omega\boldsymbol{Z}$.
\end{prop}

As is well-known, the periodic structure of $\sigma(K)$
stated in Proposition \ref{prop3.4}
has been used frequently in the previous works.
One has only to take account of the formula
$e^{-i\omega t}Ke^{i\omega t}=K+\omega$.

\begin{prop}\label{prop3.5} $(1)$ $\langle p\rangle^2\langle D_t\rangle^{-1}\langle K_0\rangle^{-1}$ and $\langle p\rangle\langle D_t\rangle^{-1/2}\langle K_0\rangle^{-1/2}$ are bounded.

\noindent
$(2)$ Suppose that $g\in C^2(\boldsymbol{R})$ satisfies
$\mathrm{supp}\,g\subset(0,\infty)$, and $g^{(k)}$, $k=0,\,1,\,2$,
are all bounded. Then $\langle p\rangle g(D_t)\langle K_0\rangle^{-1}$ and
$\langle D_t\rangle^{1/2}g(D_t)\langle K_0\rangle^{-1}$ are bounded.
\end{prop}

Proposition \ref{prop3.5} can be proved by the same argument as in the
case of the free Stark Hamiltonian $p^2/2-E\cdot x$ with $E\not=0$
(see e.g. \cite{AH},
\cite{HMS1} and Simon~\cite{S}). So we omit the proof.
\smallskip

The following proposition says the so-called local compactness
property of $K_0$.

\begin{prop}\label{prop3.6} Let $R>0$ and $z\in\boldsymbol{C}\setminus
\boldsymbol{R}$. Then
$F(|x|\le R)(K_0-z)^{-1}$ is compact on $\mathscr{K}$.
Here $F(|x|\le R)$ stands for a smoothed one of the
characteristic function of
$\bigl\{x\in X\bigm||x|\le R\bigr\}$,
whose definition is given in \S1.
\end{prop}

In the same way as
in the proof of 
Lemma 3.1 of Yajima~\cite{Ya1} and Lemma 4.6 of M{\o}ller~\cite{Mo},
the compactness of
\begin{align*}
&(F(|x|\le R)(K_0-z)^{-1})(F(|x|\le R)(K_0-z)^{-1})^*\\
={}&\frac{1}{z-\bar{z}}F(|x|\le R)\{(K_0-z)^{-1}-(K_0-\bar{z})^{-1}\}F(|x|\le R)\end{align*}
can be shown, which yields Proposition \ref{prop3.6} immediately.
So we omit the proof.

\begin{cor}\label{cor3.7} Assume $V$ satisfies $(V_\mathrm{WR})$. Then the following hold:

\noindent
$(1)$ Let $R>0$ and $z\in\boldsymbol{C}\setminus
\boldsymbol{R}$. Then 
$F(|x|\le R)(K-z)^{-1}$ is also compact on $\mathscr{K}$.

\noindent
$(2)$ Let $\mu>0$ and $z\in\boldsymbol{C}\setminus
\boldsymbol{R}$. Then $\langle x\rangle^{-\mu}(K-z)^{-1}$
as well as $\langle x\rangle^{-\mu}(K_0-z)^{-1}$ is
compact on $\mathscr{K}$.
\end{cor}

By using the second resolvent identity
\begin{equation}
(K-z)^{-1}-(K_0-z)^{-1}=-
(K_0-z)^{-1}V(K-z)^{-1},\label{3.2}
\end{equation}
Corollary \ref{cor3.7}
follows from Proposition \ref{prop3.6} immediately.
\bigskip

As is well-known, in the proof due to Froese-Herbst~\cite{FH}
of the Mourre estimate for time-independent
$N$-body Schr\"odinger operators,
the compactness of $f(H)\tilde{J}_a(x)-\tilde{J}_a(x)f(H_a)$
is one of the keys in the induction process with respect to
$N$, where $H$ is the full Hamiltonian,
$H_a$ is the cluster Hamiltonian with $a\in\mathscr{A}_0
=\mathscr{A}\setminus\{a_\mathrm{max}\}$,
and $f$ is compactly supported.

In the case of the Floquet Hamiltonians,
$f(K)\tilde{J}_a(x)-\tilde{J}_a(x)f(K_a)$
is not compact, in general. In this paper,
when $N=3$,
by replacing $\tilde{J}_a(x)$ by $\hat{J}_a=\tilde{J}_a(L\langle D_t\rangle;x)$,
which has been introduced in \S2,
we watch $f(K)\hat{J}_a-\hat{J}_af(K_a)$:

\begin{prop}\label{prop3.8}
Let $a\in\mathscr{A}_0$ and $f\in C_0^\infty(\boldsymbol{R};\boldsymbol{R})$.
Then $f(K)\hat{J}_a-\hat{J}_af(K_a)$
is compact on $\mathscr{K}$.
\end{prop}

\begin{proof}
We will prove the proposition
by the almost analytic extension method:
Take an almost analytic extension $\tilde{f}\in C_0^\infty(\boldsymbol{C})$
of $f\in C_0^\infty(\boldsymbol{R};\boldsymbol{R})$.
We first consider the case where $a\in\mathscr{A}_0^0$.
Then we see that
\begin{align*}
&f(K)\hat{J}_a-\hat{J}_af(K_a)\\
={}&\frac{1}{2\pi i}\int_{\boldsymbol{C}}\overline{\partial}_{\zeta}\tilde{f}(\zeta)\{(\zeta-K)^{-1}\hat{J}_a-\hat{J}_a(\zeta-K_a)^{-1}\}\,d\zeta\land d\overline{\zeta}\\
={}&\frac{1}{2\pi i}\int_{\boldsymbol{C}}\overline{\partial}_{\zeta}\tilde{f}(\zeta)(\zeta-K)^{-1}\\
&\times\{-i(\nabla\hat{J}_a)\cdot p-(\Delta\hat{J}_a)/2+(V\hat{J}_a-\hat{J}_aV^a)\}(\zeta-K_a)^{-1}\,d\zeta\land d\overline{\zeta}
\end{align*}
holds. By using Lemma \ref{lem2.04}, and the $K_0$-compactness of
$\langle x\rangle^{-1}$ by Corollary \ref{cor3.7} (2), one can see that 
the first operator of the right-hand side is compact.
Since  $\Delta\hat{J}_a=O(\langle x\rangle^{-2})$
by \eqref{2.09}, it follows from Corollary \ref{cor3.7} (2) that
the second operator is also compact.
Now we rewrite $V\hat{J}_a-\hat{J}_aV^a$ as
$[V,\hat{J}_a]+\hat{J}_aI_a$. By Lemma \ref{lem2.005},
$[V,\hat{J}_a]=O(\langle x\rangle^{-\rho_{1,0}''})$ holds.
By Lemma \ref{lem2.006},
$I_a\hat{J}_a=O(\langle x\rangle^{-\rho_{0,0}''})$ holds.
Hence, by virtue of Corollary \ref{cor3.7} (2),
the third operator is also compact.
The case where $a=a_\mathrm{min}$ can be shown quite similarly.
These show the proposition.
\end{proof}

On the other hand, in the case where $N=2$, by virtue of
\eqref{3.2}, the compactness of $\langle K_0\rangle\{f(K)-f(K_0)\}$ (cf. Lemma 2.2 of \cite{FH}),
can be proved easily under the following condition:
\medskip

\noindent
$(V_\mathrm{WR})^{(2)}$ Suppose $\rho_0>0$, $\rho_1>1$ and $\rho_2\ge2$.
$V_{12}(t,y)$
belongs to $C^2(\boldsymbol{R}\times
\boldsymbol{R}^d;\boldsymbol{R})$,
is $T$-periodic in $t$, 
and satisfies the decaying conditions
\begin{equation}
\sup_{t\in\boldsymbol{R}}|(\partial_t^m\partial_y^\alpha
V_{12})(t,y)|\le C_{m,\alpha}\langle y\rangle^{-\rho_m-|\alpha|},\quad
0\le m+|\alpha|\le2.\label{3.03}
\end{equation}

\noindent
As for $V_{12}$ satisfying $(V_\mathrm{WR})$,
it also satisfies $(V_\mathrm{WR})^{(2)}$ trivially.

\begin{prop}\label{prop3.9}
Assume $N=2$, and $V$ satisfies $(V_\mathrm{WR})^{(2)}$.
Let $f\in C_0^\infty(\boldsymbol{R};\boldsymbol{R})$.
Then $\langle K_0\rangle\{f(K)-f(K_0)\}$
is compact on $\mathscr{K}$.
\end{prop}

\section{Two-body case revisited}

In this section, we revisit the proof of the Mourre estimate
for $K$ with $N=2$ throughout this section. So we suppose $N=2$.
For the sake of simplicity, we will consider a regular
potential $V^{(1,2)}$ satisfying $(V_\mathrm{WR})^{(2)}$ introduced in \S3.

First we state some properties of
$A=\tilde{A}_0$ given by \eqref{1.012}.
Unlike in Adachi-Kiyose~\cite{AK},
the weight in \eqref{1.012}
is not the signed one $(\lambda_0-\delta-D_t)^{-1}$
but the positive one $\langle D_t\rangle^{-1}$.

\begin{lem}\label{lem4.1}
As for $i[K_0,A]$ and $i[i[K_0,A],A]$,
\begin{equation}
\begin{split}
&i[K_0,A]=\langle D_t\rangle^{-1}p^2=2\langle D_t\rangle^{-1}(K_0-D_t),\\
&i[i[K_0,A],A]=2\langle D_t\rangle^{-2}p^2=4\langle D_t\rangle^{-2}(K_0-D_t)
\end{split}\label{4.1}
\end{equation}
hold. Hence, $i[K_0,A]\langle K_0\rangle^{-1}$
and $i[i[K_0,A],A]\langle K_0\rangle^{-1}$ are bounded.
\end{lem}

\begin{lem}\label{lem4.2}
Assume $N=2$, and $V=V^{(1,2)}$ satisfies $(V_\mathrm{WR})^{(2)}$.
As for $i[V,A]$ and $i[i[V,A],A]$,
\begin{equation}
\begin{split}
&i[V,A]=i[V,\langle D_t\rangle^{-1}]\hat{A}_0+\langle D_t\rangle^{-1}i[V,\hat{A}_0],\\
&i[i[V,A],A]=i[i[V,\langle D_t\rangle^{-1}],\langle D_t\rangle^{-1}]\hat{A}_0^2+\langle D_t\rangle^{-1}i[i[V,\langle D_t\rangle^{-1}],\hat{A}_0]\hat{A}_0\\
&\qquad\qquad\qquad+\langle D_t\rangle^{-1}i[i[V,\hat{A}_0],\langle D_t\rangle^{-1}]\hat{A}_0+\langle D_t\rangle^{-2}i[i[V,\hat{A}_0],\hat{A}_0]
\end{split}\label{4.2}
\end{equation}
hold. Hence, $i[V,A]\langle K_0\rangle^{-1}$
and $i[i[V,A],A]\langle K_0\rangle^{-1}$ are bounded.
Moreover, $\langle K_0\rangle^{-1}i[V,A]\langle K_0\rangle^{-1}$ is compact.
\end{lem}

By simple computation, \eqref{4.1} and \eqref{4.2} can be obtained.
Therefore Lemma \ref{lem4.1} follows trivially.
Lemma \ref{lem4.2} can be shown in the same way as
in \cite{AK}: For instance, as for $i[V,A]$, by
$(\partial_tV)(t,x)=O(\langle x\rangle^{-\rho_1})$
and $(\nabla V)(t,x)=O(\langle x\rangle^{-\rho_0-1})$,
we see that
$i[V,A]\langle K_0\rangle^{-1}$ is bounded,
by virtue of Proposition \ref{prop3.5} $(1)$.
Here we used
\begin{equation}
i[V,\langle D_t\rangle^{-1}]=
\frac{1}{2\pi i}\int_{\boldsymbol{C}}\overline{\partial}_{\zeta}\tilde{g}(\zeta)(\zeta-D_t)^{-1}i[V,D_t](\zeta-D_t)^{-1}\,d\zeta\land d\overline{\zeta}\label{4.03}
\end{equation}
with $i[V,D_t]=-\partial_tV$, where
$\tilde{g}\in C^\infty(\boldsymbol{C})$ is an almost analytic extension 
of $g(\nu)=\langle \nu\rangle^{-1}$, which satisfies $|\overline{\partial}_z\tilde{g}(z)|\le C_L\langle z\rangle^{-1-1-L}|\mathrm{Im}\,z|^L$ with $L\in\boldsymbol{N}\cup\{0\}$.
Moreover, by virtue of Corollary \ref{cor3.7} $(2)$,
we also see that $\langle K_0\rangle^{-1}i[V,A]\langle K_0\rangle^{-1}$ is compact,
by $\rho_1>1$ and $\rho_0>0$.
\bigskip

One can show the following theorem and corollary:

\begin{thm}\label{thm4.3}
Suppose $N=2$. Assume $V$ satisfies $(V_\mathrm{WR})^{(2)}$. Put
$$\varTheta=\sigma_{\mathrm{pp}}(K^{a_\mathrm{min}})=\sigma_{\mathrm{pp}}(D_t)=\omega\boldsymbol{Z},\quad
\widehat{\varTheta}=\bigcup_{a\in\mathscr{A}}\sigma_{\mathrm{pp}}(K^a)=
\varTheta\cup\sigma_\mathrm{pp}(K).$$
Let $\lambda_0\in[0,\omega)$.
Then the following hold:

\noindent
$(1)$ Put
$$d_0(\lambda)=\mathrm{dist}(\lambda,\varTheta),\quad
d_1(\lambda)=\mathrm{dist}(\lambda,\varTheta\cap(-\infty,\lambda])$$
for $\lambda\in\boldsymbol{R}$. Note $d_0(\lambda)\le d_1(\lambda)$. 
Suppose $\lambda_0\in[0,\omega)\setminus\varTheta$,
that is, $\lambda_0\in(0,\omega)$.
Take $\delta$ as $0<\delta\le d_0(\lambda_0)/2$.
Then,
for any $f_{\lambda_0,\delta}\in C_0^\infty(\boldsymbol{R};\boldsymbol{R})$
supported in $[\lambda_0-\delta,\lambda_0+\delta]$,
\begin{equation}
f_{\lambda_0,\delta}(K)i[K,A]f_{\lambda_0,\delta}(K)
\ge\frac{2(d_1(\lambda_0)-\delta)}{1+\omega}
f_{\lambda_0,\delta}(K)^2+C_0
\label{4.3}
\end{equation}
holds with some compact operator
$C_0$ on $\mathscr{K}$.
For $\lambda_0\in[0,\omega)$, $d_1(\lambda_0)=\lambda_0$ holds.

Suppose $\lambda_0\in[0,\omega)\cap\varTheta$,
that is, $\lambda_0=0$.
Take $\delta$ as $0<\delta\le\omega/2$.
Then, for any $f_{\lambda_0,\delta}
\in C_0^\infty(\boldsymbol{R};\boldsymbol{R})$
supported in $[\lambda_0-\delta,\lambda_0+\delta]$,
\begin{equation}
f_{\lambda_0,\delta}(K)i[K,A]f_{\lambda_0,\delta}(K)
\ge C_0
\label{4.4}
\end{equation}
holds with some compact operator
$C_0$ on $\mathscr{K}$.

Hence, for any $\hat{\delta}$ such that
$0<\hat{\delta}<\delta$,
$\sigma_\mathrm{pp}(K)\cap(\lambda_0-\hat{\delta},\lambda_0+\hat{\delta})$
is finite, and that the eigenvalues of $K$
in $(\lambda_0-\hat{\delta},\lambda_0+\hat{\delta})$
are of finite multiplicity.

\noindent
$(2)$ Suppose $\lambda_0\in[0,\omega)\setminus\widehat{\varTheta}$.
Take $\epsilon$ as 
$0<\epsilon<d_1(\lambda_0)-d_0(\lambda_0)/2$.
Then there exists a small $\delta_{\epsilon,1}>0$ such that
$\delta_{\epsilon,1}\le d_0(\lambda_0)/2$ and
 for any $f_{\lambda_0,\delta}
\in C_0^\infty(\boldsymbol{R};\boldsymbol{R})$
supported in $[\lambda_0-\delta,\lambda_0+\delta]$
with $0<\delta\le\delta_{\epsilon,1}$,
\begin{equation}
f_{\lambda_0,\delta}(K)i[K,A]f_{\lambda_0,\delta}
(K)\ge\frac{2(d_1(\lambda_0)-\delta)-2\epsilon}{1+\omega}f_{\lambda_0,\delta_{\epsilon,1}}(K)^2
\label{4.5}
\end{equation}
holds. Suppose $s>1/2$ and $0<\hat{\delta}<\delta_{\epsilon,1}$. Then
\begin{equation}
\sup_{\substack{\mathrm{Re}\,z\in[\lambda_0-\hat{\delta},\lambda_0+\hat{\delta}]\\
\mathrm{Im}\,z\not=0}}\|
\langle A\rangle^{-s}(K-z)^{-1}\langle A\rangle^{-s}
\|_{\mathscr{B}(\mathscr{K})}<\infty
\label{4.6}
\end{equation}
holds. Moreover, $\langle A\rangle^{-s}(K-z)^{-1}\langle A\rangle^{-s}$
is a $\mathscr{B}(\mathscr{K})$-valued $\theta(s)$-H\"older continuous
function on $z\in S_{\lambda_0,\hat{\delta},\pm}$ with some $0<\theta(s)<1$.
And, there exist the norm limits
$$\langle A\rangle^{-s}(K-(\lambda\pm i0))^{-1}\langle A\rangle^{-s}=
\lim_{\varepsilon\to+0}\langle A\rangle^{-s}(K-(\lambda\pm i\varepsilon))^{-1}\langle A\rangle^{-s}$$
in $\mathscr{B}(\mathscr{K})$ for any
$\lambda\in[\lambda_0-\hat{\delta},\lambda_0+\hat{\delta}]$.
$\langle A\rangle^{-s}(K-(\lambda\pm i0))^{-1}\langle A\rangle^{-s}$
are also $\theta(s)$-H\"older continuous in $\lambda$.
\end{thm}

\begin{cor}\label{cor4.4}
Assume $V$ satisfies $(V_\mathrm{WR})^{(2)}$.
Then the following hold:

\noindent
$(1)$ The eigenvalues of $K$ in $\boldsymbol{R}\setminus\varTheta$
can accumulate only at $\varTheta$.
Moreover, $\widehat{\varTheta}$ is a countable
closed set.

\noindent
$(2)$ Let $I$ be a compact interval in
$\boldsymbol{R}\setminus\widehat{\varTheta}$.
Suppose $1/2<s\le1$. Then
\begin{equation}
\sup_{\substack{\mathrm{Re}\,z\in I\\
\mathrm{Im}\,z\not=0}}\|
\langle x\rangle^{-s}(K-z)^{-1}\langle x\rangle^{-s}
\|_{\mathscr{B}(\mathscr{K})}<\infty
\label{4.7}
\end{equation}
holds. Moreover, $\langle x\rangle^{-s}(K-z)^{-1}\langle x\rangle^{-s}$
is a $\mathscr{B}(\mathscr{K})$-valued $\theta(s)$-H\"older continuous
function on $z\in S_{I,\pm}$.
And, there exist the norm limits
$$\langle x\rangle^{-s}(K-(\lambda\pm i0))^{-1}\langle x\rangle^{-s}=
\lim_{\varepsilon\to+0}\langle x\rangle^{-s}(K-(\lambda\pm i\varepsilon))^{-1}
\langle x\rangle^{-s}$$
in $\mathscr{B}(\mathscr{K})$ for $\lambda\in I$.
$\langle x\rangle^{-s}(K-(\lambda\pm i0))^{-1}
\langle x\rangle^{-s}$ are also $\theta(s)$-H\"older continuous in $\lambda$.
\end{cor}

We will sketch the proof of the estimates \eqref{4.3} and \eqref{4.4}
only. Thus Theorem \ref{thm4.3} and Corollary \ref{cor4.4} can be shown
by the standard argument in the Mourre theory.
In particular, for the proof of Corollary \ref{cor4.4}, we use
the argument due to Perry-Sigal-Simon~\cite{PSS},
and the boundedness of 
$A(K-\lambda_0-i)^{-1}\langle x\rangle^{-1}$,
which follows from Proposition \ref{prop3.5} $(1)$.

\begin{proof}[Proof of \eqref{4.3} and \eqref{4.4}]

Let $\lambda_0\in[0,\omega)$. For a while,
take $\delta$ as $0<\delta\le\omega/2$.
Denote by $f_{\lambda_0,\delta}$
a function in $C_0^\infty(\boldsymbol{R};\boldsymbol{R})$
such that $\mathrm{supp}\,f_{\lambda_0,\delta}\subset[\lambda_0-\delta,\lambda_0+\delta]$.
Since $\langle K_0\rangle^{-1}i[V,A]\langle K_0\rangle^{-1}$ is compact by Lemma \ref{lem4.2}, we see that
\begin{equation}
f_{\lambda_0,\delta}(K)i[K,A]f_{\lambda_0,\delta}(K)
=f_{\lambda_0,\delta}(K)i[K_0,A]f_{\lambda_0,\delta}(K)+C_1
\label{4.8}
\end{equation}
holds with some compact operator $C_1$ on $\mathscr{K}$.
Since $i[K_0,A]\langle K_0\rangle^{-1}$ is bounded by Lemma \ref{lem4.1},
and $\langle K_0\rangle\{f_{\lambda_0,\delta}(K)-f_{\lambda_0,\delta}(K_0)\}$ is compact by Proposition \ref{prop3.9},
we obtain
\begin{equation}
f_{\lambda_0,\delta}(K)i[K,A]f_{\lambda_0,\delta}(K)
=f_{\lambda_0,\delta}(K_0)i[K_0,A]f_{\lambda_0,\delta}(K_0)+C_2
\label{4.9}
\end{equation}
with some compact operator $C_2$ on $\mathscr{K}$.
As in \S2, we utilize the Fourier transform $\widehat{\mathscr{F}}:L^2(\boldsymbol{T})\to\ell^2(\boldsymbol{Z})$.
Then $D_t$ can be decomposed into
the direct sum
$\bigoplus_{n\in\boldsymbol{Z}}n\omega$,
via $\widehat{\mathscr{F}}$.
By using $\widehat{\mathscr{F}}$,  
$f_{\lambda_0,\delta}(K_0)i[K_0,A]f_{\lambda_0,\delta}(K_0)$
can be decomposed into the direct sum
$$\bigoplus_{n\in\boldsymbol{Z}}
\frac{p^2f_{\lambda_0,\delta}(n\omega+p^2/2)^2}{(1+(n\omega)^2)^{1/2}}=
\bigoplus_{n\in\boldsymbol{Z}}
\frac{p^2f_{\lambda_0-n\omega,\delta}(p^2/2)^2}{(1+(n\omega)^2)^{1/2}}=
\bigoplus_{n\in\boldsymbol{Z}}I_0(n\omega).$$

We first consider the case where $\lambda_0\in[0,\omega)\setminus\varTheta$.
Take $\delta$ as $0<\delta\le d_0(\lambda_0)/2$.
If $n\le 0$, then the fibered operator
$I_0(n\omega)$ can be estimated as
\begin{align*}
I_0(n\omega)\ge{}&\frac{2\{(\lambda_0-n\omega)-\delta\}}{(1+(n\omega)^2)^{1/2}}
f_{\lambda_0-n\omega,\delta}(p^2/2)^2\\
\ge{}&\frac{2\{(\lambda_0+|n|\omega)-\delta\}}{1+\omega+|n|\omega}
f_{\lambda_0-n\omega,\delta}(p^2/2)^2
\ge\frac{2(\lambda_0-\delta)}{1+\omega}f_{\lambda_0-n\omega,\delta}(p^2/2)^2;
\end{align*}
while, if $n\ge1$, then $I_0(n\omega)=0$.
Since $1+\omega>\omega>\lambda_0$,
$(\lambda_0+|n|\omega)/(1+\omega+|n|\omega)$ is
monotone increasing in $|n|\omega$. If $n\ge1$,
then $\lambda_0-n\omega+\delta\le\lambda_0-\omega+\delta\le
\lambda_0-\omega+d_0(\lambda_0)/2<0$ holds. We used these.
Then we obtain
$$f_{\lambda_0,\delta}(K_0)
i[K_0,A]f_{\lambda_0,\delta}(K_0)
\ge\frac{2(\lambda_0-\delta)}{1+\omega}
f_{\lambda_0,\delta}(K_0)^2.$$
It follows from this, \eqref{4.9} and
the compactness of $\langle K_0\rangle\{f_{\lambda_0,\delta}(K)-f_{\lambda_0,\delta}(K_0)\}$
by Proposition \ref{prop3.9}
that \eqref{4.3} holds. Here we used
$d_1(\lambda_0)=\lambda_0$.

We next consider the case where $\lambda_0\in[0,\omega)\cap\varTheta$.
Take $\delta$ as $0<\delta\le\omega/2$. Then
$$f_{\lambda_0,\delta}(K_0)i[K_0,A]f_{\lambda_0,\delta}(K_0)
\ge0$$
holds. In the same way as above,
it can be shown easily that \eqref{4.4} holds.
\end{proof}

By using the method in the above proof, one can obtain the following:

\begin{lem}\label{lem4.05}
Suppose $f\in C_0^\infty(\boldsymbol{R};\boldsymbol{R})$ is supported in
$[0,2\omega]$. Let $k\in\boldsymbol{N}$. Then
there exists a constant $C>0$ such that
$$\|\langle D_t-k\omega\rangle^{-1}f(K_0)\|_{\mathscr{B}(\mathscr{K})}\le
C\langle k\omega\rangle^{-1}$$
holds.
\end{lem}

\begin{proof}
We will prove the lemma
by the almost analytic extension method:
Take an almost analytic extension $\tilde{f}\in C_0^\infty(\boldsymbol{C})$
of $f$
with $\mathrm{supp}\,\tilde{f}\cap\boldsymbol{R}\subset[0,2\omega]$.
Then $f(K_0)$ can be written as
$$f(K_0)=\frac{1}{2\pi i}\int_{\boldsymbol{C}}\overline{\partial}_{\zeta}\tilde{f}(\zeta)(\zeta-K_0)^{-1}\,d\zeta\land d\overline{\zeta}.$$
By using $\widehat{\mathscr{F}}$,  
$(\zeta-K_0)^{-1}$
can be decomposed into the direct sum
$\bigoplus_{n\in\boldsymbol{Z}}
((\zeta-n\omega)-p^2/2)^{-1}$.
We note that for $\zeta\in\mathrm{supp}\,\tilde{f}$, the 
following holds:
if $n\ge2$, or if $n=1$ and $\mathrm{Re}\,(\zeta-\omega)<0$,
then $\|((\zeta-n\omega)-p^2/2)^{-1}\|_{\mathscr{B}(\mathscr{K})}\le|\zeta-n\omega|^{-1}$
holds; while, if $n\le0$, or if $n=1$ and $\mathrm{Re}\,(\zeta-\omega)\ge0$,
then $\|((\zeta-n\omega)-p^2/2)^{-1}\|_{\mathscr{B}(\mathscr{K})}\le|\mathrm{Im}\,\zeta|^{-1}$
holds. When $n\ge1$, by Peetre's inequality, we have
$\langle (n-k)\omega\rangle^{-1}\le
\sqrt{2}\langle k\omega\rangle^{-1}\langle n\omega\rangle$,
which yields the estimate
$$\langle (n-k)\omega\rangle^{-1}|\zeta-n\omega|^{-1}\le
\sqrt{2}\langle k\omega\rangle^{-1}\langle n\omega\rangle|\zeta-n\omega|^{-1}\le C\langle k\omega\rangle^{-1}.$$
On the other hand, when $n\le0$, we use the estimate $\langle (n-k)\omega\rangle^{-1}\le
\langle k\omega\rangle^{-1}$ by $k\ge1$. These estimates yield the lemma,
because compactly supported $\tilde{f}$ satisfies $|\overline{\partial}_{\zeta}\tilde{f}(\zeta)|\le C_N|\mathrm{Im}\,\zeta|^N$
for $N\ge0$.
\end{proof}

\begin{cor}\label{cor4.6}
Suppose $f\in C_0^\infty(\boldsymbol{R};\boldsymbol{R})$ is supported in
$[0,2\omega]$. Let $k\in\boldsymbol{N}$. Then
there exists a constant $C>0$ such that
$$\|\langle D_t-k\omega\rangle^{-1}f(K)\|_{\mathscr{B}(\mathscr{K})}\le
C\langle k\omega\rangle^{-1}$$
holds.
\end{cor}

\begin{proof}
In the same way as in the proof of Lemma \ref{lem4.05}, one can show
$$\|\langle D_t-k\omega\rangle^{-1}\{f(K)-f(K_0)\}\|_{\mathscr{B}(\mathscr{K})}\le
C\langle k\omega\rangle^{-1},$$
by $(\zeta-K)^{-1}-(\zeta-K_0)^{-1}=(\zeta-K_0)^{-1}V(\zeta-K)^{-1}$.
This and Lemma \ref{lem4.05} yield the corollary.
\end{proof}

 This lemma will be used in the proof of Theorem \ref{thm1.1}.

\section{Proof of Theorem \ref{thm1.1}}

In this section, we prove Theorem \ref{thm1.1}. As in \S4,
we will show the estimate \eqref{1.16} only.
\eqref{1.16} yields the Mourre estimate \eqref{1.17}. Thus
Theorem \ref{thm1.1} and Corollary \ref{cor1.2} can be shown
by the standard argument in the Mourre theory.
Throughout this section, we assume
$V$ satisfies $(V_\mathrm{WR})$.
\medskip

Take the Yafaev-type partition of
unity $\{\hat{J}_a\}_{a\in\mathscr{A}}$,
and introduce the operator $A$ by \eqref{1.014}.
By using \eqref{2.016} obtained from the second representation of \eqref{1.014},
we have the following:

\begin{lem}\label{lem5.1}
Let $\lambda_0\in[0,\omega)$ and $\delta>0$.
Then
\begin{equation}
\begin{split}
&f_{\lambda_0,\delta}(K)i[K_0,A]f_{\lambda_0,\delta}(K)\\
\ge{}&f_{\lambda_0,\delta}(K)(\langle D_t\rangle^{-1}p^2)f_{\lambda_0,\delta}(K)\\
&+\sum_{a\in\mathscr{A}_0^0}f_{\lambda_0,\delta}(K)\hat{J}_a\{-\langle D_t\rangle^{-1}(p_a)^2+(p_a)^2\langle p_a/\sqrt{2}\rangle^{-2}\}\hat{J}_af_{\lambda_0,\delta}(K)\\
&\quad+f_{\lambda_0,\delta}(K)O(L^{-1/2})f_{\lambda_0,\delta}(K)+C
\end{split}\label{5.1}
\end{equation}
holds with some compact operator $C$ on $\mathscr{K}$,
where $L\ge1$ is the large parameter which comes from the definition of $\{\hat{J}_a\}_{a\in\mathscr{A}}$.
\end{lem}

\begin{proof}
Lemma \ref{lem4.1} yields
$i[K_0,\tilde{A}_0]=\langle D_t\rangle^{-1}p^2$ and
$i[K_0,-(\tilde{A}_0)_a]=-\langle D_t\rangle^{-1}$
$(p_a)^2$. On the other hand, we have
$i[K_0,\bar{A}_a]=(p_a)^2\langle p_a/\sqrt{2}\rangle^{-2}$ by simple
calculation. As mentioned in \S2,
$\bar{S}_{1,a}=S_{1,a}-S_{1,a}^\mathrm{m}$
is $K_0$-bounded such that $\langle K_0\rangle^{-1}\bar{S}_{1,a}\langle K_0\rangle^{-1}
=O(L^{-1/2})+C_{1,a}$ with some compact operator $C_{1,a}$
on $\mathscr{K}$. These yield \eqref{5.1}. 
\end{proof}

This estimate \eqref{5.1} is important for the sake of obtaining
the Mourre estimate for $K$.

On the other hand, as for $i[V,A]$, the
following can be obtained:

\begin{lem}\label{lem5.2}
Let $\lambda_0\in[0,\omega)$ and $\delta>0$.
Then
\begin{equation}
f_{\lambda_0,\delta}(K)i[V,A]f_{\lambda_0,\delta}(K)=\sum_{a\in\mathscr{A}_0^0}
f_{\lambda_0,\delta}(K)\hat{J}_ai[V^a,(\tilde{A}_0)^a]\hat{J}_af_{\lambda_0,\delta}(K)+C'
\label{5.2}
\end{equation}
holds with some compact operator $C'$ on $\mathscr{K}$.
\end{lem}

\begin{proof}
Lemma \ref{lem2.007}, in particular \eqref{2.0019}, yields
\eqref{5.2} trivially.
\end{proof}

By the facts mentioned in \S2, one can also obtain the following:

\begin{lem}
Both $i[K_0,A]\langle K_0\rangle^{-1}$
and $i[V,A]\langle K_0\rangle^{-1}$ are bounded.
This yields that $i[K,A]\langle K_0\rangle^{-1}$
is bounded. Moreover,
$\langle K_0\rangle^{-1}i[i[K,A],A]\langle K_0\rangle^{-1}$
is also bounded.
\end{lem}

Based on these, one can show 
\eqref{1.16} as follows:

\begin{proof}[Proof of \eqref{1.16}]

Suppose $\lambda_0\in[0,\omega)\setminus\varTheta$
and $\epsilon>0$.
Put
$$\varTheta=\bigcup_{a\in\mathscr{A}_0}\sigma_\mathrm{pp}(K^a),\quad
d_0(\lambda)=\mathrm{dist}(\lambda,\varTheta),\quad
d_1(\lambda)=\mathrm{dist}(\lambda,\varTheta\cap(-\infty,\lambda])
$$
for $\lambda\in\boldsymbol{R}$.
For a while,
take $\delta$ as $0<\delta\le d_0(\lambda_0)/2$.
Then both $\lambda_0-\delta\ge\lambda_0-d_0(\lambda_0)/2>0$ and $\lambda_0+\delta\le\lambda_0+d_0(\lambda_0)/2<\omega$ hold.

By the IMS localization formula for $\{\hat{J}_a\}_{a\in\mathscr{A}}$
$$p^2=\sum_{a\in\mathscr{A}}\hat{J}_ap^2\hat{J}_a+
\sum_{a\in\mathscr{A}}|\nabla \hat{J}_a|^2
$$
and 
$\sum_{a\in\mathscr{A}}|\nabla \hat{J}_a|^2=O(\langle x\rangle^{-2})$
obtained by Proposition \ref{prop2.1}, we have
\begin{align*}
&f_{\lambda_0,\delta}(K)(\langle D_t\rangle^{-1}p^2)f_{\lambda_0,\delta}(K)\\
={}&\sum_{a\in\mathscr{A}}f_{\lambda_0,\delta}(K)
\hat{J}_a(\langle D_t\rangle^{-1}p^2)\hat{J}_af_{\lambda_0,\delta}(K)+C_1\\
={}&\sum_{a\in\mathscr{A}_0}f_{\lambda_0,\delta}(K)
\hat{J}_a(\langle D_t\rangle^{-1}p^2)\hat{J}_af_{\lambda_0,\delta}(K)+C_1'
\end{align*}
with some compact operators $C_1$ and $C_1'$ on $\mathscr{K}$.
We also used the compactness of the support of $\hat{J}_{a_\mathrm{max}}$.
By this as well as Lemmas \ref{lem5.1} and \ref{lem5.2},
we have
\begin{equation}
\begin{split}
&f_{\lambda_0,\delta}(K)i[K,A]f_{\lambda_0,\delta}(K)\\
\ge{}&f_{\lambda_0,\delta}(K)\hat{J}_{a_\mathrm{min}}(\langle D_t\rangle^{-1}p^2)\hat{J}_{a_\mathrm{min}}f_{\lambda_0,\delta}(K)\\
&+\sum_{a\in\mathscr{A}_0^0}f_{\lambda_0,\delta}(K)\hat{J}_a\{i[K^a,(\tilde{A}_0)^a]+(p_a)^2\langle p_a/\sqrt{2}\rangle^{-2}\}\hat{J}_af_{\lambda_0,\delta}(K)\\
&\quad+f_{\lambda_0,\delta}(K)O(L^{-1/2})f_{\lambda_0,\delta}(K)+C_2
\end{split}\label{5.3}
\end{equation}
with some compact operator $C_2$ on $\mathscr{K}$.
Here we used $\langle D_t\rangle^{-1}p^2-\langle D_t\rangle^{-1}(p_a)^2=
\langle D_t\rangle^{-1}(p^a)^2=i[K_0^a,(\tilde{A}_0)^a]$, with
$K_0^a=D_t+(p^a)^2/2$.

We first consider $f_{\lambda_0,\delta}(K)\hat{J}_{a_\mathrm{min}}(\langle D_t\rangle^{-1}p^2)\hat{J}_{a_\mathrm{min}}f_{\lambda_0,\delta}(K)$.
Since $f_{\lambda_0,\delta}(K)\hat{J}_{a_\mathrm{min}}-\hat{J}_{a_\mathrm{min}}
f_{\lambda_0,\delta}(K_{a_\mathrm{min}})$ is compact by Proposition \ref{prop3.8},
we have
\begin{align*}
&f_{\lambda_0,\delta}(K)\hat{J}_{a_\mathrm{min}}(\langle D_t\rangle^{-1}p^2)\hat{J}_{a_\mathrm{min}}f_{\lambda_0,\delta}(K)\\
={}&\hat{J}_{a_\mathrm{min}}f_{\lambda_0,\delta}(K_{a_\mathrm{min}})(\langle D_t\rangle^{-1}p^2)f_{\lambda_0,\delta}(K_{a_\mathrm{min}})\hat{J}_{a_\mathrm{min}}+
C_{3,a_\mathrm{min}}
\end{align*}
with some compact operator $C_{3,a_\mathrm{min}}$ on $\mathscr{K}$.
In the same way as in the proof of Theorem \ref{thm4.3}, in particular
\eqref{4.3}, we have
$$f_{\lambda_0,\delta}(K_{a_\mathrm{min}})(\langle D_t\rangle^{-1}p^2)f_{\lambda_0,\delta}(K_{a_\mathrm{min}})
\ge\frac{2(\lambda_0-\delta)}{1+\omega}
f_{\lambda_0,\delta}(K_{a_\mathrm{min}})^2.$$
By using the compactness of 
$f_{\lambda_0,\delta}(K)\hat{J}_{a_\mathrm{min}}-\hat{J}_{a_\mathrm{min}}
f_{\lambda_0,\delta}(K_{a_\mathrm{min}})$ again,
\begin{equation}
\begin{split}
&f_{\lambda_0,\delta}(K)\hat{J}_{a_\mathrm{min}}(\langle D_t\rangle^{-1}p^2)\hat{J}_{a_\mathrm{min}}f_{\lambda_0,\delta}(K)\\
\ge{}&\frac{2(\widehat{d}_{1,a_\mathrm{min}}(\lambda_0)-\delta)}{1+\omega}f_{\lambda_0,\delta}(K)\hat{J}_{a_\mathrm{min}}^2
f_{\lambda_0,\delta}(K)+C_{3,a_\mathrm{min}}'
\end{split}\label{5.4}
\end{equation}
with some compact operator $C_{3,a_\mathrm{min}}'$ on $\mathscr{K}$, where
$$\widehat{d}_{1,a_\mathrm{min}}(\lambda)=\mathrm{dist}
(\lambda,\widehat{\varTheta}_{a_\mathrm{min}}\cap(-\infty,\lambda]);\quad
\widehat{\varTheta}_{a_\mathrm{min}}=\sigma_\mathrm{pp}(D_t)=\omega\boldsymbol{Z},$$
for $\lambda\in\boldsymbol{R}$. Thus $\widehat{d}_{1,a_\mathrm{min}}(\lambda_0)=
\lambda_0$ holds.

We next consider $f_{\lambda_0,\delta}(K)\hat{J}_a\{i[K^a,(\tilde{A}_0)^a]+(p_a)^2\langle p_a/\sqrt{2}\rangle^{-2}\}\hat{J}_af_{\lambda_0,\delta}(K)$ with $a\in\mathscr{A}_0^0$.
Since $f_{\lambda_0,\delta}(K)\hat{J}_a-\hat{J}_a
f_{\lambda_0,\delta}(K_a)$ is compact by Proposition \ref{prop3.8},
we have
\begin{align*}
&f_{\lambda_0,\delta}(K)\hat{J}_a\{i[K^a,(\tilde{A}_0)^a]+(p_a)^2\langle p_a/\sqrt{2}\rangle^{-2}\}\hat{J}_af_{\lambda_0,\delta}(K)\\
={}&\hat{J}_af_{\lambda_0,\delta}(K_a)\{i[K^a,(\tilde{A}_0)^a]+(p_a)^2\langle p_a/\sqrt{2}\rangle^{-2}\}f_{\lambda_0,\delta}(K_a)\hat{J}_a+C_{3,a}
\end{align*}
with some compact operator $C_{3,a}$ on $\mathscr{K}$.

As for $f_{\lambda_0,\delta}(K_a)\{(p_a)^2\langle p_a/\sqrt{2}\rangle^{-2}\}
f_{\lambda_0,\delta}(K_a)$,
via the Fourier transform $\mathscr{F}_a:\mathscr{H}_a\to\mathscr{H}_a$,
this can be decomposed into the direct integral
$$\int_{[0,\infty)}^\oplus F_{a,1}(\lambda_a)
\,d\lambda_a;\,\,\,
F_{a,1}(\lambda_a)=\tilde{F}_{a,1}(\lambda_a)
f_{\lambda_0-\lambda_a,\delta}(K^a)^2,\,\,\,
\tilde{F}_{a,1}(\lambda_a)=
\frac{2\lambda_a}{1+\lambda_a}.$$
On the other hand, as for
$f_{\lambda_0,\delta}(K_a)i[K^a,(\tilde{A}_0)^a]
f_{\lambda_0,\delta}(K_a)$,
via $\mathscr{F}_a:\mathscr{H}_a\to\mathscr{H}_a$,
this can be decomposed into the direct integral
$$\int_{[0,\infty)}^\oplus F_{a,0}(\lambda_a)
\,d\lambda_a;\quad
F_{a,0}(\lambda_a)=f_{\lambda_0-\lambda_a,\delta}(K^a)i[K^a,(\tilde{A}_0)^a]f_{\lambda_0-\lambda_a,\delta}(K^a).
$$
Then we watch the sum of the fibered operators
$F_a(\lambda_a)=F_{a,0}(\lambda_a)+F_{a,1}(\lambda_a)$.
Since $i[K_0^a,(\tilde{A}_0)^a]=\langle D_t\rangle^{-1/2}(p^a)^2\langle D_t\rangle^{-1/2}\ge0$ with
$K_0^a=D_t+(p^a)^2/2$, we have
$$F_{a,0}(\lambda_a)\ge f_{\lambda_0-\lambda_a,\delta}(K^a)i[V^a,(\tilde{A}_0)^a]f_{\lambda_0-\lambda_a,\delta}(K^a)
=F_{a,2}(\lambda_a)$$
with
$i[V^a,(\tilde{A}_0)^a]=
i[V^a,\langle D_t\rangle^{-1}](\tilde{A}_0)^a+\langle D_t\rangle^{-1}i[V^a,(\tilde{A}_0)^a]
$.
By using the floor function
$\lfloor\cdot\rfloor$, we translate 
$F_{a,2}(\lambda_a)$ into
\begin{align*}
&e^{i(\lfloor\omega^{-1}\lambda_a \rfloor+1)\omega t}F_{a,2}(\lambda_a)e^{-i(\lfloor\omega^{-1}\lambda_a \rfloor+1)\omega t}\\
={}&f_{\lambda_0+\lambda_a',\delta}(K^a)\{e^{i(\lfloor\omega^{-1}\lambda_a \rfloor+1)\omega t}i[K^a,(\tilde{A}_0)^a]e^{-i(\lfloor\omega^{-1}\lambda_a \rfloor+1)\omega t}\}f_{\lambda_0+\lambda_a',\delta}(K^a)
\end{align*}
with $\lambda_a'=(\lfloor\omega^{-1}\lambda_a \rfloor+1)\omega-\lambda_a\in(0,\omega]$,
which implies $\lambda_0+\lambda_a'\in(0,2\omega)$.
We used
$$e^{i(\lfloor\omega^{-1}\lambda_a \rfloor+1)\omega t}D_te^{-i(\lfloor\omega^{-1}\lambda_a \rfloor+1)\omega t}
=D_t-(\lfloor\omega^{-1}\lambda_a \rfloor+1)\omega.$$
Since $\langle x^a\rangle^{-1}\langle D_t\rangle^{-1/2}(\tilde{A}_0)^af_{\lambda_0+\lambda_a',\delta}(K^a)$
by Proposition \ref{prop3.5} $(1)$, $i[V^a,D_t]\langle x^a\rangle=O(\langle x^a\rangle^{-\rho_1+1})$ is bounded
by $\rho_1>1$, and $i[V^a,(\tilde{A}_0)^a]=O(\langle x^a\rangle^{-\rho_0})$ is bounded
by $\rho_0>0$, it follows from \eqref{4.03} with $V=V^a$ and Corollary \ref{cor4.6} that
\begin{align*}
&e^{i(\lfloor\omega^{-1}\lambda_a \rfloor+1)\omega t}F_{a,2}(\lambda_a)e^{-i(\lfloor\omega^{-1}\lambda_a \rfloor+1)\omega t}\\
={}&O(\langle (\lfloor\omega^{-1}\lambda_a \rfloor+1)\omega\rangle^{-1+1/2})+O(\langle (\lfloor\omega^{-1}\lambda_a \rfloor+1)\omega\rangle^{-1})\\
={}&O(\langle (\lfloor\omega^{-1}\lambda_a \rfloor+1)\omega\rangle^{-1/2}),
\end{align*} 
which implies $F_{a,2}(\lambda_a)=O(\langle (\lfloor\omega^{-1}\lambda_a \rfloor+1)\omega\rangle^{-1/2})$.
We used that for $n,\,k\in\boldsymbol{Z}$, 
$\langle n\omega\rangle^{1/2}\le
\sqrt[4]{2}\langle k\omega\rangle^{1/2}\langle (n-k)\omega\rangle^{1/2}$
holds by Peetre's inequality,
with $k=\lfloor\omega^{-1}\lambda_a \rfloor+1$.
We note that $\tilde{F}_{a,1}(\lambda_a)$ is monotone increasing in $\lambda_a$,
and $\tilde{F}_{a,1}(\omega)=2\omega/(1+\omega)>2\lambda_0/(1+\omega)$.
Therefore, there exists a large $\lambda_{a,1}\ge\omega$ such that
if $\lambda_a>\lambda_{a,1}$, then
\begin{equation}
F_a(\lambda_a)\ge
F_{a,2}(\lambda_a)+F_{a,1}(\lambda_a)\ge\frac{2\lambda_0}{1+\omega}f_{\lambda_0-\lambda_a,\delta}(K^a)^2
\label{5.5}
\end{equation}
holds.
We next consider the case where $0\le \lambda_a\le\lambda_{a,1}$.
By virtue of the arguments of \cite{FH} and the results of \S4,
we see that
there exists a small $0<\delta_{\epsilon,a}\le d_0(\lambda_0)/2$
such that if $0<\delta\le\delta_{\epsilon,a}$, then
the following holds:
When $\omega>\lambda_0-\lambda_a\ge0$ and
$\lambda_0-\lambda_a\not\in
\widehat{\varTheta}_a$,
\begin{equation}
F_{a,0}(\lambda_a)\ge\frac{2(\widehat{d}_{1,a}(\lambda_0-\lambda_a)-\delta)-\epsilon/2}{1+\omega}
f_{\lambda_0-\lambda_a,\delta}(K^a)^2\label{5.6}
\end{equation}
holds, where
$$\widehat{d}_{1,a}(\lambda)=\mathrm{dist}
(\lambda,\widehat{\varTheta}_a\cap(-\infty,\lambda]);\quad
\widehat{\varTheta}_a=\omega\boldsymbol{Z}\cup\sigma_\mathrm{pp}(K^a),$$
for $\lambda\in\boldsymbol{R}$; while,
when $\omega>\lambda_0-\lambda_a\ge0$ and
$\lambda_0-\lambda_a\in
\widehat{\varTheta}_a$,
\begin{equation}
F_{a,0}(\lambda_a)\ge\frac{-\epsilon/2}{1+\omega}
f_{\lambda_0-\lambda_a,\delta}(K^a)^2\label{5.7}
\end{equation}
holds. When $\lambda_0-\lambda_a<0$,
we will utilize the estimate \eqref{5.7}
consistently, though if $\lambda_0-\lambda_a\not\in\widehat{\varTheta}_a$, then
the estimates better than \eqref{5.7} can be obtained by the results of \S4. 
Here we used that $\lambda_0-\lambda_a$
belongs to the compact interval
$[\lambda_0-\lambda_{a,1},\lambda_0]$, which
makes the key argument of \cite{FH} available. We note that there exists a unique $\kappa_a\in\widehat{\varTheta}_a\cap
[0,\lambda_0]$ such that $\widehat{d}_{1,a}(\lambda_0)=
\lambda_0-\kappa_a>0$, since $\lambda_0\not\in\varTheta$.
When
$\lambda_0-\lambda_a>\kappa_a$, that is,
$0\le\lambda_a<\lambda_0-\kappa_a\le\lambda_0<\omega$,
$$
F_{a,0}(\lambda_a)
\ge\frac{2\{(\lambda_0-\lambda_a)-\kappa_a-\delta\}-\epsilon/2}{1+\omega}f_{\lambda_0-\lambda_a,\delta}(K^a)^2
$$
holds by \eqref{5.6}, because $\lambda_0-\lambda_a\not\in\widehat{\varTheta}_a$.
Since $\tilde{F}_{a,0}(\lambda_a)=2\lambda_a/(1+\lambda_a)\ge2\lambda_a/(1+\omega)$,
\begin{equation}
F_a(\lambda_a)
\ge\frac{2(\lambda_0-\kappa_a-\delta)-\epsilon/2}{1+\omega}f_{\lambda_0-\lambda_a,\delta}(K^a)^2
\label{5.8}
\end{equation}
holds.
When $0\le\lambda_0-\lambda_a\le\kappa_a$, that is,
$\lambda_0-\kappa_a\le\lambda_a\le\omega$, as mentioned above,
we utilize \eqref{5.7} for $F_{a,0}(\lambda_a)$ consistently. By the monotone increasing
of $\tilde{F}_{a,1}(\lambda_a)$ in $\lambda_a$,
$\tilde{F}_{a,1}(\lambda_a)\ge\tilde{F}_{a,1}(\lambda_0-\kappa_a)=2(\lambda_0-\kappa_a)/\{1+(\lambda_0-\kappa_a)\}\ge2(\lambda_0-\kappa_a)/(1+\omega)$ holds,
which implies
\begin{equation}
F_a(\lambda_a)
\ge\frac{2(\lambda_0-\kappa_a)-\epsilon/2}{1+\omega}f_{\lambda_0-\lambda_a,\delta}(K^a)^2\label{5.9}
\end{equation}
holds. When $\lambda_0-\lambda_a<0$, that is, $\lambda_a>\lambda_0$,
we also utilize \eqref{5.7} for $F_{a,0}(\lambda_a)$ consistently.
By the monotone increasing
of $\tilde{F}_{a,1}(\lambda_a)$ in $\lambda_a$,
$\tilde{F}_{a,1}(\lambda_a)\ge\tilde{F}_{a,1}(\lambda_0)=2\lambda_0/(1+\lambda_0)\ge2\lambda_0/(1+\omega)$ holds,
which implies
\begin{equation}
F_a(\lambda_a)
\ge\frac{2\lambda_0-\epsilon/2}{1+\omega}f_{\lambda_0-\lambda_a,\delta}(K^a)^2\label{5.10}
\end{equation}
holds.
From \eqref{5.5}, \eqref{5.8}, \eqref{5.9} and \eqref{5.10},
we obtain
\begin{align*}
&f_{\lambda_0,\delta}(K_a)\{i[K^a,(\tilde{A}_0)^a]+(p_a)^2\langle p_a/\sqrt{2}\rangle^{-2}\}f_{\lambda_0,\delta}(K_a)\\
\ge{}&\frac{2(\widehat{d}_{1,a}(\lambda_0)-\delta)-\epsilon/2}{1+\omega}f_{\lambda_0,\delta}(K_a)^2
\end{align*}
since
$\widehat{d}_{1,a}(\lambda_0)=
\lambda_0-\kappa_a<\lambda_0$.
By using the compactness of 
$f_{\lambda_0,\delta}(K)\hat{J}_a-\hat{J}_a
f_{\lambda_0,\delta}(K_a)$ again,
\begin{equation}
\begin{split}
&f_{\lambda_0,\delta}(K)\hat{J}_a\{i[K^a,(\tilde{A}_0)^a]+(p_a)^2\langle p_a/\sqrt{2}\rangle^{-2}\}\hat{J}_af_{\lambda_0,\delta}(K)\\
\ge{}&\frac{2(\widehat{d}_{1,a}(\lambda_0)-\delta)-\epsilon/2}{1+\omega}f_{\lambda_0,\delta}(K)\hat{J}_a^2
f_{\lambda_0,\delta}(K)+C_{3,a}'
\end{split}\label{5.11}
\end{equation}
with some compact operator $C_{3,a}'$ on $\mathscr{K}$.

By using \eqref{5.3}, \eqref{5.4} and \eqref{5.11},
one can obtain
\begin{equation}
\begin{split}
	&f_{\lambda_0,\delta}(K)i[K,A]f_{\lambda_0,\delta}(K)\\
\ge{}&\frac{2(\widehat{d}_{1,a_\mathrm{min}}(\lambda_0)-\delta)}{1+\omega}f_{\lambda_0,\delta}(K)
\hat{J}_{a_\mathrm{min}}(x)^2f_{\lambda_0,\delta}(K)\\
&+\sum_{a\in\mathscr{A}_0^0}\frac{2(\widehat{d}_{1,a}(\lambda_0)-\delta)-\epsilon/2}{1+\omega}f_{\lambda_0,\delta}(K)\hat{J}_a^2f_{\lambda_0,\delta}(K)\\
&\quad+f_{\lambda_0,\delta}(K)O(L^{-1/2})f_{\lambda_0,\delta}(K)+C_4\\
\ge{}&\frac{2(d_1(\lambda_0)-\delta)-\epsilon/2}{1+\omega}f_{\lambda_0,\delta}(K)^2
+f_{\lambda_0,\delta}(K)O(L^{-1/2})f_{\lambda_0,\delta}(K)+C_4'
\end{split}
\end{equation}
with some compact operators $C_4$ and $C_4'$ on $\mathscr{K}$,
by $d_1(\lambda_0)=\min\bigl\{\widehat{d}_{1,a}(\lambda_0)\bigm|
a\in\mathscr{A}_0\bigr\}$. We used the compactness of the
support of $\hat{J}_{a_\mathrm{max}}$.
By taking $L\ge1$ as sufficiently large, we obtain \eqref{1.16}.
\end{proof}

\section{Minimal velocity estimates}

As was shown by some works of Sigal-Soffer
(see e.g. \cite{SS1,SS2}),
by virtue of the Mourre estimate,
one can obtain the so-called minimal velocity estimate,
which is one of the useful propagation estimates
for the time evolution of scattering states.
As for the Floquet Hamiltonian $K$ under consideration,
we have obtained the Mourre estimate \eqref{1.17}.
The aim of this section is to show Theorem \ref{thm1.3},
by utilization of \eqref{1.17}.
As mentioned in \S1, Corollary \ref{cor1.4} is
a direct consequence of Theorem \ref{thm1.3},
by virtue of the arguments of Yajima-Kitada~\cite{YK}
and M{\o}ller-Skibsted~\cite{MS}.

We first state the abstract theory for getting the minimal
velocity estimate of the integral type,
by following G\'erard-{\L}aba~\cite{GL}.
The theory was initiated by Sigal-Soffer.
The following proposition is Proposition 4.4.1 of \cite{GL}
with $P=1$ and $I(\sigma)\equiv 0$:

\begin{prop}\label{prop6.1}
Let $\mathcal{K}$ be a Hilbert space.
Let $K$ and $A$ be self-adjoint operators on $\mathcal{K}$.
Assume that $K\in C^\mu(A)$ for some $\mu>1$.
Let $\Delta\subset\boldsymbol{R}$ be an open interval such that
$$E_K(\Delta)i[K,A]E_K(\Delta)\ge c_0E_K(\Delta)$$
holds for some $c_0>0$. Then for any $f\in C_0^\infty(\boldsymbol{R};
\boldsymbol{R})$ supported in $\Delta$
\begin{align*}
&\int_1^\infty
\left\|F\left(\frac{A}{\sigma}\le c_0-\varepsilon\right)
e^{-i\sigma K}f(K)u\right\|_\mathcal{K}^2\frac{d\sigma}{\sigma}
\le C\|u\|_\mathcal{K}^2,\\
&\slim_{\sigma\to\infty}
F\left(\frac{A}{\sigma}\le c_0-\varepsilon\right)
e^{-i\sigma K}f(K)=0
\end{align*}
hold for any $\varepsilon>0$.
\end{prop}

By virtue of Proposition \ref{prop6.1}
and the Mourre estimate \eqref{1.17},
one can obtain the following:

\begin{prop}\label{prop6.2}
Suppose that the hypotheses of Theorem \ref{thm1.3} are satisfied.
Then
\begin{align}
&\int_1^\infty
\left\|F\left(\frac{A}{\sigma}\le c_0'(\lambda_0)\right)
f_{\lambda_0,\delta}(K)e^{-i\sigma K}\varPhi\right\|_\mathscr{K}^2\frac{d\sigma}{\sigma}
\le C\|\varPhi\|_\mathscr{K}^2,\label{6.1}\\
&\slim_{\sigma\to\infty}F\left(\frac{A}{\sigma}\le c_0'(\lambda_0)\right)
f_{\lambda_0,\delta}(K)e^{-i\sigma K}=0
\end{align}
hold, where
\begin{align*}
c_0(\lambda_0)={}&c_0(\lambda_0;\epsilon,\delta)=
\frac{2(d_1(\lambda_0)-\delta)-2\epsilon}{1+\omega},\\
0<c_0'(\lambda_0)={}&c_0'(\lambda_0;\epsilon,\delta)=
\frac{2(d_1(\lambda_0)-\delta)-3\epsilon}{1+\omega}<c_0(\lambda_0)
\end{align*}
for sufficiently small $\epsilon>0$.
\end{prop}

However, in the study of the problem of the asymptotic completeness,
we have to translate \eqref{6.1} into a certain minimal velocity
estimate with the localization of the propagation with respect to $x\in X$.
To this end, we need the following lemma and
proposition (see Lemma 4.4.8 and Proposition 4.4.9 of \cite{GL}).
Lemma \ref{lem6.3} is used for proving Proposition \ref{prop6.4}:

\begin{lem}\label{lem6.3}
Let $\mathcal{K}$ be a Hilbert space.
Let $K$, $A$ and $B$ be self-adjoint operators on $\mathcal{K}$.
Let $\Delta\subset\boldsymbol{R}$ be an open interval. Assume that

\noindent
$(1)$ $K\in C^1(A)\cap C^1(B)$ and $B\in C^1(A)$;

\noindent
$(2)$ $[B,A](B+i)^{-1}$ and $(B+i)^{-1}A(K+i)^{-1}$ are bounded;

\noindent
$(3)$ $B\ge1/2$;

\noindent
$(4)$ $-cB\le f_1(K)Af_1(K)\le cB$ holds for some $c>0$,
and $f_1\in C_0^\infty(\boldsymbol{R};\boldsymbol{R})$
such that $f_1=1$ on $\Delta$.

Then, for any $c_0>0$, there is an $\varepsilon_0>0$ such that
for any $f\in C_0^\infty(\boldsymbol{R};\boldsymbol{R})$
supported in $\Delta$
$$F\left(\frac{B}{\sigma}\le\varepsilon_0\right)
f(K)F\left(\frac{|A|}{\sigma}\ge c_0\right)=O(\sigma^{-1})
$$
holds.
\end{lem}

\begin{prop}\label{prop6.4}
Let $\mathcal{K}$ be a Hilbert space.
Let $K$, $A$ and $B$ be self-adjoint operators on $\mathcal{K}$.
Let $\Delta\subset\boldsymbol{R}$ be an open interval.
Suppose that the hypotheses of Proposition \ref{prop6.1}
and Lemma \ref{lem6.3} are satisfied. Then,
there exists an $\varepsilon_0>0$ such that
for any $f\in C_0^\infty(\boldsymbol{R};\boldsymbol{R})$
supported in $\Delta$
\begin{align*}
&\int_1^\infty
\left\|F\left(\frac{B}{\sigma}\le\varepsilon_0\right)
e^{-i\sigma K}f(K)u\right\|_\mathcal{K}^2\frac{d\sigma}{\sigma}
\le C\|u\|_\mathcal{K}^2,\\
&\slim_{\sigma\to\infty}
F\left(\frac{B}{\sigma}\le\varepsilon_0\right)
e^{-i\sigma K}f(K)=0
\end{align*}
hold.
\end{prop}

In the case where $K$ is a time-independent Schr\"odinger operator
$p^2/2+V$, one can
take $B$ as $\langle x\rangle$.
In fact, $K$ is in $C^1(\langle x\rangle)$,
because $i[K,\langle x\rangle]=
\mathrm{Re}\,\{(x/\langle x\rangle)\cdot p\}$
is $K$-bounded. However, in the case
where $K$ is a Floquet Hamiltonian  
$D_t+p^2/2+V(t)$, $K$ is not in $C^1(\langle x\rangle)$,
even if $V(t)$ is independent of $t$.
In fact, $i[K,\langle x\rangle]=
\mathrm{Re}\,\{(x/\langle x\rangle)\cdot p\}$
is not $K$-bounded.
In our analysis, by modifying $(1+\langle D_t\rangle^{-1/2}
x^2\langle D_t\rangle^{-1/2})^{1/2}$,
we will take $B$ as \eqref{1.21}
\begin{align*}
&B=(1+\tilde{B}_0+\tilde{B}_1)^{1/2};\quad
\tilde{B}_0=\langle D_t\rangle^{-1/2}Q_0
\langle D_t\rangle^{-1/2},\\
&\tilde{B}_1=\sum_{a\in\mathscr{A}_0^0}\tilde{B}_{1,a},\quad
\tilde{B}_{1,a}=\langle p_a/\sqrt{2}\rangle^{-1}Q_{1,a}\langle p_a/\sqrt{2}\rangle^{-1},
\end{align*}
where
\begin{align*}
&Q_0=\sum_{a\in\mathscr{A}\setminus\mathscr{A}_0^0}
x^2\hat{J}_a^2
+\sum_{a\in\mathscr{A}_0^0}
(x^a)^2\hat{J}_a^2=x^2-\sum_{a\in\mathscr{A}_0^0}
(x_a)^2\hat{J}_a^2,\\
&Q_{1,a}=(x_a)^2\hat{J}_a^2,
\end{align*}
with $a\in\mathscr{A}_0^0$. Then we will show Theorem \ref{thm1.3}
as follows:

\begin{proof}[Proof of Theorem \ref{thm1.3}]
We first show that $K$ is in $C^1(B)$, that is,
$(K-i)^{-1}i[K,B](K+i)^{-1}$ is bounded,
by virtue of Proposition \ref{prop3.3}:
Let $g_{1/2}\in C^\infty(\boldsymbol{R};\boldsymbol{R})$ be
such that $g_{1/2}\ge0$, $g_{1/2}(\nu)=0$ for $\nu\le0$,
and $g_{1/2}(\nu)=\nu^{1/2}$ for $\nu\ge1$.
Take an almost analytic extension $\tilde{g}_{1/2}\in C^\infty(\boldsymbol{C})$
of $g_{1/2}$,
which satisfies $|\overline{\partial}_z\tilde{g}_{1/2}(z)|\le C_L\langle z\rangle^{1/2-1-L}|\mathrm{Im}\,z|^L$ with $L\in\boldsymbol{N}\cup\{0\}$.
Then, by virtue of the commutator expansion formula (see e.g. \cite{DG}),
$[K,B]=[K,g_{1/2}(B^2)]$ is represented as
\begin{align*}
&[K,B]=g_{1/2}^{(1)}(B^2)[K,B^2]+R_2'=\frac{1}{2}B^{-1}[K,\tilde{B}_0+\tilde{B}_1]+R_2';\\
&R_2'=
\frac{1}{2\pi i}\int_{\boldsymbol{C}}\overline{\partial}_{\zeta}\tilde{g}_{1/2}(\zeta)(\zeta-B^2)^{-2}[[K,B^2],B^2](\zeta-B^2)^{-1}\,d\zeta\land d\overline{\zeta}.
\end{align*}
We will
show that $B^{-1}i[K,\tilde{B}_0](K+i)^{-1}$, $B^{-1}i[K,\tilde{B}_{1,a}](K+i)^{-1}$ and $R_2'$ are bounded.

We first consider
$i[K,\tilde{B}_0]=S_{0,0}+S_{0,1}+S_{0,2}+S_{0,3}$
with
\begin{align*}
&S_{0,0}=(\nabla Q_0)\cdot p\langle D_t\rangle^{-1},\quad
S_{0,1}=-i(\Delta Q_0)\langle D_t\rangle^{-1}/2,\\
&S_{0,2}=i[V,Q_0]\langle D_t\rangle^{-1},\quad
S_{0,3}=Q_0i[V,\langle D_t\rangle^{-1}]
\end{align*}
Since $V$ does not commute with $\hat{J}_a^2$ generally,
we also have to deal with $S_{0,2}$.
Trivially, $(1+\tilde{B}_0)^{-1/2}(\nabla Q_0)\langle D_t\rangle^{-1/2}$ is bounded.
Since
$\langle D_t\rangle^{1/2}
p\langle D_t\rangle^{-1}\langle K_0\rangle^{-1/2}$ is bounded
by Proposition \ref{prop3.5} $(1)$,
$(1+\tilde{B}_0)^{-1/2}S_{0,0}\langle K_0\rangle^{-1/2}$
is bounded. Since $\Delta Q_0$ is bounded, $S_{0,1}$ is bounded trivially.
By applying Lemma \ref{lem2.005} to $i[V,Q_0]=-\sum_{a\in\mathscr{A}_0^0}
(x_a)^2i[V,\hat{J}_a^2]$, we have
$$S_{0,2}=\sum_{a\in\mathscr{A}_0^0}O(\langle x^a\rangle^{2-\rho_{1,0}''})\hat{J}_a\langle D_t\rangle^{-1}$$
because $i[V,\hat{J}_a]=O(\langle x\rangle^{-\rho_{1,0}''})$.
Noting that $2-\rho_{1,0}''<1$ holds by $\rho_{1,0}''>1$,
we see that
$(1+\tilde{B}_0)^{-\max\{2-\rho_{1,0}'',0\}/2}O(\langle x^a\rangle^{2-\rho_{1,0}''})\hat{J}_a\langle D_t\rangle^{-\max\{2-\rho_{1,0}'',0\}/2}$ is bounded.
In particular, this implies that $(1+\tilde{B}_0)^{-1/2}S_{0,2}$ is bounded.
For any
$b\in\mathscr{A}_0^0$,
\begin{align*}
&Q_0(\partial_tV^b)(t,x^b)\\
={}&\sum_{a\in\mathscr{A}\setminus\mathscr{A}_0^0}x^2\hat{J}_a^2
(\partial_tV^b)(t,x^b)
+\sum_{\substack{a\in\mathscr{A}_0^0\\ a\not=b}}(x^a)^2\hat{J}_a^2
(\partial_tV^b)(t,x^b)+(x^b)^2\hat{J}_b^2(\partial_tV^b)(t,x^b)\\
={}&\hat{J}_{a_\mathrm{max}}^2O(1)+
\hat{J}_{a_\mathrm{min}}^2O(\langle x\rangle^{2-\rho_{1,0}''})
+\sum_{a\in\mathscr{A}_0^0}\hat{J}_a^2O(\langle x^a\rangle^{2-\rho_{1,0}''})
\end{align*}
holds.
Here we used the compactness of the support of $\hat{J}_{a_\mathrm{max}}$, and
$\hat{J}_a^2(\partial_tV^b)(t,x^b)=O(\langle x\rangle^{-\rho_{1,0}''})$
holds for $a\in\mathscr{A}_0$ such that $b\not\subset a$ by Lemma \ref{lem2.006}
In the same way as above, we see that
$(1+\tilde{B}_0)^{-\max\{2-\rho_{1,0}'',0\}/2}Q_0(\partial_tV^b)(t,x^b)\langle D_t\rangle^{-\max\{2-\rho_{1,0}'',0\}/2}$ is bounded.
In particular, this implies that $(1+\tilde{B}_0)^{-1/2}S_{0,3}$ is also bounded.

After all, we see that $B^{-1}i[K,\tilde{B}_0](K+i)^{-1}$ is bounded.

In the same way as above, we consider $i[K,\tilde{B}_{1,a}]=
S_{1,a,0}+S_{1,a,1}+S_{1,a,2}+S_{1,a,3}$ with
\begin{align*}
&S_{1,a,0}=\langle p_a/\sqrt{2}\rangle^{-1}(\nabla Q_{1,a})\cdot p
\langle p_a/\sqrt{2}\rangle^{-1},\\
&S_{1,a,1}=-i\langle p_a/\sqrt{2}\rangle^{-1}
(\Delta Q_{1,a})\langle p_a/\sqrt{2}\rangle^{-1}/2,\\
&S_{1,a,2}=i[I_a,\tilde{B}_{1,a}],\quad
S_{1,a,3}=\langle p_a/\sqrt{2}\rangle^{-1}i[V^a,Q_{1,a}]
\langle p_a/\sqrt{2}\rangle^{-1}.
\end{align*}
Here we used $i[V^a,\langle p_a/\sqrt{2}\rangle^{-1}]=0$.
We note $\nabla Q_{1,a}=q_{1,a,1}+q_{1,a,2}$
with $q_{1,a,1}=2\hat{J}_a^2x_a$ and $q_{1,a,2}=2(x_a)^2\hat{J}_a(\nabla\hat{J}_a)$,
and that $(1+\tilde{B}_{1,a})^{-1/2}\langle p_a/\sqrt{2}\rangle^{-1}\hat{J}_aO(\langle x_a\rangle)$ is bounded.
Since $x_a\cdot p=x_a\cdot p_a$ and $p_a\langle p_a/\sqrt{2}\rangle^{-1}$ is bounded,
one can see easily that 
$(1+\tilde{B}_{1,a})^{-1/2}\langle p_a/\sqrt{2}\rangle^{-1}
q_{1,a,1}\cdot p\langle p_a/\sqrt{2}\rangle^{-1}$ is bounded. Since
$\langle x\rangle(\nabla\hat{J}_a)\cdot p\langle K_0\rangle^{-1/2}$ is bounded
by Lemma \ref{lem2.04},
$(1+\tilde{B}_{1,a})^{-1/2}\langle p_a/\sqrt{2}\rangle^{-1}
q_{1,a,2}\cdot p\langle K_0\rangle^{-1/2}$ is bounded.
These imply $(1+\tilde{B}_{1,a})^{-1/2}S_{1,a,0}\langle K_0\rangle^{-1/2}$ is bounded.
Since $(\Delta Q_{1,a})=O(1)$,
$S_{1,a,1}$ is bounded trivially.

Now we watch $S_{1,a,2}$.
We will show that $(1+\tilde{B}_{1,a})^{-1/2}S_{1,a,2}$ is bounded:
As for the commutator $[h(x),\langle p_a/\sqrt{2}\rangle^{-1}]$,
where $h(x)$ is the multiplication by $h(x)$, we use
the following commutator expansion formula:
Let $g_{-1/2}\in C^\infty(\boldsymbol{R};\boldsymbol{R})$ be
such that $g_{-1/2}\ge0$, $g_{-1/2}(\nu)=0$ for $\nu\le0$,
and $g_{-1/2}(\nu)=\nu^{-1/2}$ for $\nu\ge1$.
Take an almost analytic extension $\tilde{g}_{-1/2}\in
C^\infty(\boldsymbol{C})$
of $g_{-1/2}$, which satisfies $|\overline{\partial}_z\tilde{g}_{-1/2}(z)|\le C_L\langle z\rangle^{-1/2-1-L}|\mathrm{Im}\,z|^L$ with $L\in\boldsymbol{N}\cup\{0\}$.
By regarding $\langle p_a/\sqrt{2}\rangle^{-1}$ as
$g_{-1/2}(\langle p_a/\sqrt{2}\rangle^2)$,
$[h(x),\langle p_a/\sqrt{2}\rangle^{-1}]$
is represented as
\begin{align*}
&[h(x),\langle p_a/\sqrt{2}\rangle^{-1}]=-[h(x),(p_a)^2/2]\langle p_a/\sqrt{2}\rangle^{-3}/2+R_2'';\\
&R_2''=
\frac{-1}{2\pi i}\int_{\boldsymbol{C}}\overline{\partial}_{\zeta}\tilde{g}_{-1/2}(\zeta)(\zeta-\langle p_a/\sqrt{2}\rangle^2)^{-1}\\
&\qquad\qquad\qquad\qquad\qquad\times[[h(x),(p_a)^2/2],(p_a)^2/2](\zeta-\langle p_a/\sqrt{2}\rangle^2)^{-2}\,d\zeta\land d\overline{\zeta}.
\end{align*}
Here we used
$\langle p_a/\sqrt{2}\rangle^2=1+(p_a)^2/2$.
By virtue of this formula, we have
$\tilde{B}_{1,a}=\tilde{B}_{1,a,0}+\tilde{B}_{1,a,1}+\tilde{B}_{1,a,2}$
with 
\begin{align*}
&\tilde{B}_{1,a,0}=Q_{1,a}\langle p_a/\sqrt{2}\rangle^{-2},\quad
\tilde{B}_{1,a,1}=i(\nabla_aQ_{1,a})\cdot p_a\langle p_a/\sqrt{2}\rangle^{-4}/2,\\
&\tilde{B}_{1,a,2}=R''\langle p_a/\sqrt{2}\rangle^{-4},
\end{align*}
where $R''$ is bounded. Trivially, $i[I_a,\tilde{B}_{1,a,2}]=i(I_a\tilde{B}_{1,a,2}-
\tilde{B}_{1,a,2}I_a)$ is bounded. We focus on
$i[I_a,\tilde{B}_{1,a,0}]=(\bar{B}_{1,a,0,1}+\bar{B}_{1,a,0,2})\langle p_a/\sqrt{2}\rangle^{-2}$
with $\bar{B}_{1,a,0,1}=i[I_a,Q_{1,a}]$ and
$\bar{B}_{1,a,0,2}=Q_{1,a}\langle p_a/\sqrt{2}\rangle^{-2}i[(p_a)^2/2,I_a]$.
Using $i[I_a,\hat{J}_a]=O(\langle x\rangle^{-\rho_{1,0}''})$ by Lemma \ref{lem2.005}, 
we see that $(1+\tilde{B}_{1,a})^{-\max\{2-\rho_{1,0}'',0\}/2}
\bar{B}_{1,a,0,1}\langle p_a/\sqrt{2}\rangle^{-\max\{2-\rho_{1,0}'',0\}}$
is bounded, in the same way as above.
We watch $\bar{B}_{1,a,0,2}$.
By
$[Q_{1,a},\langle p_a/\sqrt{2}\rangle^{-2}]=
-i\langle p_a/\sqrt{2}\rangle^{-2}\{(\nabla_aQ_{1,a})\cdot p_a-i(\Delta_aQ_{1,a})/2\}\langle p_a/\sqrt{2}\rangle^{-2}$,
in the same way as above, $(1+\tilde{B}_{1,a})^{-1/2}[Q_{1,a},\langle p_a/\sqrt{2}\rangle^{-2}]$ is bounded.
Since $i[(p_a)^2/2,I_a]=(\nabla_aI_a)\cdot p_a-i(\Delta_aI_a)/2$,
$(1+\tilde{B}_{1,a})^{-1/2}[Q_{1,a},\langle p_a/\sqrt{2}\rangle^{-2}]i[(p_a)^2/2,I_a]
\langle p_a/\sqrt{2}\rangle^{-1}$ is bounded trivially. And,
since $Q_{1,a}(\nabla_aI_a)=O(\langle x\rangle^{2-\rho_{0,1}''})$
and $Q_{1,a}(\Delta_aI_a)=O(\langle x\rangle^{2-\rho_{0,2}''})$
by Lemma \ref{lem2.006},
$(1+\tilde{B}_{1,a})^{-\max\{2-\rho_{0,1}'',0\}/2}\langle p_a/\sqrt{2}\rangle^{-2}Q_{1,a}i[(p_a)^2/2,I_a]
\langle p_a/\sqrt{2}\rangle^{-1}$ is bounded.
These imply that $(1+\tilde{B}_{1,a})^{-1/2}
\bar{B}_{1,a,0,2}\langle p_a/\sqrt{2}\rangle^{-1}$
is bounded. We finally see that
$(1+\tilde{B}_{1,a})^{-1/2}i[I_a,\tilde{B}_{1,a,0}]$ is bounded.
Quite similarly, one  can show that $i[I_a,\tilde{B}_{1,a,1}]$ is bounded.
Therefore, it has been shown that $(1+\tilde{B}_{1,a})^{-1/2}S_{1,a,2}$ is bounded.

As for $S_{1,a,3}$, since $i[V^a,\hat{J}_a]=O(\langle x\rangle^{-\rho_{1,0}''})$ by Lemma \ref{lem2.005},
one can show that $(1+\tilde{B}_{1,a})^{-\max\{2-\rho_{1,0}'',0\}/2}S_{1,a,3}$ is bounded.

After all, we see that
$B^{-1}i[K,\tilde{B}_{1,a}](K+i)^{-1}$ is bounded.

In the same way as above, one can show that $B^{-2}i[i[K,B^2],B^2]$ is bounded,
which yields the boundedness of $R_2'$, by
$1/2-1-2+1=-3/2<-1$.

Summing up these, we see that $i[K,B](K+i)^{-1}$ is bounded.
In particular,
$K$ is in
$C^1(B)$. On the other hand, one can prove easily that
$B$ is in $C^1(A)$, and that
$[B,A](B+i)^{-1}$ and $(B+i)^{-1}A(K+i)^{-1}$ are bounded.
Now we will take $f_{\lambda_0,\delta_{\epsilon,1},1}\in
C_0^\infty(\boldsymbol{R};\boldsymbol{R})$ 
supported in $[\lambda_0-2\delta_{\epsilon,1},\lambda_0
+2\delta_{\epsilon,1}]$ such that
$f_{\lambda_0,\delta_{\epsilon,1},1}=1$
on $[\lambda_0-\delta_{\epsilon,1},\lambda_0
+\delta_{\epsilon,1}]$, and
put
$$c=\|B^{-1/2}f_{\lambda_0,\delta_{\epsilon,1},1}(K)
Af_{\lambda_0,\delta_{\epsilon,1},1}(K)B^{-1/2}\|_{\mathscr{B}(\mathscr{K})}$$
Then 
$$-cB\le f_{\lambda_0,\delta_{\epsilon,1},1}(K)Af_{\lambda_0,\delta_{\epsilon,1},1}(K)\le cB$$
holds.
Then, Theorem \ref{thm1.3} follows from Proposition \ref{prop6.2} immediately,
by virtue of Lemma \ref{lem6.3} and Proposition \ref{prop6.4}.
\end{proof}

For the sake of comparison, also in the case where $N=2$,
we will give the results corresponding to Theorem \ref{thm1.3} and Corollary \ref{cor1.4}, without proof.

\begin{thm}\label{thm6.5}
Suppose $N=2$. Assume $V$ satisfies $(V_\mathrm{WR})^{(2)}$. Let $\lambda_0\in[0,\omega)\setminus\widehat{\varTheta}$.
Put
\begin{equation}
B=\left(1+\langle D_t\rangle^{-1/2}x^2\langle D_t\rangle^{-1/2}\right)^{1/2}.
\end{equation}
Then there exists an $\varepsilon_0(\lambda_0)=\varepsilon_0(\lambda_0;\epsilon,\delta)>0$,
which is determined by the positive constant $\{2(d_1(\lambda_0)-\delta)-2\epsilon\}/(1+\omega)$ in \eqref{4.5},
\begin{align}
&\int_1^\infty\left\|F\left(\frac{B}{\sigma}\le\varepsilon_0(\lambda_0)\right)e^{-i\sigma K}f_{\lambda_0,\delta}(K)\varPhi\right\|_{\mathscr{K}}^2\,\frac{d\sigma}{\sigma}\le C\|\varPhi\|_{\mathscr{K}}^2,
\\
&\slim_{\sigma\to\infty}
F\left(\frac{B}{\sigma}\le\varepsilon_0(\lambda_0)\right)e^{-i\sigma K}f_{\lambda_0,\delta}(K)=0
\end{align}
hold. In particular, these yield
\begin{align}
&\int_1^\infty\left\|F\left(\frac{\langle x\rangle}{\sigma}\le\varepsilon_0(\lambda_0)\right)e^{-i\sigma K}f_{\lambda_0,\delta}(K)\varPhi\right\|_{\mathscr{K}}^2\,\frac{d\sigma}{\sigma}\le C\|\varPhi\|_{\mathscr{K}}^2,
\\
&\slim_{\sigma\to\infty}F\left(\frac{\langle x\rangle}{\sigma}\le\varepsilon_0(\lambda_0)\right)e^{-i\sigma K}f_{\lambda_0,\delta}(K)=0.
\end{align}
\end{thm}

\begin{cor}\label{cor6.6}
Suppose that the hypotheses of Theorem \ref{thm6.5} are satisfied.
Let $g_{\lambda_0,\delta}$ be
the function on $\bigl\{z\in\boldsymbol{C}\bigm||z|=1\bigr\}$
such that $g_{\lambda_0,\delta}(e^{-i2\pi\lambda/\omega})=
f_{\lambda_0,\delta}(\lambda)$ for $\lambda\in[0,\omega)$.
Then
\begin{align}
&\int_1^\infty\left\|F\left(\frac{\langle x\rangle}{t}\le\varepsilon_0(\lambda_0)\right)U(t,0)g_{\lambda_0,\delta}(U(T,0))\phi\right\|_{\mathscr{H}}^2\,\frac{dt}{t}\le C\|\phi\|_{\mathscr{H}}^2,
\\
&\slim_{t\to\infty}F\left(\frac{\langle x\rangle}{t}\le\varepsilon_0(\lambda_0)\right)U(t,0)g_{\lambda_0,\delta}(U(T,0))=0
\end{align}
hold.
\end{cor}

\begin{rem}
In the case where $N=2$,
as for the minimal velocity estimate of
the time-wise type, by virtue of the abstract theory of Skibsted~\cite{Sk},
which was also initiated by Sigal-Soffer, Yokoyama~\cite{Yo} gave the result
that for $0<s'<s$,
\begin{equation}
\left\|F\left(\frac{|x|}{t}\le c_1(\lambda_0;\epsilon,\delta)\right)
e^{-i\sigma K}f_{\lambda_0,\delta}(K)
\langle x\rangle^{-s}\right\|_{\mathscr{B}(\mathscr{K})}=
O(\sigma^{-s'})\label{6.11}
\end{equation}
holds with
$$c_1(\lambda_0;\epsilon,\delta)=\left(
\frac{2(d_1(\lambda_0)-\delta-\epsilon)}{1+2(d_1(\lambda_0)-\delta)}\right)^{1/2},$$
under the assumption that
$V_{12}(t,y)\in C^0(\boldsymbol{R};C^\infty(\boldsymbol{R}^d;\boldsymbol{R}))$
is $T$-periodic in $t$, and satisfies the decaying conditions
\begin{equation}
\sup_{t\in\boldsymbol{R}}|(\partial_y^\alpha V_{12})(t,y)|
\le C_\alpha\langle y\rangle^{-\rho-|\alpha|},
\end{equation}
by using the minimal velocity estimate like
\begin{equation}
\left\|F\left(\frac{\bar{\bar{A}}_{0,1}}{t}\le c_1(\lambda_0;\epsilon,\delta)\right)
e^{-i\sigma K}f_{\lambda_0,\delta}(K)
\langle\bar{\bar{A}}_{0,1}\rangle^{-s}\right\|_{\mathscr{B}(\mathscr{K})}=
O(\sigma^{-s})
\end{equation}
in terms of the conjugate operator $\bar{\bar{A}}_{0,1}$ in \eqref{1.06}.
The advantage of \eqref{6.11} is that $c_1(\lambda_0;\epsilon,\delta)$
can be taken as the one
nearly equal to the square root of the positive constant
of the Mourre estimate \eqref{1.08}.

Getting the minimal velocity estimate like \eqref{6.11} also in the
case where $N=3$ is one of the future tasks. We think that to this end,
we need a rather strong assumption on $V_{jk}$'s like
that each $V_{jk}\in C^\infty
(\boldsymbol{R};C^\infty(\boldsymbol{R}^d;\boldsymbol{R}))$
is $T$-periodic in $t$, and satisfies the decaying conditions
\begin{equation}
\sup_{t\in\boldsymbol{R}}|(\partial_t^m\partial_y^\alpha V_{jk})(t,y)|
\le C_{m,\alpha}\langle y\rangle^{-\rho-(m+|\alpha|)}.
\end{equation}
\end{rem}

\section{AC Stark effect case}

In this section, we consider
a quantum system of $N$ particles moving in a
given time-periodic electric field $\mathscr{E}(t)\in \boldsymbol{R}^d$.
We suppose that $\mathscr{E}(t)\in C^0(\boldsymbol{R};\boldsymbol{R}^d)$,
and that $\mathscr{E}(t)$ has a period $T>0$, that is,
$\mathscr{E}(t+T)=\mathscr{E}(t)$ for any $t\in\boldsymbol{R}$.

Let $m_j>0$, $q_j\in\boldsymbol{R}$ and $r_j \in \boldsymbol{R}^d$,
$1\le j\le N$, denote the mass, charge and
position vector of the $j$-th particle, respectively.
We suppose that the particles under consideration
interact with one another through the time-independent pair
potentials $\bar{V}_{jk}(r_j-r_k)$, $1\le j<k\le N$.
The system under consideration is governed by the total Hamiltonian
in the laboratory frame
\begin{equation}
\begin{split}
&\bar{H}_\mathrm{LF}(t)=\sum_{j=1}^N\left(-\frac{1}{2m_j}\Delta_j-q_j\langle \mathscr{E}(t),
r_j\rangle\right)+\bar{V}(r);\\
&\bar{V}(r)=\sum_{1\le j<k\le N}\bar{V}_{jk}(r_j-r_k)
\end{split}
\label{7.1}
\end{equation}
acting on $L^2(\boldsymbol{R}^{d\times N})$.
$\bar{H}_\mathrm{LF}(t)$ can be written as
\begin{equation}
\bar{H}_\mathrm{LF}(t)=\bar{H}_{\mathrm{LF},0}(t)+\bar{V}(r);\quad
\bar{H}_{\mathrm{LF},0}(t)=\frac{1}{2}(p_{\bar{X}})^2-E_{\bar{X}}(t)\cdot r
\label{7.2}
\end{equation}
acting on $L^2(\bar{X})$, where
$$E_{\bar{X}}(t)=((q_1/m_1)\mathscr{E}(t),\ldots,(q_N/m_N)\mathscr{E}(t))$$
is a $\bar{X}$-valued $T$-periodic function.
$q_j/m_j$ is called the specific charge of the $j$-th particle.

Now we would like to state the so-called
Avron-Herbst formula for the propagator
$\bar{U}_\mathrm{LF}(t,s)$ generated by $\bar{H}_\mathrm{LF}(t)$:
In the same way as in M\o ller~\cite{Mo} and Adachi~\cite{A2,A3},
introduce $\bar{X}$-valued $T$-periodic functions
$E_{\bar{X},\mathrm{os}}(t)$, $b_{\bar{X}}(t)$,
$b_{\bar{X},\mathrm{os}}(t)$ and $c_{\bar{X}}(t)$ as
\begin{align*}
&E_{\bar{X},\mathrm{m}}=\frac{1}{T}\int_0^TE_{\bar{X}}(s)\,ds,\quad
E_{\bar{X},\mathrm{os}}(t)=E_{\bar{X}}(t)-E_{\bar{X},\mathrm{m}},\\
&b_{\bar{X}}(t)=\int_0^tE_{\bar{X},\mathrm{os}}(s)\,ds,\quad
b_{\bar{X},\mathrm{m}}=\frac{1}{T}\int_0^Tb_{\bar{X}}(s)\,ds,\\
&b_{\bar{X},\mathrm{os}}(t)=b_{\bar{X}}(t)-b_{\bar{X},\mathrm{m}},\quad
c_{\bar{X}}(t)=\int_0^tb_{\bar{X},\mathrm{os}}(s)\,ds.
\end{align*}
By introducing $\boldsymbol{R}^d$-valued $T$-periodic functions
$\mathscr{E}_\mathrm{os}(t)$, $\bar{\mathscr{E}}(t)$,
$\bar{\mathscr{E}}_\mathrm{os}(t)$ and $\bar{\bar{\mathscr{E}}}(t)$ as
\begin{align*}
&\mathscr{E}_\mathrm{m}=\frac{1}{T}\int_0^T\mathscr{E}(s)\,ds,\quad
\mathscr{E}_\mathrm{os}(t)=\mathscr{E}(t)-\mathscr{E}_\mathrm{m},\\
&\bar{\mathscr{E}}(t)=\int_0^t\mathscr{E}_\mathrm{os}(s)\,ds,\quad
\bar{\mathscr{E}}_\mathrm{m}=\frac{1}{T}\int_0^T\bar{\mathscr{E}}(s)\,ds,\\
&\bar{\mathscr{E}}_\mathrm{os}(t)=\bar{\mathscr{E}}(t)-\bar{\mathscr{E}}_\mathrm{m},\quad
\bar{\bar{\mathscr{E}}}(t)=\int_0^t\bar{\mathscr{E}}_\mathrm{os}(s)\,ds,
\end{align*}
we have
\begin{align*}
&E_{\bar{X},\mathrm{m}}=((q_1/m_1)\mathscr{E}_\mathrm{m},\ldots,(q_N/m_N)\mathscr{E}_\mathrm{m}),\\
&E_{\bar{X},\mathrm{os}}(t)=((q_1/m_1)\mathscr{E}_\mathrm{os}(t),\ldots,(q_N/m_N)\mathscr{E}_\mathrm{os}(t)),\\
&b_{\bar{X}}(t)=((q_1/m_1)\bar{\mathscr{E}}(t),\ldots,(q_N/m_N)\bar{\mathscr{E}}(t)),\\
&b_{\bar{X},\mathrm{os}}(t)=((q_1/m_1)\bar{\mathscr{E}}_\mathrm{os}(t),\ldots,(q_N/m_N)\bar{\mathscr{E}}_\mathrm{os}(t)),\\
&c_{\bar{X}}(t)=((q_1/m_1)\bar{\bar{\mathscr{E}}}(t),\ldots,(q_N/m_N)\bar{\bar{\mathscr{E}}}(t)).
\end{align*}
$\mathscr{E}_\mathrm{m}$ is the time-mean of $\mathscr{E}(t)$.
Also introduce the time-dependent Hamiltonian
\begin{equation}
\begin{split}
&\bar{H}_\mathrm{MF}(t)=\bar{H}_{\mathrm{MF},0}+\bar{V}(r+c_{\bar{X}}(t));\quad
\bar{H}_{\mathrm{MF},0}=\frac{1}{2}(p_{\bar{X}})^2-E_{\bar{X},\mathrm{m}}
\cdot r,\\
&\bar{V}(r+c_{\bar{X}}(t))=\sum_{1\le j<k\le N}\bar{V}_{jk}((r_j+c_{\bar{X},j}(t))-(r_k+c_{\bar{X},k}(t)))
\end{split}
\label{7.3}
\end{equation}
acting on $L^2(\bar{X})$, which governs the system in the moving frame
accelerated by $E_{\bar{X},\mathrm{os}}(t)=E_{\bar{X}}(t)-E_{\bar{X},\mathrm{m}}$.
$c_{\bar{X},j}(t)=(q_j/m_j)\bar{\bar{\mathscr{E}}}(t)$ stands for the $j$-th component of $c_{\bar{X}}(t)$.
Here we will emphasize that the free Hamiltonian
$\bar{H}_{\mathrm{MF},0}$ is time-independent even if
$\bar{H}_{\mathrm{LF},0}(t)$ is time-dependent.
The time-independence of $\bar{H}_{\mathrm{MF},0}$ makes the dynamics of
the system governed by $\bar{H}_\mathrm{MF}(t)$ easy to handle.

If $E_{\bar{X},\mathrm{m}}=0$, then
$\bar{H}_{\mathrm{MF},0}$ is called the free Schr\"odinger operator;
while, if $E_{\bar{X},\mathrm{m}}\not=0$, then
$\bar{H}_{\mathrm{MF},0}$ is called the free DC Stark Hamiltonian.
Also, if $q_j/m_j=q_k/m_k$, then
\begin{align*}
&\bar{V}_{jk}((r_j+c_{\bar{X},j}(t))-(r_k+c_{\bar{X},k}(t)))\\
={}&\bar{V}_{jk}(r_j-r_k+((q_j/m_j)-(q_k/m_k))\bar{\bar{\mathscr{E}}}(t))
\equiv\bar{V}_{jk}(r_j-r_k)
\end{align*}
holds, that is, $\bar{V}_{jk}((r_j+c_{\bar{X},j}(t))-(r_k+c_{\bar{X},k}(t)))$
is time-independent; while, if $q_j/m_j\not=q_k/m_k$, then
the periodicity of $\bar{\bar{\mathscr{E}}}(t)$ in $t$
yields that of
$\bar{V}_{jk}((r_j+c_{\bar{X},j}(t))-(r_k+c_{\bar{X},k}(t)))$.
Hence, if there is no pair $(j,k)$ such that $q_j/m_j\not=q_k/m_k$,
then $\bar{H}_\mathrm{MF}(t)\equiv \bar{H}_{\mathrm{MF},0}+\bar{V}(r)$
holds, that is, $\bar{H}_\mathrm{MF}(t)$
is also time-independent. So,
from now on, we suppose there exists at least one pair
$(j,k)$ such that $q_j/m_j\not=q_k/m_k$.
But, even under this assumption,
$\bar{H}_\mathrm{MF}(t)$ is still time-independent
if $\mathscr{E}(t)\equiv\mathscr{E}_\mathrm{m}$,
that is, $\mathscr{E}(t)$ is constant in $t$,
because $\bar{\bar{\mathscr{E}}}(t)\equiv 0$.

As for the case where $\bar{H}_\mathrm{MF}(t)$ is time-independent,
the problem of the asymptotic completeness
of the systems governed by such Hamiltonians
was studied intently in the
1980's and 1990's; as for the case where $\mathscr{E}(t)\equiv\mathscr{E}_\mathrm{m}=0$, see e.g. Sigal-Soffer~\cite{SS1}, Graf~\cite{Gr}, Yafaev~\cite{Yaf}, Derezi\'nski~\cite{D}, and so on; while, as for the case where
$\mathscr{E}(t)\equiv\mathscr{E}_\mathrm{m}\not=0$, see e.g. 
Adachi-Tamura~\cite{AT1,AT2} and Herbst-M{\o}ller-Skibsted~\cite{HMS1,HMS2}.
So, from now on, we assume that $\mathscr{E}(t)$ is
not constant but periodic in $t$. Let
$\bar{U}_\mathrm{MF}(t,s)$ denote the propagator
generated by $\bar{H}_\mathrm{MF}(t)$.
Then the Avron-Herbst formula for $\bar{U}_\mathrm{LF}(t,s)$
\begin{equation}
\begin{split}
&\bar{U}_\mathrm{LF}(t,s)=\mathscr{T}_{\bar{X}}(t)\bar{U}_\mathrm{MF}(t,s)
\mathscr{T}_{\bar{X}}(s)^*;\\
&\mathscr{T}_{\bar{X}}(t)=e^{-ia_{\bar{X}}(t)}e^{ib_{\bar{X},\mathrm{os}}(t)\cdot r}
e^{-ic_{\bar{X}}(t)\cdot p_{\bar{X}}}
\end{split}
\label{7.4}
\end{equation}
holds, where
$$a_{\bar{X}}(t)=\int_0^t\left(\frac{1}{2}b_{\bar{X},\mathrm{os}}(s)^2-E_{\bar{X},\mathrm{m}}\cdot c_{\bar{X}}(s)\right)\,ds.
$$
By virtue of the Avron-Herbst formula \eqref{7.4}, the understanding
of $\bar{U}_\mathrm{MF}(t,s)$ yields that of $\bar{U}_\mathrm{LF}(t,s)$
immediately.

Next we would like to watch the motion of the systems in the
center-of-mass frame.
Here we note that
$\bar{V}(r)$ is independent of $x_\mathrm{cm}$, and that
$\bar{V}(r+c_{\bar{X}}(t))$
is independent of $\pi_\mathrm{cm}(r+c_{\bar{X}}(t))=x_\mathrm{cm}+c_\mathrm{cm}(t)$. Hence we will write $\bar{V}(r)$ and $\bar{V}(r+c_{\bar{X}}(t))$
as $\bar{V}(x)$ and $\bar{V}(x+c(t))$, respectively. Here
we put $c(t)=\pi c_{\bar{X}}(t)$ and $c_\mathrm{cm}(t)=\pi_\mathrm{cm}c_{\bar{X}}(t)$.
Now we introduce the Hamiltonians
\begin{equation}
\begin{split}
&\hat{H}_\mathrm{LF}(t)=\hat{H}_{\mathrm{LF},0}(t)+\bar{V}(x);\quad
\hat{H}_{\mathrm{LF},0}(t)=\frac{1}{2}p^2-E(t)\cdot x,
\\
&\hat{H}_\mathrm{MF}(t)=\hat{H}_{\mathrm{MF},0}+\bar{V}(x+c(t));\quad
\hat{H}_{\mathrm{MF},0}=\frac{1}{2}p^2-E_\mathrm{m}\cdot x
\end{split}
\label{7.5}
\end{equation}
acting on $\mathscr{H}=L^2(X)$, and
\begin{equation}
\bar{T}_{\mathrm{LF},\mathrm{cm}}(t)=\frac{1}{2}(p_\mathrm{cm})^2-E_\mathrm{cm}(t)\cdot x_\mathrm{cm},\quad
\bar{T}_{\mathrm{MF},\mathrm{cm}}=\frac{1}{2}(p_\mathrm{cm})^2-E_{\mathrm{cm},\mathrm{m}}\cdot x_\mathrm{cm}
\label{7.6}
\end{equation}
acting on $L^2(X_\mathrm{cm})$, where
$$E(t)=\pi E_{\bar{X}}(t),\quad E_\mathrm{m}=\pi E_{\bar{X},\mathrm{m}},\quad
E_\mathrm{cm}(t)=\pi_\mathrm{cm}E_{\bar{X}}(t),\quad
E_{\mathrm{cm},\mathrm{m}}=\pi_\mathrm{cm}E_{\bar{X},\mathrm{m}}.$$
Then $\bar{H}_\mathrm{LF}(t)$ and $\bar{H}_\mathrm{MF}(t)$ are represented as
\begin{equation}
\begin{split}
&\bar{H}_\mathrm{LF}(t)=\hat{H}_\mathrm{LF}(t)\otimes\mathrm{Id}+
\mathrm{Id}\otimes\bar{T}_{\mathrm{LF},\mathrm{cm}}(t),\\
&\bar{H}_\mathrm{MF}(t)=\hat{H}_\mathrm{MF}(t)\otimes\mathrm{Id}+
\mathrm{Id}\otimes\bar{T}_{\mathrm{MF},\mathrm{cm}}
\end{split}
\label{7.7}
\end{equation}
on $L^2(\bar{X})=\mathscr{H}\otimes L^2(X_\mathrm{cm})$.
Since $\bar{T}_{\mathrm{LF},\mathrm{cm}}(t)$
(resp. $\bar{T}_{\mathrm{MF},\mathrm{cm}}$) does not depend on
pair interactions, the understanding of the dynamics
of the system governed by $\hat{H}_\mathrm{LF}(t)$ (resp.
$\hat{H}_\mathrm{MF}(t)$) yields that of
the system governed by $\bar{H}_\mathrm{LF}(t)$ (resp.
$\bar{H}_\mathrm{MF}(t)$) immediately.

Now we will focus on $\hat{H}_\mathrm{LF}(t)$ and
$\hat{H}_\mathrm{MF}(t)$. Let
$\hat{U}_\mathrm{LF}(t,s)$ and $\hat{U}_\mathrm{MF}(t,s)$
denote the propagators
generated by $\hat{H}_\mathrm{LF}(t)$ and $\hat{H}_\mathrm{MF}(t)$, respectively. Then, in the same way as in \eqref{7.4},
the Avron-Herbst formula for $\hat{U}_\mathrm{LF}(t,s)$
\begin{equation}
\hat{U}_\mathrm{LF}(t,s)=\hat{\mathscr{T}}(t)\hat{U}_\mathrm{MF}(t,s)
\hat{\mathscr{T}}(s)^*;\quad
\hat{\mathscr{T}}(t)=e^{-ia(t)}e^{ib(t)\cdot x}
e^{-ic(t)\cdot p}
\label{7.8}
\end{equation}
holds, where
$$b(t)=\pi b_{\bar{X},\mathrm{os}}(t),\quad
a(t)=\int_0^t\left(\frac{1}{2}b(s)^2-E_{\mathrm{m}}\cdot c(s)\right)\,ds.
$$
By virtue of the Avron-Herbst formula \eqref{7.8}, the understanding
of $\hat{U}_\mathrm{MF}(t,s)$ yields that of $\hat{U}_\mathrm{LF}(t,s)$
immediately. Hence, we would like to focus on the dynamics
of the system governed by $\hat{H}_\mathrm{MF}(t)$ in the center-of-mass
frame.

The case where $\mathscr{E}_\mathrm{m}=0$
is that of the so-called AC Stark effect.
In the case where $\mathscr{E}_\mathrm{m}\not=0$,
there are some desirable results on the asymptotic completeness
of many body systems governed by such Hamiltonians. In fact,
in \cite{A1,A2},
the author obtained the result
of the asymptotic completeness for the system under consideration,
both in the short-range
and the long-range cases,
by introducing the Floquet Hamiltonian
\begin{equation}
K=D_t+\hat{H}_{\mathrm{MF}}(t)=D_t+\frac{1}{2}p^2-E_\mathrm{m}\cdot x
+\bar{V}(x+c(t))
\end{equation}
associated with $\hat{H}_{\mathrm{MF}}(t)$.
Because of $E_\mathrm{m}\not=0$,
\begin{equation}
A=\frac{E_\mathrm{m}}{|E_\mathrm{m}|}\cdot p
\end{equation}
can be taken as a conjugate operator for $K$ in the standard Mourre theory.
Here we emphasize that in the two-body case, M\o ller~\cite{Mo}
first proposed this
operator as a conjugate operator for $K$.

Now we will focus on the case where $\mathscr{E}_\mathrm{m}=0$.
Because of $E_\mathrm{m}=0$,
we will deal with an $N$-body Schr\"odinger operator with
time-periodic potentials
\begin{equation}
\hat{H}_{\mathrm{MF}}(t)=\frac{1}{2}p^2+V(t)
;\quad V(t)=\bar{V}(x+c(t))=\sum_{1\le j<k\le N}V_{jk}(t,r_j-r_k),
\label{7.11}
\end{equation}
where
$$V_{jk}(t,y)=
\bar{V}_{jk}(y+\tilde{e}_{jk}\bar{\bar{\mathscr{E}}}(t));
\quad\tilde{e}_{jk}=(q_j/m_j)-(q_k/m_k).$$
Here we note that $V_{jk}(t,y)$'s are $T$-periodic in $t$.
Now we assume $\mathscr{E}(t)\in C^2(\boldsymbol{R};\boldsymbol{R}^d)$ additionally,
and impose the following condition $(V_\mathrm{ST})_{\bar{\rho}}$ on $\bar{V}_{jk}$'s with $\bar{\rho}>0$:
\medskip

\noindent
$(V_\mathrm{ST})_{\bar{\rho}}$ $\bar{V}_{jk}(y)$, $(j,k)\in\mathscr{A}$,
belongs to $C^4(\boldsymbol{R}^d;\boldsymbol{R})$,
is independent of $t$, 
and satisfies the decaying conditions
\begin{equation}
|(\partial_y^\alpha
\bar{V}_{jk})(y)|\le C_{\alpha}\langle y\rangle^{-\bar{\rho}-|\alpha|},\quad
0\le|\alpha|\le4.
\label{7.12}
\end{equation}

Since
\begin{align*}
(\partial_tV_{jk})(t,y)={}&
\sum_{\ell=1}^d
\tilde{e}_{jk}\bar{\mathscr{E}}_{\mathrm{os},\ell}(t)
(\partial_{y_\ell}\bar{V}_{jk})(y+\tilde{e}_{jk}\bar{\bar{\mathscr{E}}}(t)),\\
(\partial_t^2V_{jk})(t,y)={}&
\sum_{\ell=1}^d
\tilde{e}_{jk}\mathscr{E}_{\mathrm{os},\ell}(t)
(\partial_{y_\ell}\bar{V}_{jk})(y+\tilde{e}_{jk}\bar{\bar{\mathscr{E}}}(t))\\
&+\sum_{\ell_2=1}^d\sum_{\ell_1=1}^d
\tilde{e}_{jk}\bar{\mathscr{E}}_{\mathrm{os},\ell_1}(t)\tilde{e}_{jk}\bar{\mathscr{E}}_{\mathrm{os},\ell_2}(t)
(\partial_{y_{\ell_1}}\partial_{y_{\ell_2}}\bar{V}_{jk})(y+\tilde{e}_{jk}\bar{\bar{\mathscr{E}}}(t)),
\end{align*}
we see that $V_{jk}(t,y)$'s satisfy
\begin{align*}
&|(\partial_y^\alpha
V_{jk})(t,y)|\le C_{0,\alpha}\langle y\rangle^{-\bar{\rho}-|\alpha|},\quad
0\le|\alpha|\le4,\\
&|(\partial_t\partial_y^\alpha
V_{jk})(t,y)|\le C_{1,\alpha}\langle y\rangle^{-\bar{\rho}-1-|\alpha|},\quad
0\le|\alpha|\le3,\\
&|(\partial_t^m\partial_y^\alpha
V_{jk})(t,y)|\le C_{m,\alpha}\langle y\rangle^{-\bar{\rho}-1-|\alpha|},\quad
2\le m\le 4,\quad
0\le |\alpha|\le 4-m.
\end{align*}
What we emphasize here is that
the decaying rate $O(\langle y\rangle^{-\bar{\rho}-1})$ of $(\partial_t^2
V_{jk})(t,y)$ is more moderate than 
$O(\langle y\rangle^{-\bar{\rho}-2})$.
If $V_{jk}(t,y)$ satisfies $(V_\mathrm{WR})$,
then the decaying rate of $(\partial_t^2
V_{jk})(t,y)$ is $O(\langle y\rangle^{-\rho_2})$.
But, if $\bar{\rho}\ge1$, then one can regard 
$V_{jk}(t,y)=
\bar{V}_{jk}(y+\tilde{e}_{jk}\bar{\bar{\mathscr{E}}}(t))$
as a time-periodic potential satisfying
the condition $(V_\mathrm{WR})$ with $\rho_0=\bar{\rho}\ge1>0$,
$\rho_1=\bar{\rho}+1\ge2>1$ and $\rho_2=\rho_3=\rho_4=\bar{\rho}+1\ge2$.
Therefore, in the case where $N=3$, Theorem \ref{thm1.5}
is a direct consequence of Theorem \ref{thm1.1},
Corollary \ref{cor1.2}, Theorem \ref{thm1.3},
and Corollary \ref{cor1.4}.

\end{document}